# Agent-based simulations for protecting nursing homes with prevention and vaccination strategies

**Authors**

*Jana Lasser[1,2*], Johannes Zuber[4], Johannes Sorger[2], Elma Dervic[3], Katharina Ledebur[3], Simon David Lindner[3], Elisabeth Klager[5], Maria Kletečka-Pulker[5,6], Harald Willschke[5], Katrin Stangl[7], Sarah Stadtmann[7], Christian Haslinger[8], Peter Klimek[2,3*], Thomas Wochele-Thoma[5,7*]*

**Affiliations**

(1) Graz University of Technology; Institute for Interactive Systems and Data Science, Inffeldgasse 16C, 8010 Graz
(2) Complexity Science Hub Vienna, Josefstädterstraße 39, 1080, Vienna, Austria
(3) Medical University Vienna, Section for Science of Complex Systems; Center for Medical Statistics, Informatics and Intelligent Systems, Spitalgasse 23, 1090 Vienna, Austria
(4) Research Institute of Molecular Pathology, Campus-Vienna-Biocenter 1, 1030 Vienna, Austria
(5) Ludwig Boltzmann Institute for Digital Health and Patient Safety, Spitalgasse 23, BT86, 1090 Vienna, Austria
(6) University Vienna, Institut für Ethik und Recht in der Medizin, Spitalgasse 2-4, 1090 Vienna, Austria
(7) Caritas Erzdiözese Wien, Albrechtskreithgasse 19-21, 1160 Vienna, Austria
(8) Hygienefachkraft-unlimited, Oswaldgasse 14-22/2/11, 1120 Vienna, Austria

* Corresponding authors: Thomas.Wochele-Thoma@caritas-wien.at, peter.klimek@meduniwien.ac.at, jana.lasser@tugraz.at

## Abstract

Due to its high lethality amongst the elderly, the safety of nursing homes has been of central importance during the COVID-19 pandemic. With test procedures becoming available at scale, such as antigen or RT-LAMP tests, and increasing availability of vaccinations, nursing homes might be able to safely relax prohibitory measures while controlling the spread of infections (meaning an average of one or less secondary infections per index case). Here, we develop a detailed agent-based epidemiological model for the spread of SARS-CoV-2 in nursing homes to identify optimal prevention strategies. The model is microscopically calibrated to high-resolution data from nursing homes in Austria, including detailed social contact networks and information on past outbreaks.

We find that the effectiveness of mitigation testing depends critically on the timespan between test and test result, the detection threshold of the viral load for the test to give a positive result, and the screening frequencies of residents and employees. Under realistic conditions and in absence of an effective vaccine, we find that preventive screening of employees only might be sufficient to control outbreaks in nursing homes, provided that turnover times and detection thresholds of the tests are low enough. If vaccines that are moderately effective against infection and transmission are available, control is achieved if 80% or more of the inhabitants are vaccinated, even if no preventive testing is in place and residents are allowed to have visitors. Since these results strongly depend on vaccine efficacy against infection, retention of testing infrastructures, regular voluntary screening and sequencing of virus genomes is advised to enable early identification of new variants of concern.

# Introduction

Nursing homes and other long-term care facilities are the ground zero of the COVID 19 pandemic (Barnett & Grabowski, 2020). Around the globe, a disproportionate number of confirmed deaths has been attributed to nursing home residents. For instance, as of July 2020, nursing homes accounted for 37% of the 719 confirmed COVID deaths in Austria (52% of all female, 25% of all male fatalities) (BMSGPK (Hg.), 2020, p.). With 923 confirmed cases in nursing homes during this period of time, this results in a case fatality rate of 28%, in line with reported high case fatality rates in the age group above 80 years old (Dowd et al., 2020).

Due to this extreme severity, in most countries stringent non-pharmaceutical interventions have been suggested for nursing homes, such as bans on visitors, individual movement restrictions and other quarantine policies (Ouslander & Grabowski, 2020; Wang et al., 2020). COVID 19, therefore, severely affects the quality of life of all nursing home residents, not just the infected ones (Fallon et al., 2020).

The widespread availability of novel rapid testing procedures, e.g. antigen tests (Dinnes et al., 2020) or tests based on the RT-LAMP procedure (Park et al., 2020) might turn out to be a, important innovation to control the spread of COVID in nursing homes in the absence of effective vaccines. With cheap unit costs (orders of 1USD per test) and durations of less than an hour, such tests enable the design of "testing for mitigation" strategies. The aim of mitigation testing is to use widespread testing within a specific setting to quickly identify and isolate infectious individuals but, in contrast to diagnostic testing, not necessarily infected individuals which don't have a high enough viral load to infect others. As vaccines become available, nursing homes naturally have become priority targets for the vaccination of as many employees and residents as possible (Dooling et al., 2020; European Centre for Disease Prevention and Control, 2020).

At the moment, non-pharmaceutical measures, testing strategies and vaccines exist side-by-side. In the intermediate future, vaccines will most likely replace non-pharmaceutical measures and reduce the need for testing. Yet, at this point in time it is unclear how to design optimal mitigation strategies involving testing and vaccination. Moreover, a transition from purely non-pharmaceutical interventions over a mix of vaccines and interventions towards widespread vaccine proliferation is expected to occur for potential future immune escape variants of SARS-CoV-2 and any other pandemic threat yet to emerge, as vaccines will require time to be developed and rolled out on a large scale.

The design of optimal mitigation testing strategies for a specific facility like nursing homes is challenging due to the following and sometimes interrelated factors that determine the effectiveness of a given strategy (Hatfield et al., 2020): Next to the epidemiological contagion dynamics, the optimal testing strategy also depends on the structure of the contact networks of the employees and residents, personal protective and physical distancing measures already in place, as well as the characteristics of the test. Two of these test characteristics are of particular importance for mitigation testing, namely (i) the turnover time (time span between test and availability of test result) and (ii) the detection threshold (viral load necessary for a positive result). For instance, RT-PCR tests, the current gold standard, have a turnover time of one or two days and a detection threshold that is typically much lower than the threshold above which an infected individual becomes infectious. Antigen tests have a turnover of less than an hour but a substantially higher detection threshold than PCR tests. Finally, RT-LAMP tests combine a same-day-turnover with a low detection threshold. For all these tests, sensitivity and specificity are close to 100% above the corresponding detection thresholds (Aziz et al., 2020; Kellner et al., 2020; Liotti et al., 2020).

As vaccinations become more widely available, nursing homes increasingly face the situation where parts of their residents and employees are vaccinated. First evidence shows that vaccinations can be highly effective in preventing severe courses of the disease (Aran, 2021), infection (Amit et al., 2021; Dagan et al., 2021; Hall et al., 2021; Jones et al., 2021; Shrotri et al., 2021) and even in reducing onward transmission (Harris et al., 2021; Levine-Tiefenbrun et al., 2021). Given the impact of measures such as a reduction in visits and physical distancing on the mental and emotional wellbeing of residents (O'Caoimh et al., 2020), ethical questions concerning the necessity of non-pharmaceutical interventions alongside vaccinations arise. In the context of substantial vaccine hesitancy among employees (Gharpure et al., 2021; Kwok et al., 2021), an assessment of the necessary level of non-pharmaceutical intervention measures given different levels of vaccination prevalence and effectiveness is crucial.

Here, we aim to design optimal mitigation testing and vaccination strategies for nursing homes for different testing technologies and levels of vaccination prevalence by means of network-based epidemiological modelling (Pastor-Satorras et al., 2015). In particular, we use a SEIRX model that is calibrated with individual-level data from actual Austrian nursing homes. Within this model, individuals are initially in a susceptible state (S). After exposure (E), they turn infectious (I). Depending on the test technology, individuals can be tested positive either before or after turning infectious. Vaccinations lower the chance of becoming infected after exposure to another infected individual and of transmission in case of infection of a vaccinated individual. Infected individuals either recover (R) or they are identified by a test and isolated (X). Individual-based epidemiological models with a similar structure have already been used to investigate the spread of SARS-CoV-2 in nursing homes under different prevention measures such as routine testing (Holmdahl et al., 2021; Nguyen et al., 2020; Roselló et al., 2021; Smith et al., 2020), and the combination of testing with vaccination strategies (Gómez-Vázquez et al., 2021; Kahn et al., 2021; Love et al., 2021; Vilches et al., 2021). Overall, these previous studies have shown that routine testing needs to be frequent enough to mitigate outbreaks and that this frequency can be reduced in proportion to vaccination rates in employees and residents. Contagions occur on a network of social contacts (Wasserman & Faust, 1994) that is modeled after the actual living conditions in a nursing home (see fig. 1) and fully calibrated using observational outbreak data (see SI note 3). We consider residents and employees with different types of social interaction such as shared rooms and lunch tables, or living and working in the same ward of a nursing home. Furthermore, we use infection data from four recorded outbreaks to calibrate the transmission risk associated with different types of interactions. Simulations were calibrated using data of outbreaks involving the wild-type SARS-CoV-2 strain that was dominant in Austria in spring 2020. Since then, the B.1.1.7 variant has become dominant, which is reported to have a 50% increased transmissibility (Davies et al., 2021; Fort, 2021; Institute of Social and Preventive Medicine, 2021; Leung et al., 2021; Statens Serum Institut, 2021). We model this variant in all results presented in this work by increasing the transmissibility that was calibrated using wild-type outbreak data by 50%. For results of outbreaks with the wild-type strain see SI note 6.

Mitigation testing strategies are parameterized by test technology and screening frequency. We consider RT-PCR, antigen, and RT-LAMP tests as well as their specific turnover times and detection thresholds. Furthermore, the strategies are determined by the frequencies by which all residents and employees are screened, respectively. Vaccination prevalence is parameterized by the ratio of vaccinated employees and residents. We assume that vaccinations are 60% effective in preventing infection and 30% effective in preventing transmission. These are conservative estimates of the vaccine efficacies reported three or more weeks after the first dose of the BNT162b2 (Biontech-Pfizer) and ChAdOx1 nCOV-19 (AstraZeneca) vaccines, which are currently most prevalent in Austrian nursing homes (Amit et al., 2021; Dagan et al., 2021; Hall et al., 2021; Harris et al., 2021; Jones et al., 2021; Levine-Tiefenbrun et al., 2021).

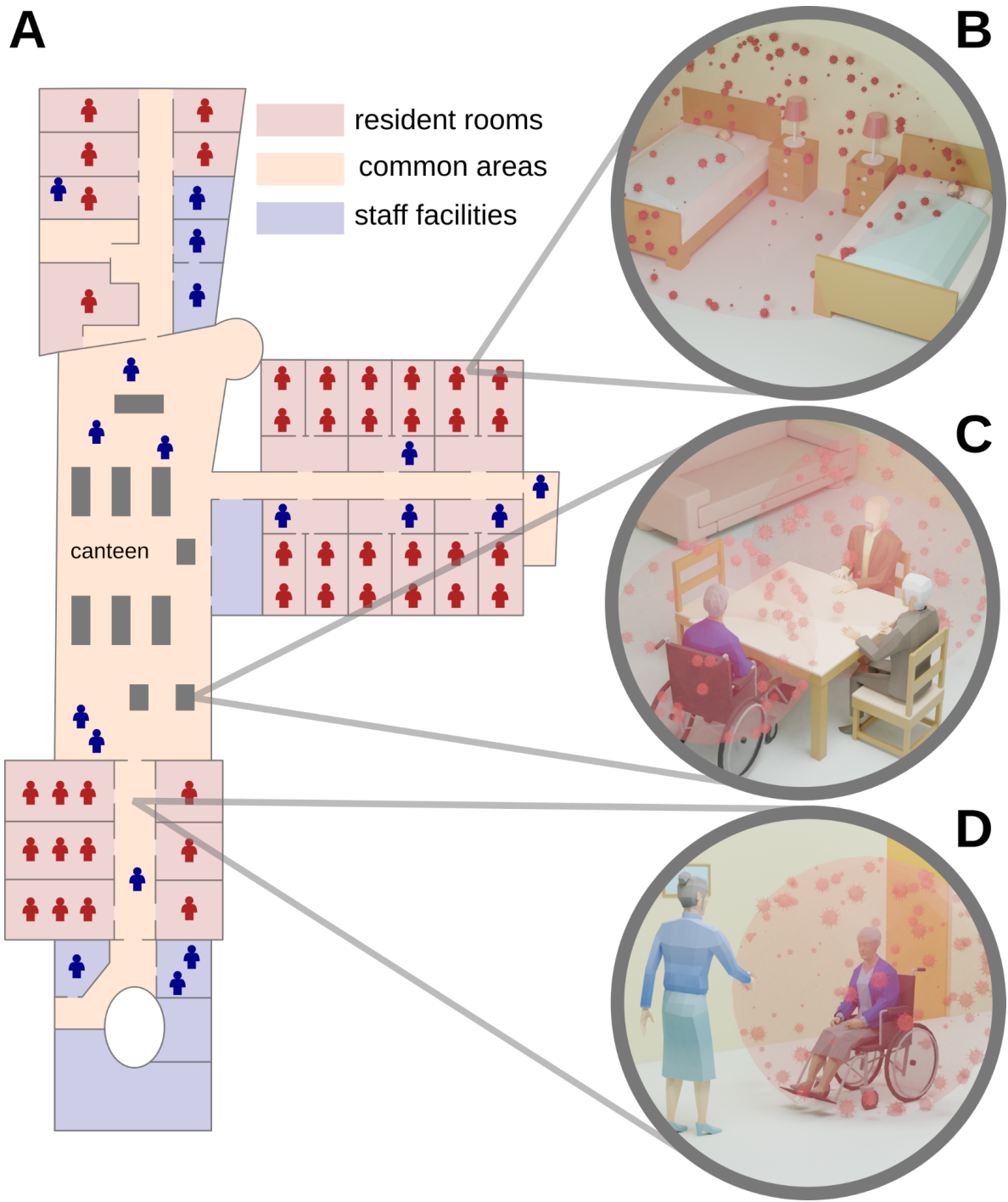

*Figure 1: Living conditions in a ward of an Austrian nursing home. (A) Simplified floor plan of the ward with resident rooms (red), common areas (orange) and staff facilities (blue). The ward houses up to 42 residents (red figures), is staffed by 18 employees (blue figures) and corresponds to the homes described in case studies 2 and 3 (see SI note 5). Contact networks for the simulations were extracted from such floor plans and information about shared tables in the canteen. (B) Rooms: up to two residents share a room and up to two rooms share a bathroom. (C) Shared table: up to six residents share a table during joint meals. (D) Shared common areas: residents living in the same ward of the home can move freely within the hallways, canteen and other common areas and regularly meet other residents. Spread of the virus by means of aerosols (Santarpia et al., 2020) is indicated as red clouds.*

For a given prevention scenario, we compute the expected distributions of outbreak sizes, i.e., the average number of infected residents after introducing a single index case into the home and simulating the ensuing outbreak. In particular, we assume that index cases are introduced either through an employee or a resident and differentiate between the two situations in the reporting of our results. The former reflects a situation where visitors are not allowed and resident's movement is restricted. The latter reflects a situation where residents can introduce an infection into the home, for example through contacts with visitors. In the absence of vaccinations, the optimal testing strategy is then identified as the test technology and test frequency for residents and personnel that minimizes the outbreak size for a given number of tests being performed. If an index case causes fewer than one infection among residents, we label the situation as "controlled". For scenarios in which a number of employees and residents are vaccinated, we investigate how much testing is necessary to achieve the same level of security as the optimal testing strategy.

## Results

We simulate epidemic spread in a nursing home in a range of different scenarios: (i) introduction of index cases through either employees or residents, (ii) different testing technologies used for preventive screening, (iii) different intervals for the preventive screens, (iv) different prevalence of vaccinations and (v) combinations of different testing strategies and vaccinations.

We report the outbreak size as the mean number of follow-up cases among residents caused by a single index case over the course of the simulation, and its standard deviation. We chose to report outbreak size among residents as residents are the most vulnerable group due to the high lethality of COVID-19 in the > 80 years old age group (Dowd et al., 2020).

In addition, we report the average number of transmissions from the index case $R_{eff}$, to approximate the reproduction number $R_0$ of the virus in the model system (Delamater et al., 2019). For scenarios that implement some sort of testing strategy (TTI and/or preventive screening), we also report the number of tests per day per person needed to implement the testing strategy (test rate), where the number of tests is the sum of the tests used for diagnostic testing and preventive screening.

### Effectiveness of test-trace-isolate

In the absence of non-pharmaceutical interventions or containment measures and infection with the strain B.1.1.7, $R_{eff}$ = 2.25 ± 1.86 if an employee is the index case and $R_{eff}$ = 2.97 ± 2.23 if a resident is the index case. Mean outbreak sizes are 25.2 ± 13.9 if an employee is the index case and 26.8 ± 11.7 if a resident is the index case. In a scenario in which only TTI is implemented, our model yields reproduction numbers of $R_{eff}$ = 1.90 ± 1.69 if an employee is the index case, and $R_{eff}$ = 1.82 ± 1.73 if a resident is the index case and mean outbreak sizes of 13.8 ± 11.9 and 12.6 ± 11.7 for employee and resident index cases, respectively.

## Effectiveness of different testing strategies

For the assessment of the merit of different testing technologies and strategies, we report our main results as outbreak-heatmaps in fig. 2 for scenarios in which employees (top row) or residents (bottom row) present as index cases. In these heatmaps, rows and columns indicate the preventive screening frequency for residents and employees, respectively.

The mean outbreak sizes for a given test strategy (screening frequencies) are colour-coded in the cells. As shown in the outbreak-heatmaps for the different scenarios displayed in fig. 2, mean outbreak sizes range between 0.1 ± 0.5 and 14.0 ± 11.9, depending on the scenario. Higher screening frequencies and lower test turnover times always reduce the size of outbreaks. Intuitively, prioritising the agent group that is more likely to introduce index cases in the preventive screening strategy considerably reduces the size of outbreaks. The lowest outbreak sizes of 0.1 ± 0.5 are achieved if index cases are predominantly introduced by employees and residents and employees are tested three times a week with PCR tests with same-day turnover. The highest outbreak sizes of 14.0 ± 11.9 are recorded if no regular preventive screening happens and an employee is the index case. Nevertheless, it is noteworthy that only reactive diagnostic testing with PCR tests with a two-day turnover is sufficient to contain outbreak sizes (i.e., the infection is stopped in the majority of cases before all 35 residents are infected) in these scenarios.

In a realistic scenario, where index cases are introduced by employees (because residents are not allowed to have visitors), employees are screened twice per week while residents are not screened at all, and PCR tests achieve a turnover time of one day, RT-LAMP and PCR tests perform similarly well, with outbreak sizes of 1.5 ± 4.6 for an employee index case and 6.1 ± 7.7 for a resident index case (RT-LAMP) and 1.5 ± 4.7 for an employee index case and 6.1 ± 7.6 for a resident index case (PCR). Antigen tests perform considerably worse, with outbreak sizes of 3.1 ± 6.3 and 7.1 ± 8.3 for employee and resident index cases, respectively. Nevertheless, if the logistics around PCR tests are optimized such that a same-day turnover can be achieved, PCR tests outperform all other tests with outbreak sizes of 0.9 ± 3.7 (employee) and 5.6 ± 7.3 (resident). If employees are screened only once a week, outbreak sizes increase to 3.0 ± 6.6 and 7.0 ± 8.4 (PCR, same-day turnover), 4.2 ± 7.7 and 7.7 ± 8.9 (RT-LAMP), and 5.8 ± 9.0 and 8.9 ± 9.5 (antigen) for employee and resident index cases, respectively. $R_{eff}$ for each scenario is shown in SI fig. A7. For mean and median outbreak sizes alongside the 10th and 90th percentile outbreak size range, $R_{eff}$, and test rates for each of the three test technologies with same-day turnover, as well as one-day and two-day turnover for PCR tests, screening of employees never, once, two times or three times a week and screening of residents never or once a week see SI note 4[1]. If, in addition to the testing, employees wear protective equipment, this offsets the increased transmissibility of the B.1.1.7 variant and reduces outbreak sizes to sizes comparable to the ones observed for the wild-type (see SI notes 6 and 7).

The base rate of tests needed for reactive diagnostic testing in all scenarios is approximately 0.003 ± 0.002 tests per day per person. Implementation of regular preventive screening of only employees two times a week increases this rate to 0.09 ± 0.013, independent of test technology and index case. Implementation of regular screening twice per week for only residents increases the rate to 0.18 ± 0.02. Implementation of screens three times a week increases the rate to 0.27 ± 0.03. The test rate for each scenario is visualised in SI fig. A8.

Next to the detection threshold, another important parameter for regular testing is the test turnover time, i.e., the time it takes for tests to return results. The turnover time determines how quickly contact tracing can start and contacts of infected people are quarantined. In fig. 3 we investigated different turnover times between same-day and two days for PCR tests in the same model setting as described above. These turnover times are realistic for PCR tests, especially in congested testing systems close to overload, as was the case in Austria in spring 2020 (Scherndl, 2020). In the same scenario as described above (employee screening twice per week, employee as index case), and in case of PCR tests with a turnover rate of two days, mean outbreak sizes increase significantly to 3.0 ± 6.5 (employee index case) and 6.9 ± 8.2 (resident index case) – very similar to the performance of the less accurate antigen tests. Only if employees are screened three times a week and residents are screened at least twice a week, outbreak sizes drop to 0.5 ± 1.4 (employee index case) and 0.9 ± 1.9 (resident index case). PCR tests with same-day turnover are obviously the best option and reduce outbreak sizes to 0.8 ± 2.0 (employee index case) and 0.9 ± 2.0 (resident index case), even if employees and residents are screened only once a week. $R_{eff}$ and test rates for different PCR test turnover times are visualized in SI figures A9 and A10.

## Effectiveness of vaccinations

To assess the impact of vaccination prevalence on outbreak sizes, we simulate scenarios with different vaccination rates for employees and residents. Additionally, nursing homes implement only TTI, i.e., testing and isolating symptomatic agents and their contacts, but no preventive tests. For a scenario in which vaccinations are scarce and employees are prioritised for vaccinations (50% of employees are vaccinated), outbreak sizes range from 9.4 ± 9.8 (employee index case) to 7.0 ± 9.3 (resident index case). If residents are prioritised instead, outbreak sizes are reduced to 3.1 ± 4.1 and 3.1 ± 4.0, respectively. In a situation in which the vaccine is broadly available but employees are hesitant to get vaccinated (50% of employees and 90% of residents vaccinated), outbreak sizes are reduced to 0.2 ± 0.5, independent of the index case. Interestingly, further increasing the ratio of vaccinated employees only slightly reduces the number of follow-up cases among residents, compared to the scenario in which employees show vaccine hesitancy: If 90% of residents and employees are vaccinated, outbreak sizes are 0.1 ± 0.5, independent of the index case. In fig. 4, we show outbreak sizes for a wide range of vaccination rates. In general, if 80% or more of the resident population is vaccinated, outbreak sizes stay <1, independent of the number of vaccinated employees and outbreaks are controlled. In SI note 6 we report similar simulation results for the strain which was dominant in Austria during data collection. For this less infectious strain, it is sufficient if 60% or more of the resident population is vaccinated to keep outbreak sizes below one. In SI note 7 we report the same results for a scenario in which B1.1.7 is introduced to the nursing home and employees are required to wear protective gear. We find that in this scenario a vaccination rate of 70% amongst employees is sufficient to control the spread of the more transmissible variant. If a high number of residents is vaccinated, vaccinating an increasing number of employees is still beneficial, since it further lowers the number of infected residents towards zero (see SI fig. A11).

---

[1] Tables are also available in the corresponding data repository at https://doi.org/10.17605/OSF.IO/HYD4R

To assess the impact of vaccination rates on the merits of different testing strategies, we simulate scenarios in which parts of the nursing home population are vaccinated while different testing strategies are employed at the same time. We specifically analyse three scenarios: (i) employees are subject to frequent preventive screening (two times per week) with cheap, fast but insensitive antigen tests on top of TTI, (ii) employees are subject to frequent preventive screening with expensive, fast and very sensitive PCR tests on top of TTI, and (iii) there are no preventive tests and homes rely solely on vaccines and TTI to prevent the spread of infections.

If 50% of the resident population is vaccinated, same-day turnover antigen and PCR tests perform similarly well, with outbreak sizes of 0.5 ± 1.6 (antigen, employee index case) and 1.6 ± 2.5 (antigen, resident), and 0.9 ± 3.7 (PCR, employee) and 1.3 ± 2.1 (PCR, resident). If no residents are vaccinated, only preventive screening with highly sensitive tests with a fast result turnover can keep outbreak sizes in check: if employees are not vaccinated as well, we observe outbreak sizes of 0.9 ± 3.7 (employee) and 5.6 ± 7.3 (resident) if employees are screened twice a week with same-day turnover PCR tests. Using the same test setup, if 50% of employees are vaccinated, outbreak sizes are only mildly reduced, with 0.8 ± 3.3 and 5.2 ± 6.9 for employee and resident index cases, respectively. If 50% of employees and 90% of residents are vaccinated, additional testing only slightly reduces outbreak sizes further, since outbreak sizes are already very low: no testing yields outbreak sizes of 0.2 ± 0.5 (independent of the index case), testing employees twice a week with antigen tests yields outbreak sizes of 0.0 ± 0.3 and 0.1 ± 0.4 and testing employees twice a week with PCR tests yields outbreak sizes of 0.0 ± 0.2 and 0.1 ± 0.5 for employee and resident index cases, respectively – ensuring almost a complete stop of any transmissions. We show outbreak size distributions for all three selected testing strategies and all four vaccination scenarios in fig. 5. In SI fig. A12 results for all previously discussed testing frequencies and vaccination scenarios are shown.

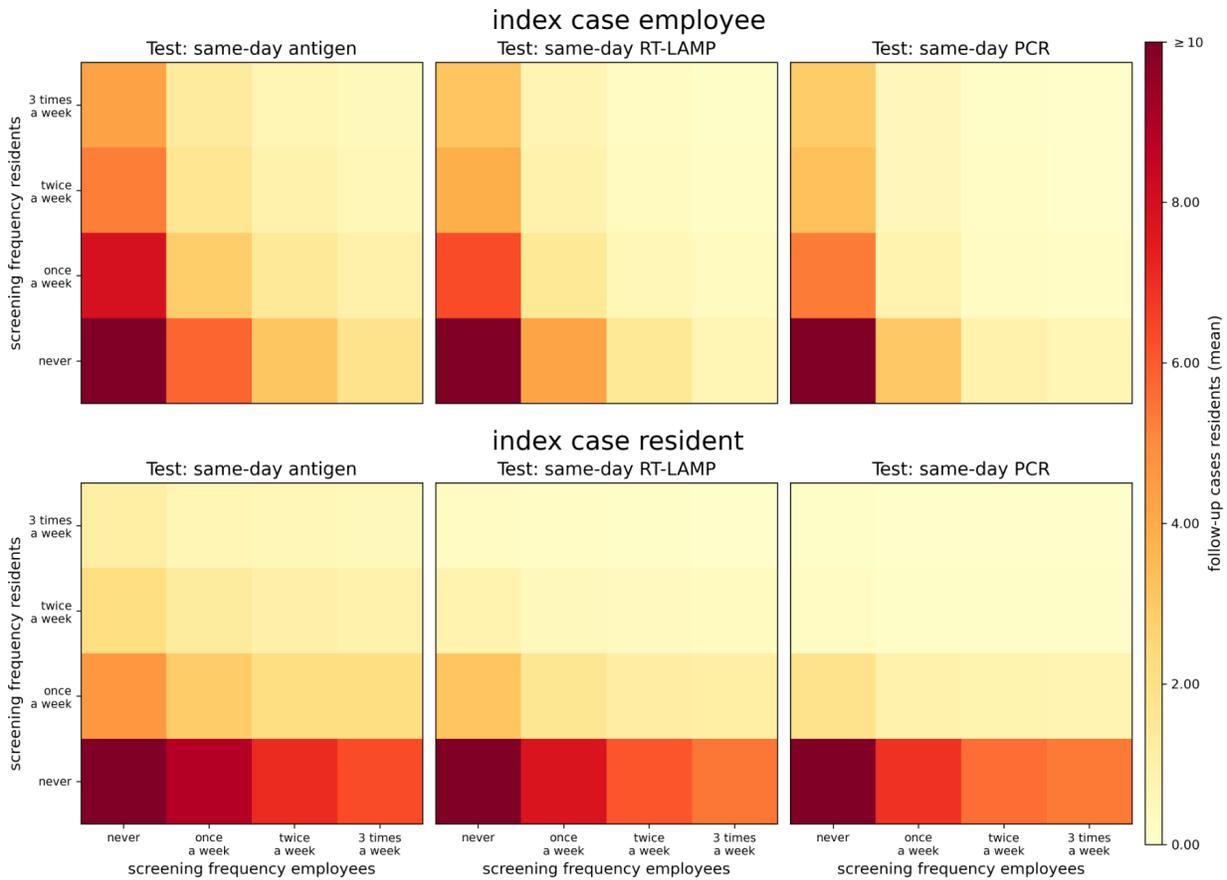

*Figure 2: effectiveness of different test technologies.* *Mean outbreak sizes for a range of testing scenarios in nursing homes, investigating different testing technologies with their characteristic turnover times. Index cases: in the first row, infections are introduced by personnel, in the second row, by residents (typically after seeing visitors). Testing technology: in the first column, antigen tests with same-day turnover are used, in the middle column, RT-LAMP test with same-day turnover, and in the third column, PCR tests with same-day turnover. Preventive screening frequency: in each heatmap, preventive screening frequency of employees (x-axis) and residents (y-axis) is varied between no screening and one screening every two days. Results represent mean values of 5000 simulation runs per unique configuration.*

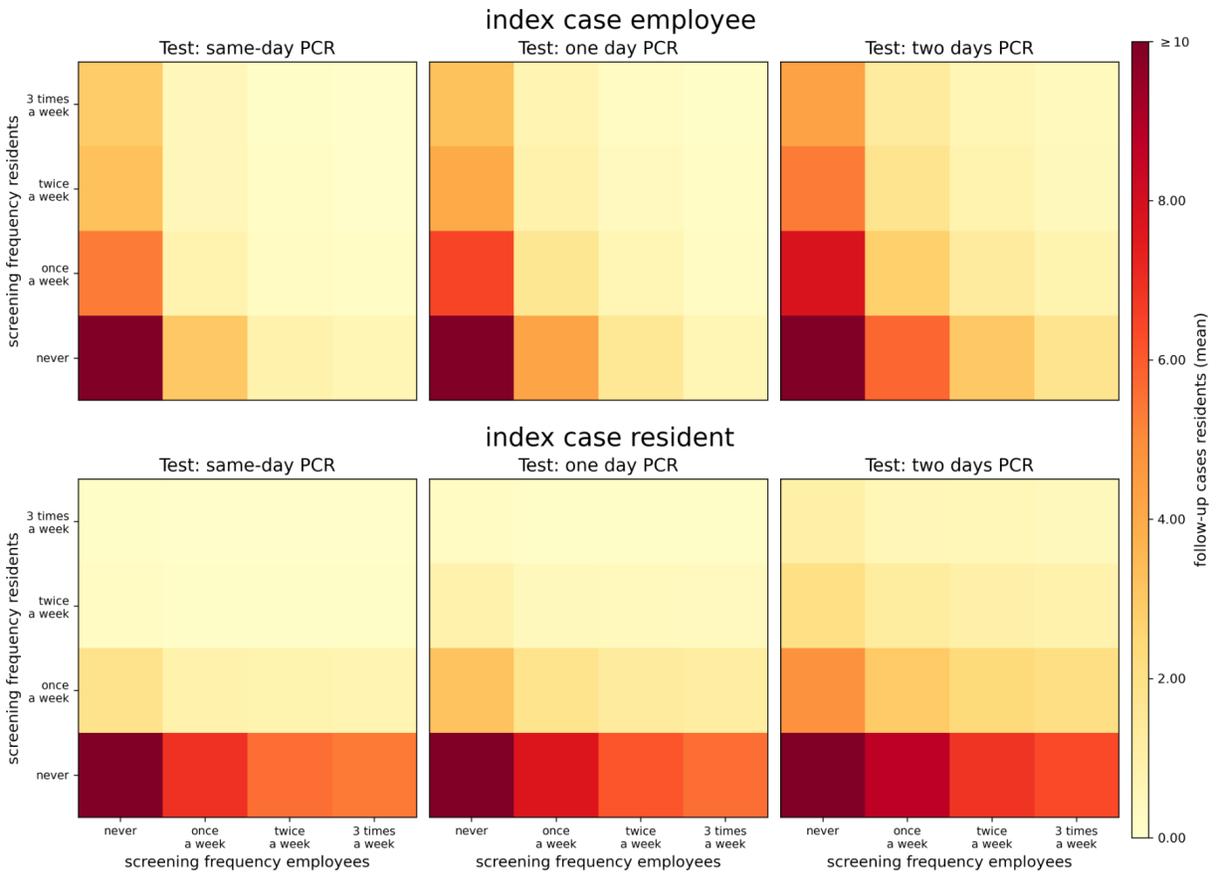

*Figure 3: Influence of result turnover times. Mean outbreak sizes for a range of testing scenarios in nursing homes investigating different test turnover times. Index cases: in the first row, infections are introduced by personnel, in the second row, by residents (resembling visitors). Test turnover time: in all scenarios, PCR tests are used. In the first column, tests have same-day turnover, in the middle column, tests have one-day turnover, and in the third column, tests have two-day turnover. Preventive screening frequency: in each heatmap, preventive screening frequency of employees (x-axis) and residents (y-axis) is varied between no screens and one screen every two days. Results represent mean values of 5000 simulation runs per unique configuration.*

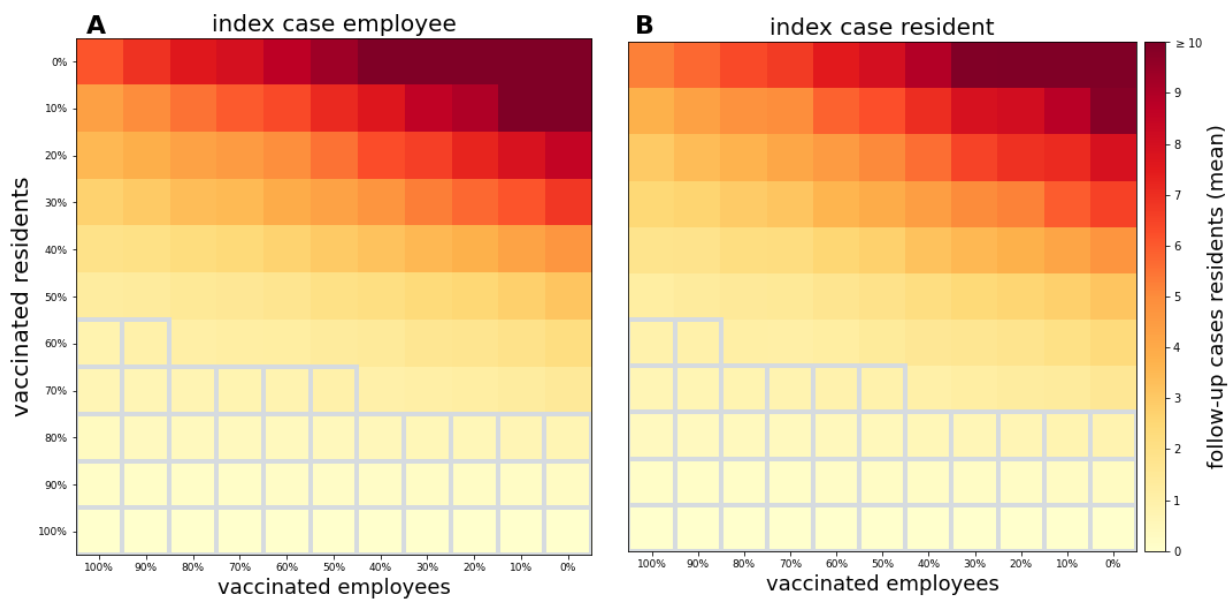

*Figure 4: Outbreak sizes for different ratios of vaccinated employees and residents.* Outbreak sizes are indicated from low (yellow) to high (red) for **(A)** employee index cases and **(B)** resident index cases. Vaccination ratios for which the mean number of resident follow-up cases is < 1 are indicated with grey borders. Outbreak sizes for each combination of (employee, resident) vaccination ratio are averages over 5000 randomly initialised simulation runs each. In addition to vaccinations, the model also implements TTI.

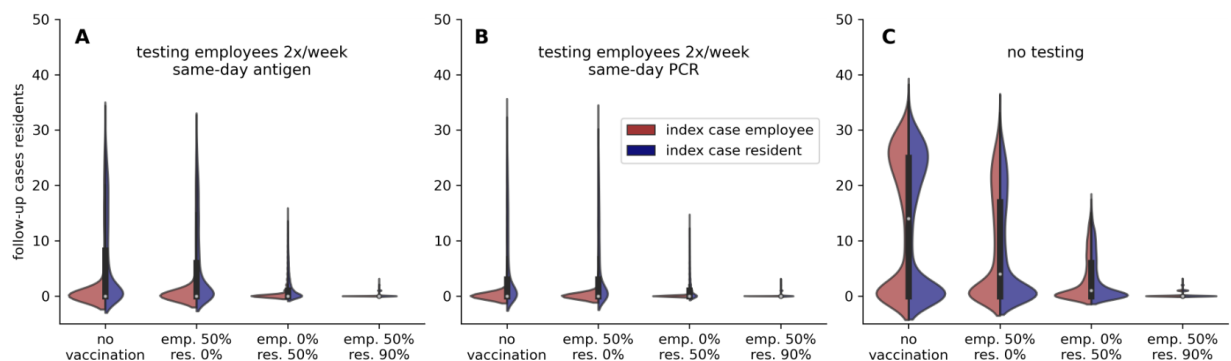

*Figure 5: Distributions of the number of infected residents for different vaccination scenarios and testing strategies.* The left (red) part of the violins indicates employee index cases, the right (blue) part of the violins indicates resident index cases. **(A)** The testing strategy consists of TTI and preventive testing of employees two times a week with antigen tests with a same-day turnover. **(B)** The testing strategy consists of TTI and preventive testing of employees two times a week with PCR tests with same-day turnover. **(C)** The testing strategy consists of TTI only. For every testing strategy, four vaccination scenarios (no vaccination, 50% of employees, 50% of residents, 50% of employees and 90% of residents) are shown. The plot shows distributions of outbreak sizes from 5000 randomly initialised simulation runs per testing strategy and vaccination scenario.

## Discussion

In this study, we aimed to design optimal prevention measures for nursing homes by means of an agent-based epidemiological model. The model has been calibrated to individual-level contact networks modeled after the living conditions in an actual nursing home and the epidemic dynamics have been calibrated to recorded outbreak events therein. By considering three different testing technologies, we identified testing frequencies for residents and employees that result in the minimal average outbreak size at a given maximal capacity to perform tests. By considering three realistic vaccination scenarios together with various testing strategies, we identify the optimal prevention strategy depending on vaccine availability.

In brief, our simulations confirm that reactive screening of residents and employees combined with quarantine of close contacts of positive cases (test-trace-isolate, TTI) according to current recommendations for nursing homes (Dumyati et al., 2020) limits outbreak sizes in nursing homes to approximately 13 follow-up cases per index case in situations where the more transmissible B.1.1.7 strain is dominant. Compared to the baseline TTI scenario, more frequent testing, faster turnover of the test results, and a lower detection threshold for the tests are always beneficial to reduce the average outbreak size. However, the extent to which these individual factors contribute to an outbreak size reduction is non-trivial.

For scenarios in which contacts between residents and visitors or other external people are drastically reduced and we can assume that infections are introduced into the home solely through employees, we find that the marginal effectiveness (outbreak size reduction per performed test) of personnel screening strongly outperforms the marginal effectiveness of resident screening. This means that screening only the personnel two or three times per week can have an equal or even higher protective effect than screening all residents once per week. However, in cases where infections can be introduced through residents, i.e. residents frequently have visitors, these visits take place without other precautionary measures and the visitors have a high risk to be infected themselves, also screening of residents becomes increasingly important.

All of our results are strongly sensitive to the turnover time between the test being performed and the arrival of the test result. Reducing this timespan from two days to a same-day-turnover might reduce the average outbreak size from around 3.0 ± 6.5 follow-up cases per index case to 0.9 ± 3.6 cases, in a scenario where personnel is regularly tested twice a week with PCR tests and index cases are introduced by employees. In a more realistic scenario, where employees are screened twice a week and PCR tests achieve a turnover of one day, RT-LAMP tests outperform both PCR tests and antigen tests due to their lower detection threshold and faster turnover. Depending on the scenario, antigen tests yield between 0.1 ± 0.3 and 0.4 ± 0.8 false negative tests (per outbreak), which has dramatic consequences, since the false-negative person is not isolated and is able to freely spread the infection. As RT-LAMP tests are accurate, fast and comparatively cheap, they perform best amongst the testing technologies for regular preventive screening strategies considered in this work. If same-day turnover of PCR tests can be achieved, PCR tests perform similar to RT-LAMP tests but have the added benefit of allowing for pool-testing due to their superior detection thresholds (Ben-Ami et al., 2020).

In our model we do not consider how convenient it is to be tested with a given method. Many PCR and antigen tests require a throat swab that can become quite a nuisance, particularly if employees have to undergo this procedure twice a week. In addition, staff in Austrian nursing homes reports that elderly, often demented residents do not respond well to the often-painful testing. On the other hand, RT-LAMP and PCR tests can be performed by gargling a tasteless liquid which might be beneficial for long-term compliance with the testing regimen.

To simulate the effect that varying amounts of vaccinated residents and employees have on the infection dynamics in homes, we assumed vaccine efficacies of 60% to prevent infection and of 30% to prevent transmissions. Given these rather conservative estimates of vaccine efficacy, if 80% of residents are vaccinated an index case in a nursing home leads to less than one follow-up case, even if no employees are vaccinated and homes only perform diagnostic testing and isolation of symptomatic agents and their contacts. In a scenario where 90% of residents are already vaccinated, further increasing the ratio of vaccinated employees is still beneficial, as it further reduces the average number of infected residents: if 90% of employees are vaccinated, the number of infected residents is reduced to 0.13 ± 0.39, as compared to 0.17 ± 0.46 if only 50% of employees are vaccinated (employee index case).

If high numbers of residents are vaccinated, preventive testing only slightly reduces outbreak sizes further and there is no significant difference between the introduction of the index case by an employee or resident. This means that if the resident population is vaccinated to a high degree (more than 80%), expensive and logistically complex preventive testing schemes at scale can be discontinued without risking large outbreaks, and there is no justification to further disallow visits in nursing homes. Nevertheless, this result depends on the efficacy of vaccines next to our other modelling assumptions. If at some point in the future a strain emerges that escapes the immune response elicited by vaccines, only regular screening with fast and highly sensitive tests is able to keep outbreak sizes relatively low. We therefore strongly advise to retain testing capacities at scale, even if they are not needed at a given point in time. In addition, vaccines are highly effective in reducing severe and symptomatic courses of the disease (Aran, 2021; Polack et al., 2020; Voysey et al., 2021). Therefore, in situations in which high numbers of employees and residents are vaccinated, purely symptomatic testing within a TTI strategy might lead to a very low number of tests. A point can be made to keep up voluntary preventive screening in nursing homes, combined with sequencing of samples from positive tests to facilitate the identification of and reaction to novel variants of concern. Nursing homes lend themselves for such screening and sequencing activity, since they already have established testing infrastructure and medically trained personnel, and new variants of concern are likely to quickly find their way into nursing homes.

Our model has several limitations: The contact networks in our simulations are models, based on the architecture of nursing home wards, insights of practitioners, and data about occupancy, shared rooms and shared lunch tables at the time of outbreaks. They are not based on measurements of actual contacts between persons in the homes. In addition, we had to make assumptions about the contact patterns between employees and residents as well as employees and other employees. We assume that the contact patterns do not change depending on the testing strategy (e.g., contacts to take swab samples). While these assumptions are based on insights from practitioners, these contact patterns are not based on observational data. We do not model the potential replacement of isolated employees by new employees. Neither do we include the possibility of dying from the disease. Therefore the number of agents in the simulation stays constant throughout the simulation: no agents leave the system and no new agents enter. As simulation durations are rather short (on the order of 6 weeks in a simulation where only TTI is employed), these simplifications seem warranted. In addition, as the number of agents is rather small, finite size effects will occur, limiting the size of larger outbreaks. Furthermore, though most model parameters have been calibrated using individual-level observational data, some simplifying assumptions had to be made: For instance, all contacts of a given type (e.g., room mates) are assumed to have the same transmission probability, independent of other environmental factors like occasional ventilation. The viral load dynamics reported in the literature that are translated into a time-dependent transmission risk in the model are approximated in a piece-wise linear way. One could think of test strategies, in which the time resolution of our model would need to be increased from days to hours to more accurately assess their effectiveness. We also do not differentiate between agents that have received a single or several vaccination doses and we do not consider the time-dependence of vaccination efficacy. Finally, there are first reports from Austrian nursing homes (unpublished results) that indicate a potentially lower immune response in elderly people. If these results are substantiated, our already conservative assumptions regarding vaccine efficacy in nursing home residents might have to be reconsidered.

In summary, our results indicate that personnel screening twice a week with RT-LAMP or PCR tests can severely reduce outbreak sizes even without a screening of residents and vaccines, provided that other precautionary measures are taken for social interactions of the residents. Given the same testing strategy, antigen tests provide less protection than RT-LAMP or PCR tests due to their higher detection threshold. On the other hand, vaccines that are moderately effective in preventing infection and transmission render other prevention measures obsolete if at least 80% of residents are vaccinated. Nevertheless, retainment of testing infrastructure, voluntary screening and regular sequencing of positive cases is still beneficial and is advised, in case novel virus variants emerge.

## Methods

We simulate the infection dynamics using an agent-based model. The model includes two types of agents (residents and employees) that live and work in nursing homes, respectively. Infections are introduced from outside the home either through an employee or a resident (see SI note 1 "Index cases"). In every simulation, a single index case is introduced and the ensuing outbreak simulated until no agents are exposed or infected any longer. This way we are able to investigate whether the measures a nursing home implements are sufficient to control an outbreak if an index case is introduced, independent of the prevalence of the disease in the general population.

Inhabitants have individual networks of social contacts. The contact network defines interactions between residents in one of three ways, in decreasing order of infection transmission risk: Two residents might have social contacts due to a shared room, a shared meal table, or a shared ward. While occasional contacts between different wards of the same facility are possible, we assume wards to function independently from each other. Introduction of an infection from another ward therefore resembles the introduction of an index case from outside. This assumption is grounded in the efforts nursing homes undertake, to limit contacts of residents between individual wards and to specifically assign personnel to one ward only. The contact network used in our simulation is a model of social contacts in a nursing home ward, based on data about occupancy and staffing during a number of observed outbreaks in homes, and information from practitioners (see SI note 1 "Contact networks"). It is not based on direct measurement of contacts between agents. In our model, one ward includes 35 residents and 18 employees, resembling a typical ward in an Austrian nursing home (see SI note 5 for details). Only considering a single ward also makes the model independent of the general layout of the whole facility. As such, the model can be applied to facilities around the world that feature such a ward structure.

At the first day of a simulation, a random resident or employee is chosen to become the index case and the agent's state is set to "exposed". At every step (day) of the simulation, agents interact according to their interaction rules and infectious agents can transmit the virus to susceptible individuals. Depending on an agent's individual exposure duration, incubation duration, infection duration and probability to develop symptoms (see SI note 1 "Agents"), each agent is in one of twelve states: susceptible (*S*), exposed (*E*), infectious presymptomatic (*I*), infectious asymptomatic ($I_1$), infectious symptomatic ($I_2$) or recovered (*R*) (depicted in fig. 6 **B**). Each of these states also exists in an isolated/quarantined (*X*) version). States *S*, *E*, *I* and *R* also translate to viral load, as depicted in fig. 6 **A**, which is important for the ability of different test technologies to detect an infection (see SI note 1 "test technologies"). Once an agent has become infected, the agent stays exposed for an average of 5 days, matching the latent time reported for SARS-CoV-2 (Lauer et al., 2020; Linton et al., 2020). After five days on average, agents become infectious and stay infectious for on average 11 days (Walsh et al., 2020; Wölfel et al., 2020).

The risk of transmission to a contact person is particularly high during the first days of the infectious phase and then decreases as the infection progresses (He et al., 2020; Walsh et al., 2020). Not all agents develop symptoms and the probability to develop symptoms depends on age (Poletti et al., 2021). We assume that the age of employees is uniformly distributed between 20 and 59 years. Therefore, employees have an average probability to develop symptoms of 26.46% (Poletti et al., 2021). For residents, we assume a probability to develop symptoms of 64.52%, corresponding to the value reported by Poletti et al. for people aged 80 and above. We do not consider age to be a relevant factor for susceptibility, since evidence for this effect is still inconclusive (Madewell et al., 2020). If an agent develops a symptomatic course of the disease, symptoms start to appear shortly after becoming infectious (He et al., 2020). Transmissibility of asymptomatic agents is reduced by 40% (Byambasuren et al., 2020). Vaccinations reduce the susceptibility by 60% (Amit et al., 2021; Dagan et al., 2021; Hall et al., 2021; Jones et al., 2021), and the transmissibility by 30% (Harris et al., 2021; Levine-Tiefenbrun et al., 2021). These are conservative estimates of the values reported in the literature for the vaccines BNT162b2 and ChAdOx1 nCOV-19 that are predominantly used in Austrian nursing homes.

In our model, each time-step (day) of the simulation is associated with an independent Bernoulli trial for disease transmission between susceptible and infectious agents given a contact (Laskowski & Moghadas, 2014; Mostaço-Guidolin et al., 2011):

$$P = 1 - [1 - \gamma\beta(1 - q_1(c))(1 - q_2(t))(1 - q_3)(1 - q_4)(1 - q_5)],$$

Where *β* is the transmission probability per person per day, calibrated to reflect the secondary attack rate for household contacts between adults of 28.3% (Madewell et al., 2020) (see SI note 3). The modifier $\gamma = 1.5$ reflects the 50% increase in transmissibility reported for the virus variant B.1.1.7 (Davies et al., 2021; Fort, 2021; Institute of Social and Preventive Medicine, 2021; Leung et al., 2021; Statens Serum Institut, 2021). The $q_i$ are reductions of transmission risk due to the various factors described above: $q_1(c)$ modifies transmission risk depending on contact type *c*, $q_2(t)$ reflects the reduction of transmission risk due to lower viral loads as the infection progresses, $q_3$ reflects reduced transmissibility due to asymptomatic presentation, and $q_4$ and $q_5$ represent reduced susceptibility and transmissibility due to vaccinations (see also SI note 1 "Transmission probability"). We consider the use of masks by employees in a separate model described below.

We calibrate $β$ and $q_1$ by means of an iterated grid search such that the transmission risk for close contacts reflects the household secondary attack rate, and such that outbreak sizes produced by our model correspond to observed outbreak sizes in nursing homes. All other $q_i$ are chosen to correspond to values reported in the literature. See SI note 1 for details on all aspects of model design, implementation and assumptions, SI note 2 for an overview of all model parameters and their sources and SI note 3 for more details about the calibration.

Exposed or infectious agents can be testable – meaning that their virus load is high enough for a given test to detect the infection – depending on the period of time they have already been infected and the test being used (see fig. 6 **A**). Different types of tests also have their specific turnover times (delay between making the test and knowing its result). Tests return positive or negative results depending on whether the agent was testable at the time of testing. For sake of simplicity, we assume the sensitivity and specificity of all simulated tests to be 100% for the time period in which they are able to detect an infection (see SI note 1 "Test technologies" for details).

Next to the transmission of the infection, we simulate containment measures (quarantine and isolation) and a testing and tracing strategy implemented by the nursing home to curb the spread of the virus. In the baseline scenario, without any preventive screening, only diagnostic testing takes place: Symptomatic cases are immediately isolated and tested using a PCR test with a two-day turnover time. Once a positive test result is returned, all close and intermediate contacts of the positive agent are immediately quarantined. Only residents can have close or intermediate contacts, namely being roommates or table neighbours. We summarize this strategy as "test-trace-isolate" or TTI (see SI note 1 "Intervention measures" for details).

In addition to the TTI scenario, we simulate several scenarios in which the nursing home implements a preventive screening strategy using different testing technologies, independent of reported positive cases. Preventive screens are conducted in the employee group, resident group or both groups once, twice or three times per week. To simulate the scenarios, a single index case is introduced either via an employee or via a resident and simulations are terminated if no agents are either exposed or infected anymore. We simulate scenarios with all possible combinations of preventive screening frequencies for three types of testing technologies: (1) the PCR test has a result turnover time of same-day, one day or two days and detects an infection at around one virus copy / µl or on average four days after transmission (i.e., one day before agents become infectious) and until the infection has subsided (Aziz et al., 2020). (2) The antigen test has same-day turnover and detects an infection two days after an agent has become infectious up until three days before an agent stops being infectious, due to the higher viral load needed for the test to yield a positive result (Liotti et al., 2020). (3) The RT-LAMP test also has same-day turnover and detects an infection for the same period of time an agent is infectious (Kellner et al., 2020).

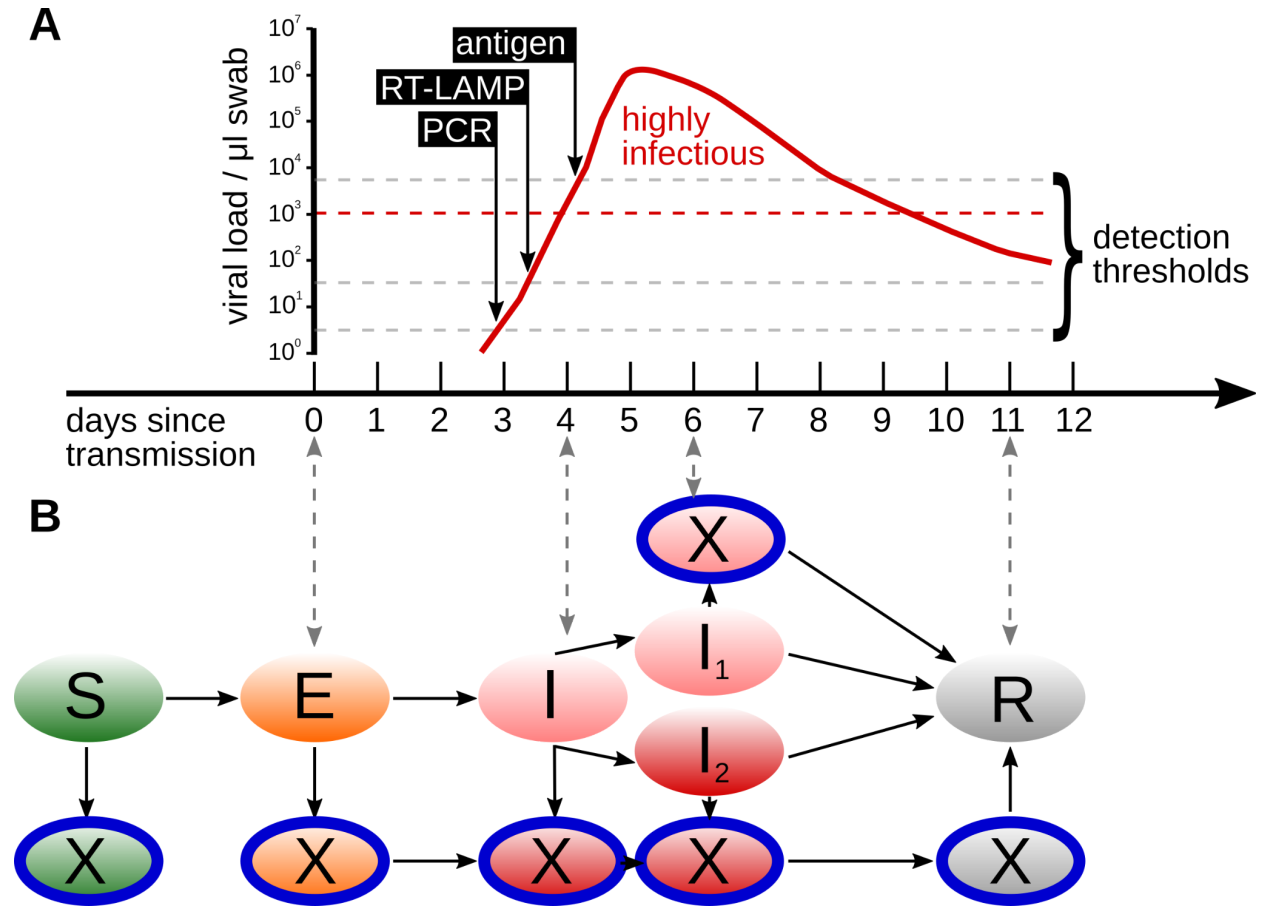

*Figure 6: Testability and agent states of the agent-based epidemiological model. A Illustration of viral load over time and detection thresholds of PCR, RT-LAMP and antigen tests reproduced after Kellner et al., 2020; Larremore et al., 2020; Wölfel et al., 2020: in our model, PCR tests can detect an infection one day before an agent becomes infectious, RT-LAMP tests on the day an agent becomes infectious and antigen tests one day after an agent becomes infectious. Individuals with > $10^3$ virus copies per µl swab are considered infectious (Wölfel et al., 2020). B Agents in the epidemiological model can be in the states (circles) susceptible (S), exposed (E), infectious presymptomatic (I), infectious asymptomatic ($I_1$), infectious symptomatic ($I_2$) and recovered (R). Possible state transitions are shown by arrows. Each of these states also exists in an isolated/quarantined version (X), preventing an agent from interacting with other agents. Transitions between states follow the individual agent's exposure durations, incubation times and infection durations.*

To model vaccinations, for a given vaccination ratio in an agent group (resident or employee), a number of agents from that group corresponding to the ratio are picked at random at the start of the simulation and assigned a vaccination state. Being vaccinated reduces both the probability for a successful transmission to a vaccinated and susceptible agent ($q_4$), and from a vaccinated but infected agent to another susceptible agent ($q_5$). We define four likely vaccination scenarios: (i) vaccines are scarce and residents are prioritised: 50% of residents and 0% of employees are vaccinated, (ii) vaccines are scarce and employees are prioritised: 0% of residents and 50% of employees are vaccinated, (iii) vaccines are ubiquitous but employees are hesitant to get vaccinated: 90% of residents and 50% of employees are vaccinated, (iv) vaccines are ubiquitous and employees are not hesitant to get the vaccine: 90% of residents and employees are vaccinated.

There are a range of other mitigation measures worth considering, such as the reduction of contacts, physical distancing, room ventilation and wearing of protective equipment. Nursing homes in Austria early on tried to reduce contacts by limiting staff and resident contacts to single living areas. We model this by limiting the contact network of employees and residents to a ward. The ward includes resident rooms, common rooms, a canteen and staff rooms (see fig. 1), and is a unit of the nursing home that can function independently from other wards. Employees regularly have to come into close proximity with residents to care for them, therefore physical distancing is not practical. As many residents in such facilities suffer from an advanced state of dementia, enforcing physical distancing between residents is also not possible. To our knowledge, regular ventilation of rooms was and is not practiced in the nursing homes we considered for this study. This might be due to the cold temperatures during European autumn and winter and the lack of good room ventilation systems in nursing homes. In spring 2020, the period of time in which our calibration data was collected, nursing home staff in the homes considered in this study did not consistently wear face masks, therefore the calibration was performed without protective equipment for employees. This has changed since but we nevertheless chose to not include face masks as a protective measure in our model. The reasoning behind this choice is that we wanted to limit the number of measures introduced on top of our calibrated system, to clearly see the effects of single measures and not compounds of measures. Since, for the same reasons as for physical distancing, only employees could be required to wear protective gear, we also estimate the impact of such prevention measures to be limited, once an index case has been introduced into the home.

We nevertheless investigated the effect of masks in a scenario where the B.1.1.7 variant is dominant (resembling the current situation in Austria), and repeated all simulations and analyses described above (see SI note 7 for results). To this end, we extended the equation describing the probability of successful transmission by two additional factors: the reduction of transmissibility due to wearing a mask $q_6$, and the reduction of susceptibility due to wearing a mask, $q_7$, thus yielding

$$P = 1 - [1 - \beta\gamma(1 - q_1(c))(1 - q_2(t))(1 - q_3)(1 - q_4)(1 - q_5)(1 - q_6)(1 - q_7)].$$

Based on values reported in the literature for various mask types (Pan et al., 2020), we chose conservative estimates of $q_6$ = 0.5 and $q_7$ = 0.3, resulting in a combined 85% reduction in transmission risk if both the infectious and susceptible agent are wearing a mask. This is in line with more recent reports about mask efficacy (Howard et al., 2021).

For each testing and vaccination scenario described above, we compute 5,000 randomly initialised simulation runs in a simulation modelling a nursing home with 35 residents and 18 employees. The number of residents and their contact network as well as the number of employees correspond to the empirically observed situation in a typical ward in Austrian nursing homes. All model parameters are summarised in SI note 2.

## Data availability

All data used to calibrate the simulations as well as simulation results and the results of this work in tabular format and all figures are available in the data repository corresponding to this publication at https://doi.org/10.17605/OSF.IO/HYD4R.

## Code availability

The simulation package used to model the infection dynamics is openly available (Lasser, 2021).

All code used to calibrate the simulations, generate simulation results and generate all figures in this work is available at https://github.com/JanaLasser/nursing_home_SEIRX, the exact code version used for this manuscript has been frozen and is available at https://doi.org/10.5281/zenodo.4898849. Together with the data provided in the data repository at https://doi.org/10.17605/OSF.IO/HYD4R, this allows a full reproduction of the results presented in this work.

# Acknowledgements

We thank Wolfgang Knecht for help with the visualizations and the dedicated and motivated staff of Caritas for supporting this project.

# Author contributions

**JL** programmed the agent based simulation, conducted the numerical experiments, analysed the simulation data, created the visualizations and contributed to writing the original draft of the manuscript.
**PK** conceptualised the project, supervised the work on the project and contributed to writing the original draft of the manuscript.
**TWT** conceptualised the project and curated the nursing home outbreak data.
**ED** contributed to running the simulations necessary for the model calibration.

**SDL** and **KL** contributed to the implementation of vaccination states in the agent based simulation and to conducting the numerical experiments regarding vaccinations.

**JZ** conceptualised the project, contributed to reviewing and editing the manuscript and curated data on test technologies.

**JS** contributed to reviewing and editing the manuscript and created the illustrations of infection transmissions in nursing homes.

**EK** conceptualised the project and contributed to reviewing and editing the manuscript.

**MKP** conceptualised the project, managed ethical aspects of the project and contributed to reviewing and editing the manuscript.

**HW** contributed to conceptualising the project and reviewing the manuscript.

**KS** contributed to collecting outbreak data in nursing homes.

**SS** contributed to collecting outbreak data in nursing homes.

**CH** contributed to collecting outbreak data in nursing homes.

All authors contributed to reviewing and editing the manuscript.

## Ethics

## Funding

The work presented in this article was funded by the Austrian Science Promotion Agency, FFG project under 882184, the WWTF under COV 20-017 and MA16-045 and the Medizinisch Wissenschaftlicher Fonds des Buergermeisters der Bundeshauptstadt Wien under CoVid004.

## Competing interests

The Authors declare no competing interests.

# Supporting information: Agent-based simulations for protecting nursing homes with prevention and vaccination strategies

**Authors**

*Jana Lasser[1,2,*], Johannes Zuber[4], Johannes Sorger[2], Elma Dervic[3], Katharina Ledebur[3], Simon David Lindner[3], Elisabeth Klager[5], Maria Kletečka-Pulker[5,6], Harald Willschke[5], Katrin Stangl[7], Sarah Stadtmann[7], Christian Haslinger[8], Peter Klimek[2,3,*], Thomas Wochele-Thoma[5,7,*]*

**Affiliations**

(1) Graz University of Technology; Institute for Interactive Systems and Data Science, Inffeldgasse 16C, 8010 Graz
(2) Complexity Science Hub Vienna, Josefstädterstraße 39, 1080, Vienna, Austria
(3) Medical University Vienna, Section for Science of Complex Systems;  Center for Medical Statistics, Informatics and Intelligent Systems, Spitalgasse 23, 1090 Vienna, Austria
(4) Research Institute of Molecular Pathology, Campus-Vienna-Biocenter 1, 1030 Vienna, Austria
(5) Ludwig Boltzmann Institute for Digital Health and Patient Safety, Spitalgasse 23, BT86, 1090 Vienna, Austria
(6) University Vienna, Institut für Ethik und Recht in der Medizin, Spitalgasse 2-4, 1090 Vienna, Austria
(7) Caritas Erzdiözese Wien, Albrechtskreithgasse 19-21, 1160 Vienna, Austria
(8) Hygienefachkraft-unlimited, Oswaldgasse 14-22/2/11, 1120 Vienna, Austria

\* Corresponding authors: Thomas.Wochele-Thoma@caritas-wien.at, peter.klimek@meduniwien.ac.at, jana.lasser@tugraz.at

# SI note 1: Model implementation

Design choices are based on daily life and practice in Austrian nursing homes. To develop the model, we interviewed people responsible for nursing home management and COVID-19 prevention measures and adapted the design choices accordingly. In addition, we included resident contact networks based on the real living conditions of residents in nursing homes and calibrated the model to reproduce the characteristics of outbreaks of SARS-CoV-2 in Austrian nursing homes (see SI note 3). The model therefore offers the possibility to explore the effectiveness of various testing and vaccination strategies in the context of nursing homes.

## Agents

Our agent based model follows an SEIRX approach (see main text) , building on the agent based simulation framework mesa (Project Mesa, 2020), written in Python. All code (https://doi.org/10.5281/zenodo.4898849, Lasser, 2021) and data (https://doi.org/10.17605/OSF.IO/HYD4R) is publicly available. We simulate two types of agents (residents and employees) that live and work in nursing homes. Agents do not leave the model and no new agents enter the model after setup. Agents in the model can be susceptible (*S*), exposed (*E*), presymptomatic infectious (*I*), asymptomatic infectious ($I_1$), symptomatic infectious ($I_2$) and recovered (R). Each of these states also exists in a quarantined/isolated version (*X*). Agents remain in these states for variable time periods.

Every agent has an individual exposure duration, *l* (i.e., time until they become infectious), incubation time, *m* (i.e., time until they may show symptoms), and infection duration, *n* (i.e., time from exposure until an agent ceases to be infectious), as depicted in fig. 6 **A**. For every agent, we draw values for *l*, *m*, and *n* from previously reported distributions of these epidemiological parameters for SARS-CoV-2. Exposure duration, *l*, is distributed according to a Weibull distribution with a mean and SD of 5.0 ± 1.9 days (Ferretti et al., 2020; Lauer et al., 2020; Linton et al., 2020). Incubation time, *m*, is distributed according to a Weibull distribution with a mean of 6.4 ± 0.8 (Backer et al., 2020; He et al., 2020) days, with the additional constraint of *m* ≥ *l*. Infection duration, *n*, is distributed according to a Weibull distribution with a mean of 10.9 ± 4.0 days (Backer et al., 2020; He et al., 2020), with the additional constraint of *n* > *l*. The probability to develop a symptomatic course depends on the agent's age (see "Asymptomatic infections" below). Distributions are shown in fig. A1.

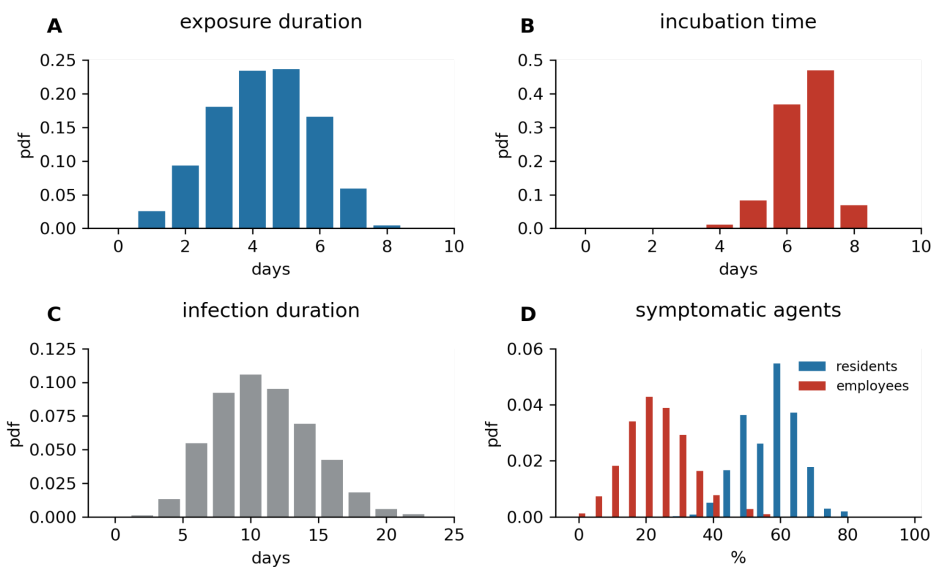

*Figure A1: Distributions of parameters used in the agent based simulation. (A) exposure duration,* l*, (B) incubation time,* m*, (C) infection duration,* n*, and (D) symptom probability*.

## Contact networks

Employees and residents interact by means of networks of contacts specific to the nursing home setting. At every step (day) of the simulation, agents interact with other agents in the neighborhood of their contact network. Infected agents can transmit an infection to susceptible agents, unless one of them is quarantined. Quarantined (and isolated) agents are represented by isolated nodes in the contact network.

Residents have an explicitly defined contact network that is defined through shared rooms (close contacts), shared lunch tables (intermediate contacts), and a shared ward (loose contacts) in the nursing home. We differentiate between these three types of contacts between residents based on their duration and physical closeness, based on insights from practitioners in nursing homes. We illustrate an exemplary floor plan and from which we extracted the contact network alongside exemplary contact situations in fig. 1, which resembles the ward described in cases 2 and 3 (see SI note 5). The different contact types have different weights, which modify the transmission risk through contacts of that type. By definition, contacts of type "close" have weight 1. The weights for contacts of type "intermediate" and "loose" are calibrated using observational data. (see SI note 3 for details). As a result, intermediate and loose contacts have a weight of 0.13. The weighted degree of employee nodes in the network is 6.76 ± 0.00, whereas the weighted degree of resident nodes is 7.41 ± 0.64 and the weighted contact matrix is given as follows:

|  | resident | employee |
|---|---|---|
| **resident** | 0.15 | 0.13 |
| **employee** | 0.13 | 0.13 |

While observational data of contact networks in nursing homes is scarce, one study confirms the assumption that residents tend to have more intense (longer) contacts to other residents, while contacts between employees and residents are more fleeting (Champredon et al., 2018).

We assume that employees have loose contact with all other employees in the same ward, as they take breaks and have lunch together, and generally move through the same ward throughout the day. In addition, employees also have loose contact with all residents in the ward, as they are involved in different care-taking activities throughout the day. While these assumptions are based on insights from practitioners, the actual contact patterns between employees and other employees, and between employees and residents are not based on observational data.

The contact network defines the strength of a contact of residents which other residents, and different contact venues modulate infection transmission risk (for example infection risk is drastically increased for roommates). Simulations in this work are run on a contact network corresponding to one ward of an Austrian nursing home, which housed 35 residents and 18 staff at the time. The contact network extracted from the floor plan in fig. 1 is displayed in fig. A2. The nursing home ward modeled by this contact network is representative for the wards for which we have outbreak data and, more generally, for wards of nursing homes run by Caritas, a care provider that runs a total of 31 homes with over 2650 residents[1] in Austria: wards in these homes have on average 30 residents and between 15 and 20 employees, depending on the care level of residents. A single home has between two and nine wards.

---

[1] https://www.caritas.at/fileadmin/storage/global/document/Positionspapiere/positionspapier_pflege.pdf

While it is hard to find comparable statistics about ward sizes and staffing levels for other countries, nursing home sizes reported for example in the Netherlands (Verbeek et al., 2010) and Japan (Igarashi et al., 2018) seem to be comparable in size, while nursing home sizes in the US seem to be considerably larger, with 75 ± 14 residents per home[2]. In the data repository[3], we provide several additional exemplary contact networks, representing different architectures of nursing homes with different numbers of wards and sparse loose contacts between wards. We did not explore infection dynamics on these contact networks in the present work but they might be a good source to explore infection dynamics in nursing homes with multiple wards. Contact networks are stored as a networkx graph (Hagberg et al., 2008) where the edge weight depends on the type of contact.

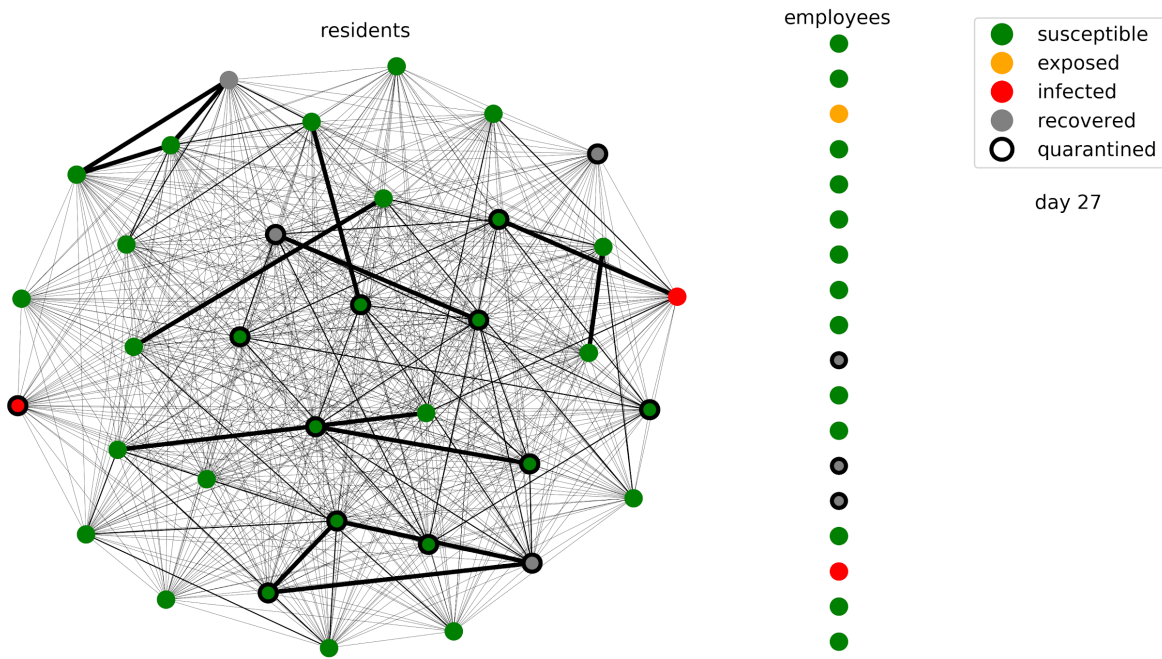

*Figure A2: Exemplary contact network of 35 residents (left) with contact intensity indicated by edge thickness.* *The 18 employees are displayed to the right and their contacts are not shown. Agent states at the given time-step of the simulation (day 27) are indicated by color: green = susceptible, orange = exposed, red = infected, gray = recovered. A black border corresponds to a quarantined or isolated agent.*

## Index cases

An index case is either introduced through a single resident or a single employee at the start of the simulation. No other infections are introduced for the remainder of the simulation. The agent type from which the index case is chosen is specified at the beginning of the simulation. The initially infected agent is then chosen at random from all agents of the given group. A situation in which infections are only introduced by employees resembles the situation in Austrian nursing homes in the time between the start of the pandemic and autumn 2020: residents were not allowed to have visitors and were strongly discouraged from leaving the home. The outbreaks that we used to design and calibrate our model all occurred in this period of time and in three out of four instances the infection was introduced by an employee. In the fourth case, the exact route of introduction into the home remains uncertain. A case in which a resident is the index case resembles a situation in which residents are allowed to have visitors or go outside and are therefore at risk of becoming infected.

---

[2] https://www.statista.com/topics/3982/nursing-homes-in-the-us/
[3] https://doi.org/10.17605/OSF.IO/HYD4R

## Transmission probability

In every step (day) of the simulation, an infected agent can transmit the infection to the agents they are in contact with according to their interaction rules. Transmission is modelled as a Bernoulli trial with a probability of success, $P$. (Laskowski & Moghadas, 2014; Mostaço-Guidolin et al., 2011)

$$P = 1 - [1 - \beta(1 - q_1(c))(1 - q_2(t))(1 - q_3)(1 - q_4)(1 - q_5)(1 - q_6)(1 - q_7)],$$

where the $q_i$ are the reduction of transmission risk due to various factors described below. A transmission risk of $\beta=0.073$ corresponds to the calibrated unmitigated transmission between two household contacts. See SI note 2 for an overview over all model parameters and their sources and SI note 3 for details about the calibration process.

## Contact type

To model the reduction of transmission risk due to the type of contact $c$ between agents, $q_1(c)$, we classify contacts into three categories. "Close" contacts are characterised by very long and physically very close interactions and occur only between roommates. "Intermediate" contacts are characterised by long and/or physically close interactions, such as between residents that share a table. "Loose" contacts are characterised by short and more distant interactions, for example between residents that live in the same ward but do not share a room or dining table and might only meet each other for short periods of time in the hallways. This classification of contacts into different categories is based on insights from practitioners at the nursing homes we characterise in this study (see also SI note 5). We calibrate $\beta$ and $q_1(c)$ such that the outbreak sizes produced by our model correspond to observed outbreak sizes in nursing homes (see SI note 3).

## Viral load over the course of an infection

The modification of transmission risk due to a changing viral load over the course of an infection, $q_2(t)$ approximates the reported dynamics of viral load after an infection with SARS-CoV-2 (He et al. 2020, Walsh et al. 2020). It is modelled as a trapezoid function that depends on the time an agent has already been exposed to the virus, $t$, given the exposure duration, $l$, incubation time, $m$, and infection duration, $n$, of the infected agent:

$$q_2(t) = 0 \text{ if } l < t \leq m; \qquad q_2(t) = 1 - \frac{t-m}{n-m+1} \text{if } t > m \text{ and } t \leq n; \qquad q_2(t) = 1 \text{ else.}$$

## Asymptomatic infections

The modification of transmission risk due to an asymptomatic course of the disease is chosen to be $q_3 = 0.4$, reducing the probability of an agent with an asymptomatic course to infect another agent by 40% (Byambasuren et al., 2020). Nevertheless, since viral load does not seem to significantly decrease for asymptomatic cases (Walsh et al., 2020), in our model asymptomatic cases have the same test detection thresholds as symptomatic cases. As reported by a large cohort study from Italy (Poletti et al., 2021), the probability to develop a symptomatic course of the disease strongly depends on age. To include this dependence into our model, we assume that residents have a probability of 64.52% to develop a symptomatic course (the probability reported by Poletti et al. for people aged 80 and above). For employees, we assume a probability of 26.46% to develop a symptomatic course. This is the average of the probabilities for people aged 20 to 39 years (22.41%) and people aged 30 to 59 years (30.51%). We therefore assume that the age of employees is uniformly distributed between the ages of 20 and 59, which is a simplification but roughly consistent with the age distribution reported for Austrian nurses (Rappold & Juraszovich, 2019). Lastly, we assume that vaccinated agents that become infected will not develop symptomatic courses. While this is a simplification, it reflects the high efficacy of the BNT162b2 and ChAdOx1 nCOV-19 vaccines against symptomatic infections (Aran, 2021; Polack et al., 2020; Voysey et al., 2021).

## Effectiveness of vaccines against infection

For vaccinated agents, both the probability to become infected, as well as the probability to transmit an infection to another agent are reduced. Therefore, the vaccination status of both the infected as well as the susceptible agents influence the transmission probability. The evidence for the effectiveness of vaccinations regarding the prevention of asymptomatic or mildly symptomatic infections is still incomplete. Early studies in healthcare workers suggests that a single dose of the BNT162b2 vaccine (Biontech/Pfizer) is between 60% and 92% effective, depending on the delay after the dose was administered (Amit et al., 2021; Dagan et al., 2021; Hall et al., 2021; Harris et al., 2021; Jones et al., 2021; Levine-Tiefenbrun et al., 2021b). A recent preprint that studied vaccine effectiveness for a single dose of both BNT162b2 and ChAdOx1 nCOV-19 (AstraZeneca) in long term care facility residents finds that both vaccines are similarly effective and effectiveness against infection ranges from 56% to 62% after 28-34 days and 35-48, respectively (Shrotri et al., 2021). Both of the vaccines investigated in these studies are highly relevant for the context of the nursing homes modelled in this study, as residents in Austria have been vaccinated with BNT162b2 and nursing home staff has been vaccinated with both BNT162b2 and ChAdOx1 nCOV-19. Considering the significant uncertainty associated with the number of doses and the dependence of vaccine effectiveness on time after vaccination, we therefore choose a conservative estimate of $q_4 = 0.6$. Therefore, if the susceptible agent is vaccinated, the transmission probability is reduced by 60%.

### Effectiveness of vaccines against transmission

First evidence suggests reduced viral loads of infected people that had been vaccinated before (Harris et al., 2021; Levine-Tiefenbrun et al., 2021b). In this study, a single dose of BNT162b2 is associated with a decrease of 2.8-4.5 fold in viral load. Viral load is associated with infectivity (Walsh et al., 2020) but the functional form of the association is unclear. Given the high uncertainty associated with this effect, we choose a conservative estimate of $q_5 = 0.3$. Therefore, if the infected agent is vaccinated, the transmission probability is reduced by 30%.

### Protective equipment

Since the wearing of protective equipment, including first and foremost masks that cover the mouth and nose, is among the intervention measures most frequently employed in the context of health care facilities, we decided to include the wearing of face masks by employees as a supplement to this study. To this end, we use a conservative estimate for the reduction of transmission risk by face masks: a mask worn by an infected person reduces that person's transmission risk by 50%, whereas a mask worn by a susceptible person reduces that person's risk of getting infected by 30% (Pan et al., 2021). We model these two transmission risk reductions as $q_6 = 0.5$ and $q_7 = 0.3$, respectively. To reflect the circumstances in nursing homes, where it is next to impossible to make inhabitants consistently wear masks, we simulate only scenarios in which all employees wear masks, but not the residents (see SI note 7 for results).

## Test technologies

Different test technologies can be selected in the simulations. Agents can be testable, depending on the time passed since transmission and the test used. Testability and detection thresholds in relation to viral loads are illustrated in fig. 6 **A**. PCR tests can detect an infection after $l$ - 1 days and for a duration of $n$, where $l$ and $n$ are the agent specific exposure duration and infection duration, respectively. Antigen tests detect an infection after $l$ + 1 days and for a duration of $n$ - 1 days, and LAMP tests detect an infection after $l$ days and for $n$ days.

Tests take a certain amount of time to return results (turnover time), depending on the chosen test technology. Accordingly, agents can have a pending test result, which will prevent them from getting tested again before the pending result arrives. This corresponds to the daily practice in diagnostic testing in the nursing homes we model.

We also implement different test sensitivities and specificities for different test technologies, but use sensitivities and specificities of 100% for the time window in which a given test technology can detect an infection, for the sake of simplicity in this work. Therefore, tests return positive or negative results, depending on whether the agent was testable at the time of testing of the chosen test. Under the assumption that the likelihood of being tested within the scope of screening by use of AG tests is uniformly distributed over the entire course of the infection and with a mean infection duration of 10.9 days (Walsh et al., 2020; You et al., 2020), this results in an effective sensitivity of AG tests of 0.45 in our model.

Depending on the containment strategy (see below), different tests are used for diagnostic purposes if an agent shows symptoms, and for preventive screening measures.

## Intervention measures

### Test-trace-isolate (TTI)

Next to the transmission of the infection, the nursing home implements containment measures (quarantine and isolation) and a testing and tracing strategy to curb the spread of the virus among its residents and employees. Symptomatic cases are immediately isolated and tested, using PCR tests. They remain isolated for 10 days subsequent to their test result. Once a positive test result is returned, all close and intermediate contacts of the positive agent (i.e. roommates and table neighbours) are immediately quarantined for 10 days. Since employees do not have specified close or intermediate contacts, employees can be isolated if they are symptomatic or tested positive, but they will not be quarantined due to contact tracing. According to nursing home practitioners, quarantined residents are transferred to a separate part of the nursing home where they are isolated and cared for by employees in full PPE. We assume that these contacts between quarantined residents and employees in full PPE have a negligible risk of transmission and therefore do not include them in the model. Quarantined employees stay at home and do not come to work for the duration of the quarantine. Isolated or quarantined agents are therefore assumed to not have any contacts with other agents for the duration of their isolation or quarantine. Isolated or quarantined agents are not replaced by new agents in the model. The strategy described above is summarised as "test-trace-isolate" (TTI) strategy and is always followed by the facilities, independent of potential additional preventive measures.

Preventive testing

Next to the TTI strategy, the nursing home can implement a preventive screening strategy, where employees and/or residents are tested in set intervals: if the interval is set to "once a week", preventive tests occur on Mondays. If the interval is set to "twice a week", preventive tests occur on Mondays and Thursdays. If the interval is set to "three times a week", preventive tests occur on Mondays, Wednesdays and Fridays. The intervals for these screens can be specified and can be chosen differently for the residents and employees. The simulation is initialised at a random day of the week, to avoid artificial interactions of the testing schedule and the exposure duration of the index case. The tests used for preventive screening can be different from the tests used within the scope of the TTI strategy. This is useful since cheaper and faster tests with higher detection thresholds like RT-LAMP or antigen tests might be preferred over PCR tests for preventive screening. Figure A3 showcases an exemplary timeline of agent states and timepoints of screens for a simulation with 35 residents and 18 employees using one-day turnover PCR tests for diagnostic testing and same-day turnover antigen tests for preventive screening. Residents are screened with antigen tests every seven days, employees are screened every three days.

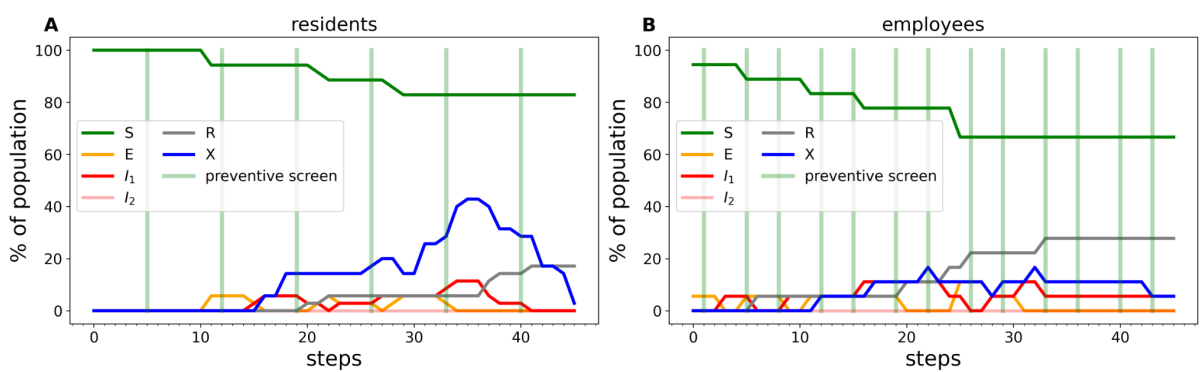

*Figure A3 timelines of agent states:* percentage of susceptible (S), exposed (E), asymptomatic and symptomatic infected ($I_1$, $I_2$), quarantined or isolated (X) and recovered (R) residents **(A)** and employees **(B)**. Preventive resident and employee screens occur every 7 and every 3 days, and are indicated as green vertical lines in panel **(A)** and **(B)**, respectively.

## Vaccinations

To model different vaccination prevalences in nursing homes, the model allows for the specification of a ratio of vaccinated agents per agent type. If for example 70% of residents in a given home are vaccinated, of all resident agents, 70% are picked at random at model initialisation and are given the status "vaccinated". Being vaccinated reduces an agent's chance of getting infected by 60% (see section "Effectiveness of vaccines against infection" above for details). If a vaccinated agent gets infected nevertheless, their chance of transmitting the infection is reduced by 30% (see section "Effectiveness of vaccines against transmission" above for details). Vaccination status does not change throughout the simulation but vaccinated agents that are infected will recover at the end of their infection and will subsequently be completely immune to reinfection after recovery.

# SI note 2: Table of parameter values

| Parameter | Symbol | Value | Source |
|---|---|---|---|
| Exposure duration | $l$ | 5.0 ± 1.9 days | (Ferretti et al., 2020; Lauer et al., 2020; Linton et al., 2020) |
| Incubation time | $m$ | 6.4 ± 0.8 | (Backer et al., 2020; He et al., 2020) |
| Infection duration | $n$ | 10.9 ± 4.0 | (Backer et al., 2020; He et al., 2020) |
| Household transmission risk | $\beta$ | 0.073 | Calibrated against Madewell et al., 2020) |
| Transmissibility increase of B.1.1.7 compared to the strain dominant in Austrian in spring 2020 | $\gamma$ | 1.5 | (Davies et al., 2021; Fort, 2021; Institute of Social and Preventive Medicine, 2021; Leung et al., 2021; Statens Serum Institut, 2021) |
| Transmission risk reduction for intermediate and loose contacts | $q_1$ | 0.87 | Calibrated against observational outbreak data. |
| Transmission risk reduction over the course of an infection | $q_2(t)$ | $0$ if $l < t \leq m$; $1 - \frac{t-m}{n-m+1}$ if $t > m$ and $t \leq n$; $1$ else | (He et al., 2020; Walsh et al., 2020) |
| Transmission risk reduction for asymptomatic courses | $q_3$ | 0.4 | (Byambasuren et al., 2020) |
| Vaccine efficacy against infection | $q_4$ | 0.6 | (Amit et al., 2021; Dagan et al., 2021; Hall et al., 2021; Jones et al., 2021) |
| Vaccine efficacy against transmission | $q_5$ | 0.3 | (Harris et al., 2021; Levine-Tiefenbrun et al., 2021a) |
| Transmission risk reduction for an infected person wearing a mask (exhaling) | $q_6$ | 0.5 | (Pan et al., 2021) |
| Transmission risk reduction for a susceptible person wearing a mask (inhaling) | $q_7$ | 0.3 | (Pan et al., 2021) |

*Table A1: simulation parameters. Main simulation parameters, their values and sources.*

# SI note 3: Calibration

## Calibration of household transmission risk

According to a recent meta study (Madewell et al., 2020), the probability of an index case infecting an adult member of the same household (secondary attack rate) was 28.3%; (95% CI, 20.2%-37.1%), for the dominant variants at the time of data collection. This value includes both symptomatic and asymptomatic index cases. We calibrate the base transmission risk, $β$, between two agents in our model such that it reflects this secondary attack rate. For household transmissions between adults, the only relevant factors that modify the base transmission risk are the dependence of the transmissibility on the progression of the disease, $q_2(t)$, and the reduction of transmissibility in case of an asymptomatic course, $q_3$. The values for both of these factors are taken from the literature (Byambasuren et al., 2020; He et al., 2020; Walsh et al., 2020). Therefore, for a contact to a household member on day $t$ after the exposure, the probability of a successful transmission is given as

$$P(t) = 1 - [\ 1 - β(1 - q_2(t))\ (1 - q_3)\ ].$$

Our goal is to calibrate $β$, modified by $q_2$ and $q_3$, such that the cumulative probability of infecting a household member over the whole infection duration of an infected agent is equal to the secondary attack rate reported in the literature.

The probability of having a symptomatic course depends on age. According to a large cohort study from Italy (Poletti et al., 2021), the probability to develop a symptomatic course stratified by age groups is 18.09% [95% CI, 13.93-22.89] (0-19 years) 22.41% [95% CI, 18.93-26.2] (20 to 39 years), 30.54% [95% CI, 27.7-33.49] (40 to 59 years), 35.46% [95% CI, 32.2-38.83] (60 to 79 years), and 64.56% [95% CI, 56.56-71.99] (80 and above years).

Since the household secondary attack rates reported by in the meta-study by Madewell et al. (2020) are averaged for index cases over all age brackets, we need to make an assumption about the underlying age distribution of the population and match the age-distribution of the agents in our calibration simulations to it. The studies incorporated in the meta-study are predominantly from populations in Europe (17327 index cases), China (6194 index cases) and South Korea (6125 index cases), as these were the regions primarily affected by the spread of SARS-CoV-2 at the time the study was published. Studies in other countries amount to a total of 2627 index cases. We therefore use an average of the age distributions in these three countries[4] to construct the distribution of age brackets in the overall study population. The average is weighted by the number of contributed index cases to the meta-study. This is only a coarse approximation to the weighting employed in the meta-study to combine results of different studies, since we do not consider other factors such as possible bias or uncertainties when calculating the adjusted population ratios. The final population ratios used in our calibration simulations the ins 20.82% (0 to 19 years), 26.36% (20-39 years), 29.41% (40 to 59 years), 19.19% (60-79 years) and 4.22% (80 years and above).

---

[4] Population age distributions for Europe, China and South Korea for the year 2020 were obtained from the United Nations Department of Economic and Social Affairs
https://population.un.org/wpp/Download/Standard/Population/

To calibrate $\beta$, we create pairs of agents and let one of them be infected. For every agent, we draw relevant epidemiological parameters (exposure duration, infection duration, probability of a symptomatic course) from corresponding distributions (Backer et al., 2020; Ferretti et al., 2020; He et al., 2020; Lauer et al., 2020; Linton et al., 2020; Walsh et al., 2020; You et al., 2020). Age is drawn from the averaged age distribution described above and determines the probability to develop a symptomatic course, as reported by Poletti et al. 2020. We then simulate the whole course of the infection (from day $t = 0$ to the end of the infection duration $n$) and perform a Bernoulli trial for the infection with a probability of success $P(t)$ on every day $t$. We minimize the difference between the expected number of successful infections given the reported secondary attack rate (28.3%) and the simulated number of successful infections by varying $\beta$, as shown in fig. A4. This results in an optimal value of $\beta = 0.073$ or an average risk of 7.3% per day for a household member to be infected.

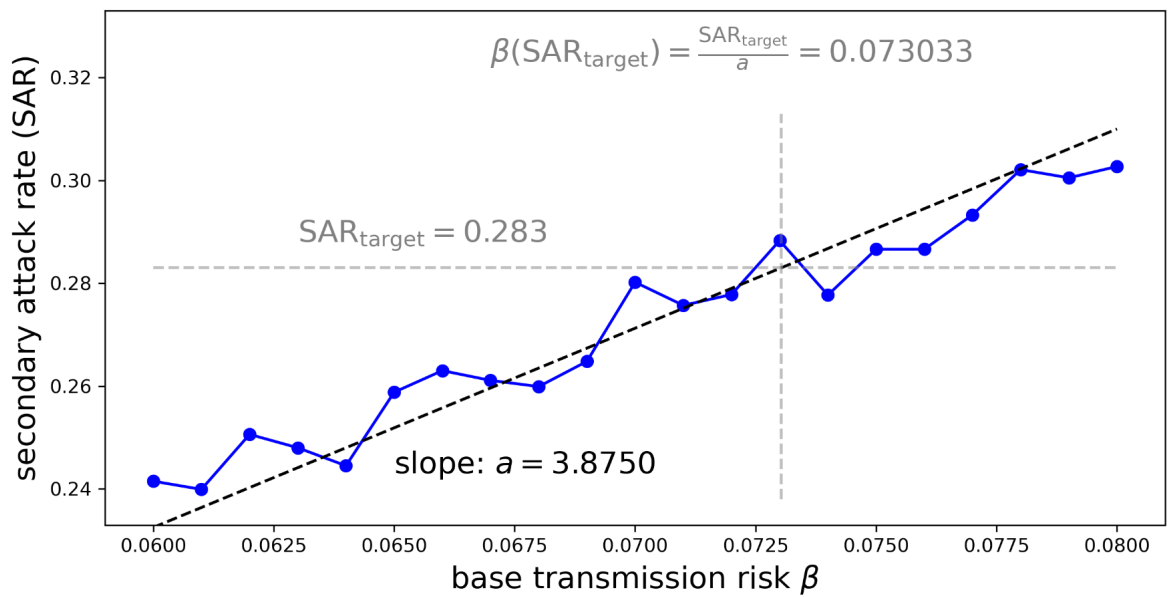

*Figure A4 calibration of base transmission risk: The base transmission risk $\beta$ is calibrated by minimising the difference between the percentage of transmissions in simulations in a household setting, and the secondary attack rate of 28.3% reported in the literature for transmissions to adults in household settings (Potelli et al. 2020).*

## Calibration of contact weights

In our model, a household contact is equivalent to a contact of type "close". After calibrating $\beta$, the values of $q_1(c)$ for contacts $c$ of type "intermediate" and "loose", respectively, remain the only two free parameters in the simulation. To calibrate $q_1(c)$ for these two contact types, we used data of empirically observed outbreaks in four Austrian nursing homes (see "Case studies of outbreaks in Austrian nursing homes" below and https://doi.org/10.17605/OSF.IO/HYD4R). Specifically, we optimize the sum of squared distances between the empirically observed cumulative number of infected employees over time and the simulated cumulative number of employees over time, $e_1$; and the sum of squared distances between the empirically observed cumulative number of infected residents over time and the simulated cumulative number of residents over time, $e_2$

$$E = \sum_{i=1}^{4} e_{1,i} + e_{2,i},$$

where $i$ denotes the $i$th outbreak. Here, if $x_j$ is the cumulative number of infected employees in an empirically observed outbreak on day $j$, and $y_j$ is the cumulative number of infected employees in a given simulation run on day $j$, then

$$e_1 = \sum_{j=1}^{d} (x_j - y_j)^2.$$

We choose $d = 27$ days, which corresponds to the duration of the longest empirically observed outbreak. The duration is defined as the number of days between the first reported positive test result, and the last newly reported positive test result that is causally associated with the given cluster of infections.

To run the calibration simulations, we choose a set of prevention measures that were in place in Austrian nursing homes at the time when the empirically observed outbreaks occurred. These homes had implemented a TTI strategy, using PCR tests with a two-day results turnover but no additional preventive testing. Neither residents nor employees were consistently wearing masks and there was no additional room ventilation to prevent accumulation of aerosols. We use a model that is similar to the nursing home wards described in cases 2 and 3 (see SI note 5), housing 35 residents and 18 employees.

Using this model and the described settings for prevention measures, to find optimal values for the free parameters, we first conduct a random search in the parameter grid spanned by the following ranges ([start : stop : step]): $q_1$(intermediate): [0 : 1 : 0.1] and $q_1$(loose) [0 : 1 : 0.1], with the additional constraint that $q_1$(loose) > $q_1$(intermediate), i.e., the probability of transmission *failure* is always greater for a contact of type "loose" than for a contact of type "intermediate". For each parameter combination, we run an ensemble of 1000 randomly initialised simulations and calculate the distances $e_1$ and $e_2$ between each empirically observed cumulative number of infected and the simulated number of infected. Since the empirical observations start with the first *confirmed positive* test result on the first day, we truncate the simulation results such that the first day also corresponds to the first positive test result. We note that this also means that we discard simulations in which there is no positive test result. This can be the case if the index case does not infect any other agents and does not show symptoms themselves, or in (very rare) cases where the asymptomatic index case infects a small number of other agents which all also do not show symptoms. Approximately 38% of simulation runs result in no positive tests. We think this treatment is warranted, since such cases would also not be observable in the real world, as there would never be an indication of an ongoing outbreak.

We find that, although there is a clearly defined optimum of 0.9 for $q_1$(loose), values for $q_1$(intermediate) do not converge to a stable optimum. This is possibly due to the case that the number of intermediate contacts is small compared to the number of loose contacts in the system, and the value of $q_1$(intermediate) does not influence the infection dynamics to a great extent. We therefore decided to set $q_1$(intermediate) = $q_1$(loose) going forward.
We proceeded to scan a finer grid of values for $q_1$ in the range [0.11 : 0.25 : 0.01], running 5000 simulations per parameter value, and find an optimal value of $q_1$ = 0.87. Therefore, contacts of type "loose" (and "intermediate") are 87% less likely to transmit the virus than contacts of type "close" (household contact). In fig. A5 (**A**) we show the relative contribution of the error terms $e_1$ (infected employees) and $e_2$ (infected residents) to the overall error $E$ (values for $q_1$ > 0.5 not shown). In fig. A5 (**B**) we show an ensemble of 5000 randomly initialised runs, using the calibrated values for $\beta$ and $q_1$, in comparison to the empirically observed cumulative numbers of infected employees and residents. We note that the ensemble values shown in fig. A5 (**B**) does not exclude simulations in which no infections were detected. It seems that three out of the four observed outbreaks follow a similar pattern that is well-captured by the infection dynamics in the simulation, whereas the second recorded outbreak (stars) is characterised by a very sharp increase in infected residents during the first week. This could for example be caused by a (rare) super-spreading event at the beginning of the outbreak: as illustrated in fig. A6, in approximately 10% of the simulations that have at least one positive test result, the virus spreads to 20 or more residents. Another explanation for the rapid spread in the beginning of the outbreak could be the introduction of the virus into the home through multiple source cases at the same time.

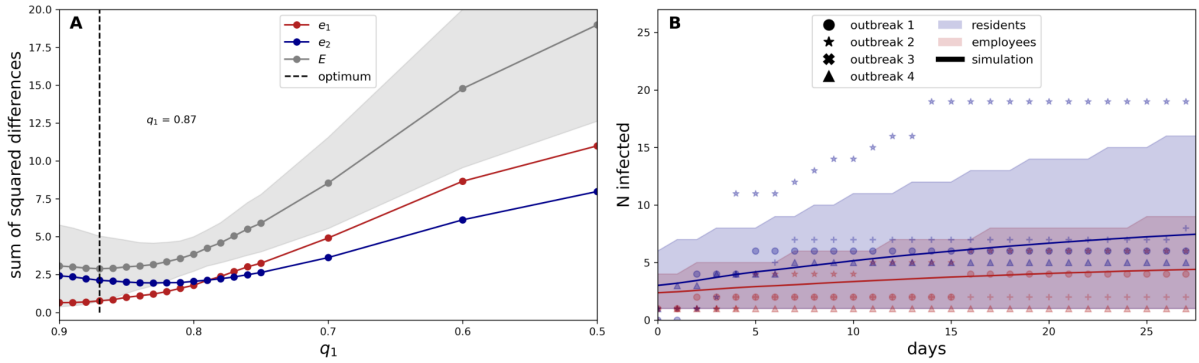

*Figure A5 calibration of the simulation by means of comparing to empirically observed outbreak data. (A) contribution of the employee error term $e_1$ (red) and resident error term $e_2$ (blue) to the overall error E (grey, standard deviation shown as shaded area), and optimal value for $q_2$ (dotted line). (B) cumulative number of infected employees (red) and residents (blue) in empirically observed outbreaks (dots, stars, crosses and triangles) and simulations, using the calibrated values for β = 0.073 and $q_1$ = 0.87. Solid lines show ensemble averages of 5000 runs, shaded areas show the [10%; 90%] quantiles around the ensemble averages.*

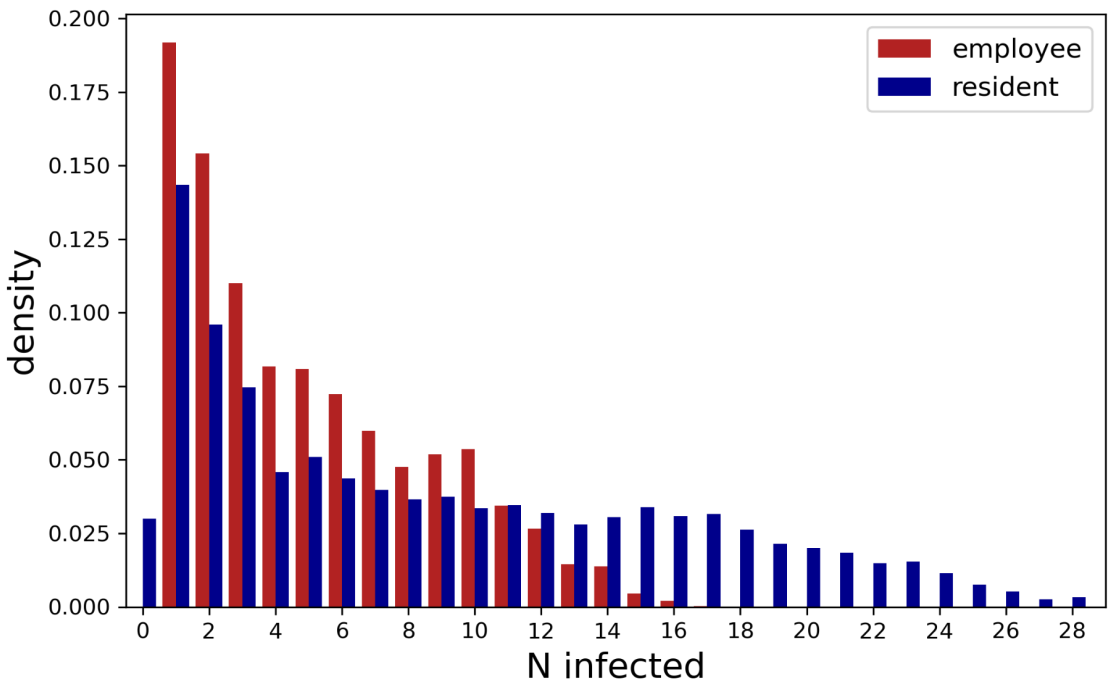

*Figure A6 simulated distribution of outbreak sizes in nursing homes. The distribution of the overall number of infected employees (red) and residents (blue), i.e., the size of the outbreak shows a bimodal distribution of outbreak sizes: a large number of outbreaks stop after one or two transmissions. A small number of outbreaks spreads through the whole nursing home and infects >20 residents. The histogram shows probability densities for an ensemble of 5000 randomly initialised runs at β = 0.073 and $q_1$ = 0.87.*

# SI note 4: detailed results tables and additional figures

## Result tables B.1.1.7 variant

The tables below are also available in the data repository associated with this work at https://doi.org/10.17605/OSF.IO/HYD4R in expanded form.

| Staff screens per week | Resident screens per week | Outbreak size mean ± std [infected residents] | Outbreak size [10th; 90th] percentile | $R_{eff}$ mean ± std | Test rate mean ± std [tests / day / person] |
|---|---|---|---|---|---|
| Antigen tests with same-day turnover | | | | | |
| 3 | 1 | 0.4 ± 1.4 | 0.0 [0.0; 1.0] | 0.29 ± 0.62 | 0.23 ± 0.03 |
| 3 | never | 0.8 ± 2.6 | 0.0 [0.0; 2.0] | 0.29 ± 0.63 | 0.14 ± 0.02 |
| 2 | 1 | 0.7 ± 1.8 | 0.0 [0.0; 2.0] | 0.41 ± 0.75 | 0.19 ± 0.03 |
| 2 | never | 1.2 ± 3.3 | 0.0 [0.0; 4.0] | 0.41 ± 0.76 | 0.09 ± 0.02 |
| 1 | 1 | 1.2 ± 2.6 | 0.0 [0.0; 4.0] | 0.67 ± 0.96 | 0.14 ± 0.04 |
| 1 | never | 2.2 ± 4.7 | 0.0 [0.0; 8.0] | 0.66 ± 0.96 | 0.05 ± 0.02 |
| never | 1 | 3.2 ± 4.5 | 1.0 [0.0; 10.0] | 1.30 ± 1.33 | 0.09 ± 0.02 |
| never | never | 6.0 ± 7.7 | 2.0 [0.0; 19.0] | 1.31 ± 1.32 | 0.00 ± 0.01 |
| LAMP tests with same-day turnover | | | | | |
| 3 | 1 | 0.2 ± 0.8 | 0.0 [0.0; 0.0] | 0.13 ± 0.46 | 0.23 ± 0.03 |
| 3 | never | 0.3 ± 1.7 | 0.0 [0.0; 0.0] | 0.12 ± 0.45 | 0.14 ± 0.02 |
| 2 | 1 | 0.3 ± 1.1 | 0.0 [0.0; 1.0] | 0.24 ± 0.59 | 0.19 ± 0.04 |
| 2 | never | 0.6 ± 2.3 | 0.0 [0.0; 1.0] | 0.23 ± 0.58 | 0.09 ± 0.02 |
| 1 | 1 | 0.8 ± 1.8 | 0.0 [0.0; 3.0] | 0.52 ± 0.87 | 0.14 ± 0.04 |
| 1 | never | 1.5 ± 3.8 | 0.0 [0.0; 5.0] | 0.49 ± 0.83 | 0.05 ± 0.02 |
| never | 1 | 2.9 ± 4.0 | 1.0 [0.0; 9.0] | 1.33 ± 1.34 | 0.09 ± 0.02 |
| never | never | 6.0 ± 7.6 | 2.0 [0.0; 19.0] | 1.34 ± 1.34 | 0.00 ± 0.01 |
| PCR tests with same-day turnover | | | | | |
| 3 | 1 | 0.1 ± 0.7 | 0.0 [0.0; 0.0] | 0.11 ± 0.48 | 0.23 ± 0.04 |
| 3 | never | 0.2 ± 1.5 | 0.0 [0.0; 0.0] | 0.11 ± 0.44 | 0.14 ± 0.02 |
| 2 | 1 | 0.2 ± 0.7 | 0.0 [0.0; 0.0] | 0.14 ± 0.47 | 0.19 ± 0.04 |
| 2 | never | 0.4 ± 1.9 | 0.0 [0.0; 0.0] | 0.14 ± 0.49 | 0.09 ± 0.02 |
| 1 | 1 | 0.5 ± 1.3 | 0.0 [0.0; 2.0] | 0.40 ± 0.78 | 0.14 ± 0.04 |
| 1 | never | 1.1 ± 3.1 | 0.0 [0.0; 3.0] | 0.39 ± 0.77 | 0.05 ± 0.02 |
| never | 1 | 2.4 ± 3.4 | 1.0 [0.0; 7.0] | 1.32 ± 1.36 | 0.09 ± 0.02 |
| never | never | 5.7 ± 7.5 | 2.0 [0.0; 18.0] | 1.30 ± 1.34 | 0.00 ± 0.01 |
| colspan=6 | PCR tests with one day turnover | | | | |
| 3 | 1 | 0.2 ± 0.9 | 0.0 [0.0; 0.0] | 0.12 ± 0.46 | 0.23 ± 0.04 |
| 3 | never | 0.3 ± 1.8 | 0.0 [0.0; 0.0] | 0.13 ± 0.47 | 0.14 ± 0.02 |
| 2 | 1 | 0.3 ± 1.2 | 0.0 [0.0; 1.0] | 0.24 ± 0.61 | 0.18 ± 0.04 |
| 2 | never | 0.7 ± 2.6 | 0.0 [0.0; 1.0] | 0.24 ± 0.60 | 0.09 ± 0.02 |
| 1 | 1 | 0.8 ± 1.8 | 0.0 [0.0; 3.0] | 0.50 ± 0.86 | 0.14 ± 0.04 |
| 1 | never | 1.7 ± 3.9 | 0.0 [0.0; 6.0] | 0.54 ± 0.87 | 0.05 ± 0.02 |
| never | 1 | 2.7 ± 3.9 | 1.0 [0.0; 8.0] | 1.30 ± 1.34 | 0.09 ± 0.02 |
| never | never | 6.1 ± 7.7 | 2.0 [0.0; 19.0] | 1.35 ± 1.35 | 0.00 ± 0.01 |
| colspan=6 | PCR tests with two day turnover | | | | |
| 3 | 1 | 0.4 ± 1.5 | 0.0 [0.0; 1.0] | 0.26 ± 0.60 | 0.23 ± 0.03 |
| 3 | never | 0.7 ± 2.4 | 0.0 [0.0; 2.0] | 0.26 ± 0.59 | 0.14 ± 0.02 |
| 2 | 1 | 0.6 ± 1.7 | 0.0 [0.0; 2.0] | 0.37 ± 0.70 | 0.18 ± 0.04 |
| 2 | never | 1.1 ± 3.2 | 0.0 [0.0; 3.0] | 0.40 ± 0.73 | 0.09 ± 0.02 |
| 1 | 1 | 1.2 ± 2.6 | 0.0 [0.0; 4.0] | 0.66 ± 0.97 | 0.14 ± 0.04 |
| 1 | never | 2.1 ± 4.6 | 0.0 [0.0; 8.0] | 0.65 ± 0.98 | 0.05 ± 0.02 |
| never | 1 | 3.4 ± 4.7 | 1.0 [0.0; 10.0] | 1.33 ± 1.31 | 0.09 ± 0.02 |
| never | never | 6.2 ± 7.9 | 2.0 [0.0; 19.0] | 1.31 ± 1.32 | 0.00 ± 0.01 |

*Table A2: employee index cases. Mean outbreak sizes alongside median, 10th and 90th percentile outbreak ranges, $R_{eff}$ and test rate for scenarios, in which employees undergo preventive testing never, once, two times or three times a week, and residents undergo preventive testing never or once a week. Preventive testing is performed using either antigen tests, RT-LAMP tests or PCR tests, all with same-day results turnover. Values are calculated from simulations with 5000 randomly initialized runs per scenario.*

| Staff screens per week | Resident screens per week | Outbreak size mean ± std [infected residents] | Outbreak size [10th; 90th] percentile | $R_{eff}$ mean ± std | Test rate mean ± std [tests / day / person] |
|---|---|---|---|---|---|
| **Antigen tests with same-day turnover** | | | | | |
| 3 | 1 | 4.3 ± 5.3 | 2.0 [1.0; 12.0] | 1.40 ± 1.71 | 0.23 ± 0.03 |
| 3 | never | 7.9 ± 8.2 | 3.0 [1.0; 21.0] | 2.02 ± 1.99 | 0.14 ± 0.02 |
| 2 | 1 | 5.0 ± 5.9 | 2.0 [1.0; 15.0] | 1.44 ± 1.74 | 0.18 ± 0.03 |
| 2 | never | 9.0 ± 9.0 | 4.0 [1.0; 23.0] | 2.03 ± 1.98 | 0.09 ± 0.02 |
| 1 | 1 | 6.2 ± 7.3 | 2.0 [1.0; 19.0] | 1.43 ± 1.72 | 0.13 ± 0.03 |
| 1 | never | 10.7 ± 10.2 | 6.0 [1.0; 26.0] | 1.99 ± 1.98 | 0.05 ± 0.01 |
| never | 1 | 7.4 ± 8.8 | 2.0 [1.0; 22.0] | 1.41 ± 1.71 | 0.09 ± 0.02 |
| never | never | 13.1 ± 11.4 | 12.0 [1.0; 28.0] | 2.04 ± 1.97 | 0.01 ± 0.01 |
| **LAMP tests with same-day turnover** | | | | | |
| 3 | 1 | 2.1 ± 2.4 | 1.0 [1.0; 5.0] | 0.82 ± 1.24 | 0.23 ± 0.03 |
| 3 | never | 6.5 ± 7.1 | 3.0 [1.0; 18.0] | 1.83 ± 1.76 | 0.14 ± 0.02 |
| 2 | 1 | 2.2 ± 2.5 | 1.0 [1.0; 5.0] | 0.85 ± 1.24 | 0.19 ± 0.03 |
| 2 | never | 7.1 ± 7.7 | 3.0 [1.0; 20.0] | 1.81 ± 1.73 | 0.09 ± 0.02 |
| 1 | 1 | 2.6 ± 3.2 | 1.0 [1.0; 7.0] | 0.84 ± 1.25 | 0.14 ± 0.03 |
| 1 | never | 8.7 ± 8.9 | 4.0 [1.0; 23.0] | 1.82 ± 1.72 | 0.05 ± 0.01 |
| never | 1 | 4.1 ± 5.8 | 1.0 [1.0; 14.0] | 0.85 ± 1.25 | 0.09 ± 0.02 |
| never | never | 13.4 ± 11.6 | 13.0 [1.0; 29.0] | 1.82 ± 1.72 | 0.00 ± 0.01 |
| **PCR tests with same-day turnover** | | | | | |
| 3 | 1 | 1.8 ± 1.7 | 1.0 [1.0; 4.0] | 0.66 ± 1.12 | 0.23 ± 0.03 |
| 3 | never | 6.4 ± 7.1 | 3.0 [1.0; 18.0] | 1.79 ± 1.71 | 0.14 ± 0.02 |
| 2 | 1 | 1.8 ± 1.7 | 1.0 [1.0; 4.0] | 0.70 ± 1.15 | 0.19 ± 0.03 |
| 2 | never | 6.6 ± 7.3 | 3.0 [1.0; 19.0] | 1.82 ± 1.74 | 0.09 ± 0.02 |
| 1 | 1 | 1.9 ± 2.0 | 1.0 [1.0; 4.0] | 0.66 ± 1.14 | 0.14 ± 0.04 |
| 1 | never | 8.0 ± 8.4 | 3.0 [1.0; 21.0] | 1.83 ± 1.72 | 0.05 ± 0.01 |
| never | 1 | 2.9 ± 4.2 | 1.0 [1.0; 8.0] | 0.67 ± 1.13 | 0.09 ± 0.02 |
| never | never | 13.5 ± 11.7 | 14.0 [1.0; 29.0] | 1.87 ± 1.77 | 0.00 ± 0.01 |
| **PCR tests with one day turnover** | | | | | |
| 3 | 1 | 2.2 ± 2.4 | 1.0 [1.0; 5.0] | 0.88 ± 1.28 | 0.23 ± 0.03 |
| 3 | never | 6.6 ± 7.3 | 3.0 [1.0; 19.0] | 1.84 ± 1.74 | 0.14 ± 0.02 |
| 2 | 1 | 2.3 ± 2.6 | 1.0 [1.0; 5.0] | 0.90 ± 1.31 | 0.18 ± 0.03 |
| 2 | never | 7.1 ± 7.6 | 3.0 [1.0; 20.0] | 1.82 ± 1.75 | 0.09 ± 0.02 |
| 1 | 1 | 2.8 ± 3.4 | 1.0 [1.0; 7.0] | 0.91 ± 1.30 | 0.14 ± 0.03 |
| 1 | never | 8.7 ± 8.9 | 4.0 [1.0; 23.0] | 1.81 ± 1.67 | 0.05 ± 0.01 |
| never | 1 | 4.2 ± 5.9 | 1.0 [1.0; 14.0] | 0.91 ± 1.30 | 0.09 ± 0.02 |
| never | never | 13.4 ± 11.6 | 13.0 [1.0; 29.0] | 1.83 ± 1.73 | 0.00 ± 0.01 |
| **PCR tests with two day turnover** | | | | | |
| 3 | 1 | 3.1 ± 3.6 | 1.0 [1.0; 8.0] | 1.12 ± 1.41 | 0.23 ± 0.03 |
| 3 | never | 7.3 ± 7.8 | 3.0 [1.0; 20.0] | 1.84 ± 1.71 | 0.14 ± 0.02 |
| 2 | 1 | 3.3 ± 4.0 | 1.0 [1.0; 9.0] | 1.15 ± 1.43 | 0.18 ± 0.03 |
| 2 | never | 7.9 ± 8.2 | 3.0 [1.0; 21.0] | 1.81 ± 1.74 | 0.09 ± 0.02 |
| 1 | 1 | 4.0 ± 4.9 | 1.0 [1.0; 11.0] | 1.12 ± 1.40 | 0.14 ± 0.03 |
| 1 | never | 9.7 ± 9.5 | 5.0 [1.0; 24.0] | 1.87 ± 1.78 | 0.05 ± 0.01 |
| never | 1 | 5.8 ± 7.3 | 2.0 [1.0; 19.0] | 1.14 ± 1.40 | 0.09 ± 0.02 |
| never | never | 13.6 ± 11.7 | 13.0 [1.0; 29.0] | 1.83 ± 1.73 | 0.00 ± 0.01 |

*Table A3 resident index cases:* Mean outbreak sizes alongside median, 10th and 90th percentile outbreak ranges, $R_{eff}$ and test rate for scenarios, in which employees undergo preventive testing never, once, two times or three times a week, and residents undergo preventive testing never or once a week. Preventive testing is performed using either antigen tests, RT-LAMP tests or PCR tests, all with same-day results turnover. Values are calculated from simulations with 5000 randomly initialized runs per scenario.

# Test technology: $R_{eff}$ and test rates

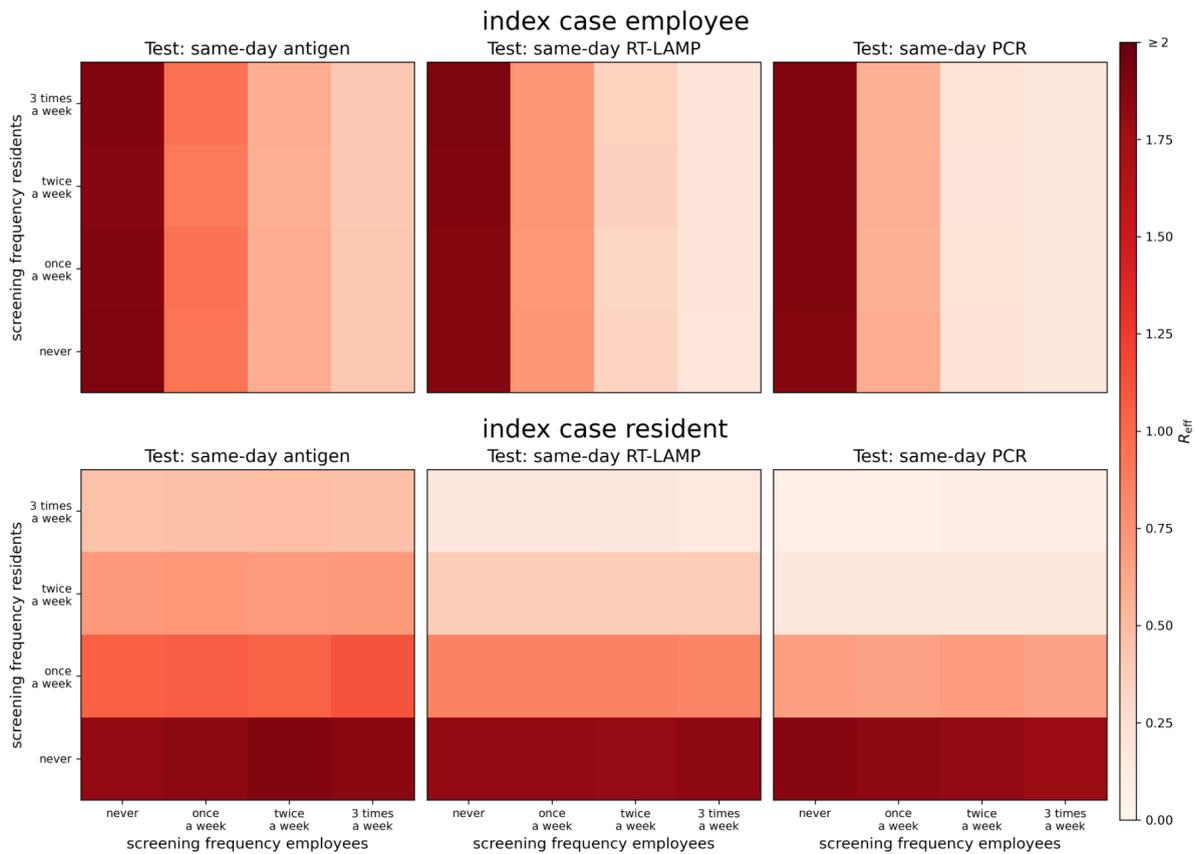

*Figure A7: effectiveness of different test technologies.* Values of $R_{eff}$ for a range of testing scenarios in nursing homes, investigating different testing technologies with their characteristic turnover times. Index cases: in the first row, infections are introduced by personnel, in the second row, by residents (typically after seeing visitors). Testing technology: in the first column, antigen tests with same-day turnover are used, in the middle column, RT-LAMP test with same-day turnover, and in the third column, PCR tests with same-day turnover. Preventive screening frequency: in each heatmap, preventive screening frequency of employees (x-axis) and residents (y-axis) is varied between no screening and one screening every two days. Results represent mean values of 5000 simulation runs per unique configuration.

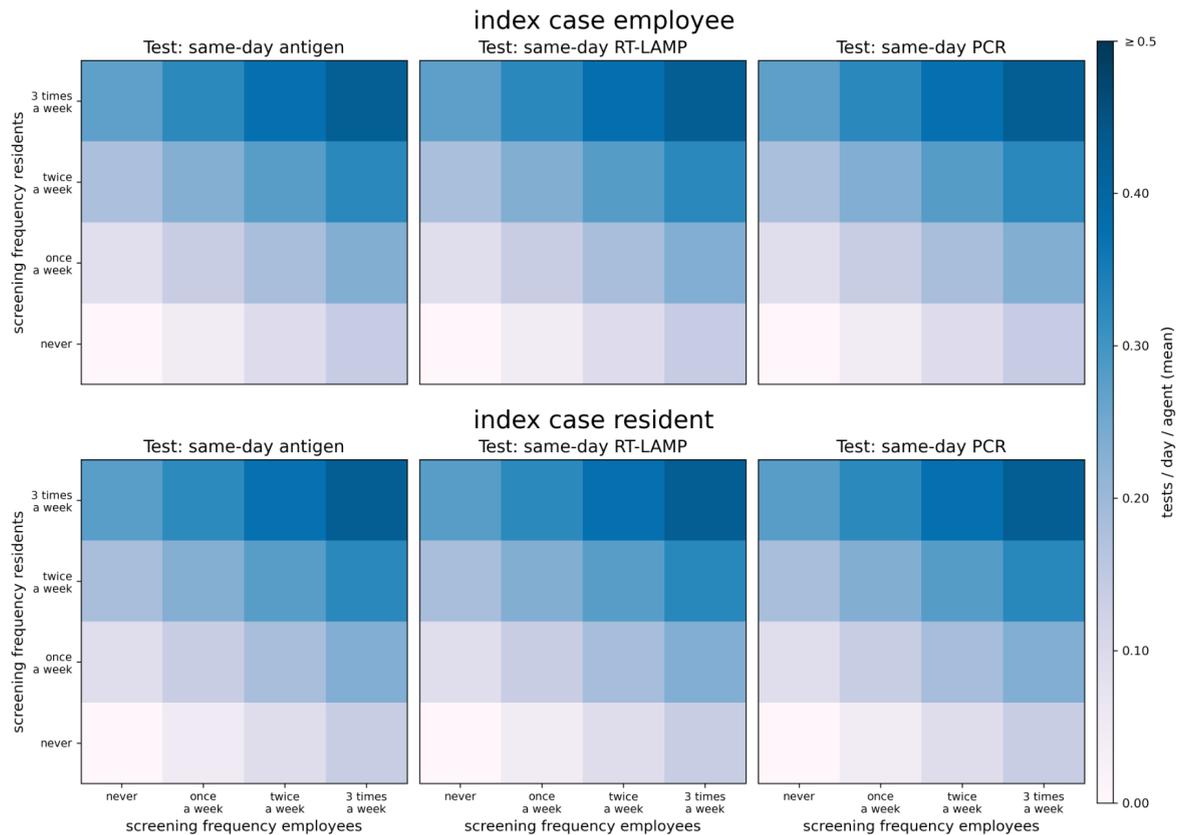

*Figure A8: test rate for different test technologies. Test rates (tests per agent per day) for a range of testing scenarios in nursing homes, investigating different testing technologies with their characteristic turnover times. Index cases: in the first row, infections are introduced by personnel, in the second row, by residents (typically after seeing visitors). Testing technology: in the first column, antigen tests with same-day turnover are used, in the middle column, RT-LAMP test with same-day turnover, and in the third column, PCR tests with same-day turnover. Preventive screening frequency: in each heatmap, preventive screening frequency of employees (x-axis) and residents (y-axis) is varied between no screening and one screening every two days. Results represent mean values of 5000 simulation runs per unique configuration.*

# Test turnover time: $R_{eff}$ and test rates

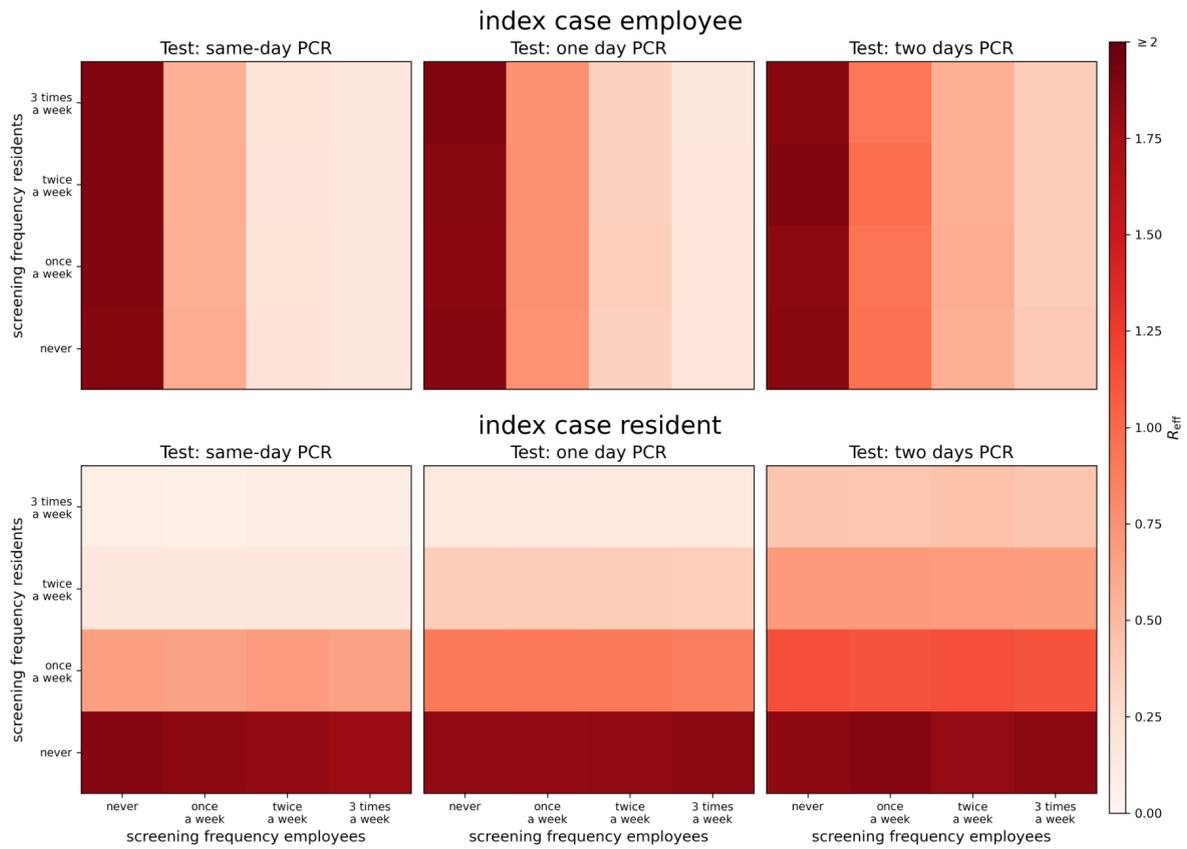

**Figure A9: effectiveness of different test turnover times.** *Values of $R_{eff}$ for a range of testing scenarios in nursing homes, investigating different test turnover times. Index cases: in the first row, infections are introduced by personnel, in the second row, by residents (resembling visitors). Test turnover time: in all scenarios, PCR tests are used. In the first column, tests have same-day turnover, in the middle column, tests have one-day turnover, and in the third column, tests have two-day turnover. Preventive screening frequency: in each heatmap, preventive screening frequency of employees (x-axis) and residents (y-axis) is varied between no screens and one screen every two days. Results represent mean values of 5000 simulation runs per unique configuration.*

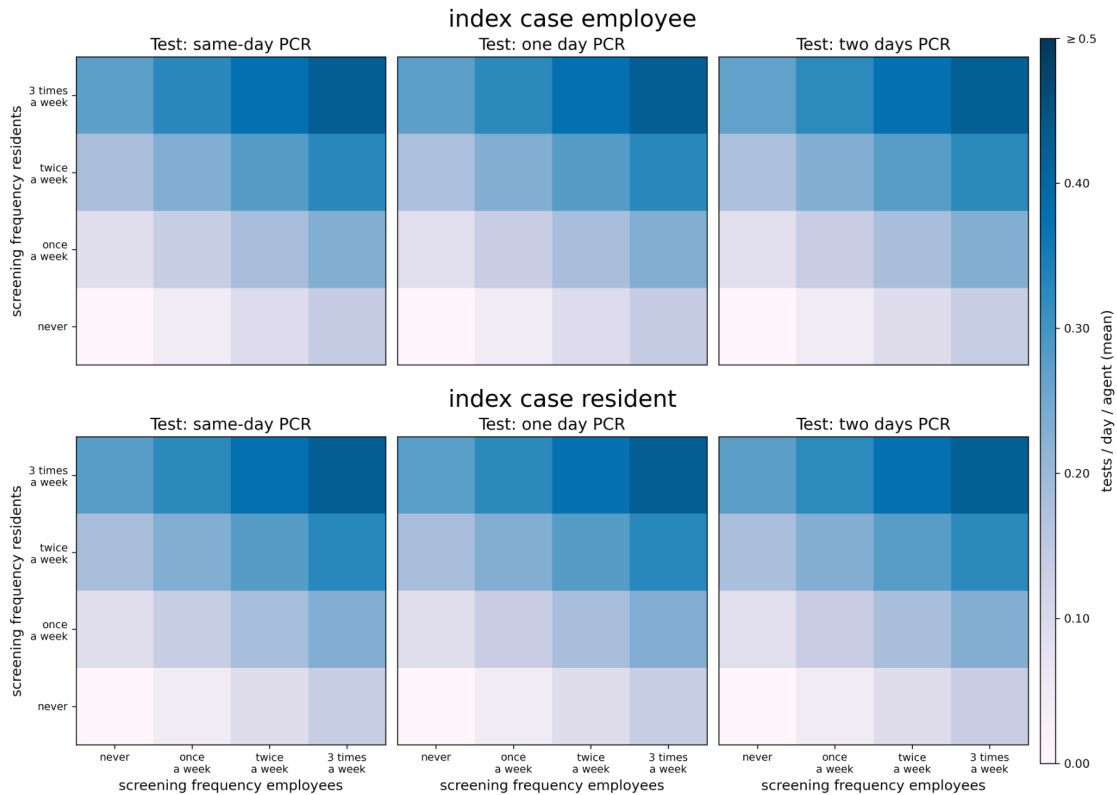

*Figure A10: test rate for different test turnover times. Test rates (tests per agent per day) for a range of testing scenarios in nursing homes, investigating different test turnover times. Index cases: in the first row, infections are introduced by personnel, in the second row, by residents (resembling visitors). Test turnover time: in all scenarios, PCR tests are used. In the first column, tests have same-day turnover, in the middle column, tests have one-day turnover, and in the third column, tests have two-day turnover. Preventive screening frequency: in each heatmap, preventive screening frequency of employees (x-axis) and residents (y-axis) is varied between no screens and one screen every two days. Results represent mean values of 5000 simulation runs per unique configuration.*

## Impact of increasing employee vaccination rates

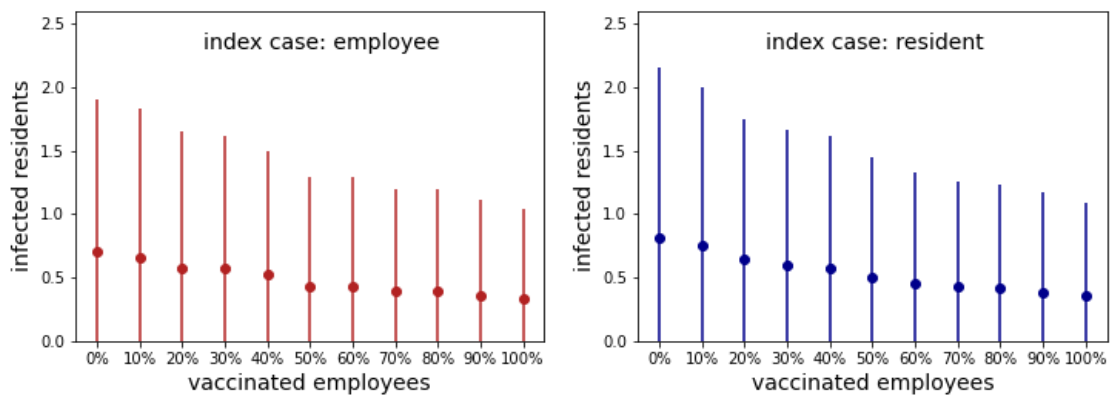

*Figure A11: impact of increasing employee vaccination rates. Average number of infected residents at a fixed resident vaccination ratio of 80% if employee vaccination ratios are increased from 0% to 100% in 10%-increments. Every data point shows the average outbreak size of 5000 randomly initiated simulation runs and its standard deviation.*

# Detailed testing strategy and vaccination rate results

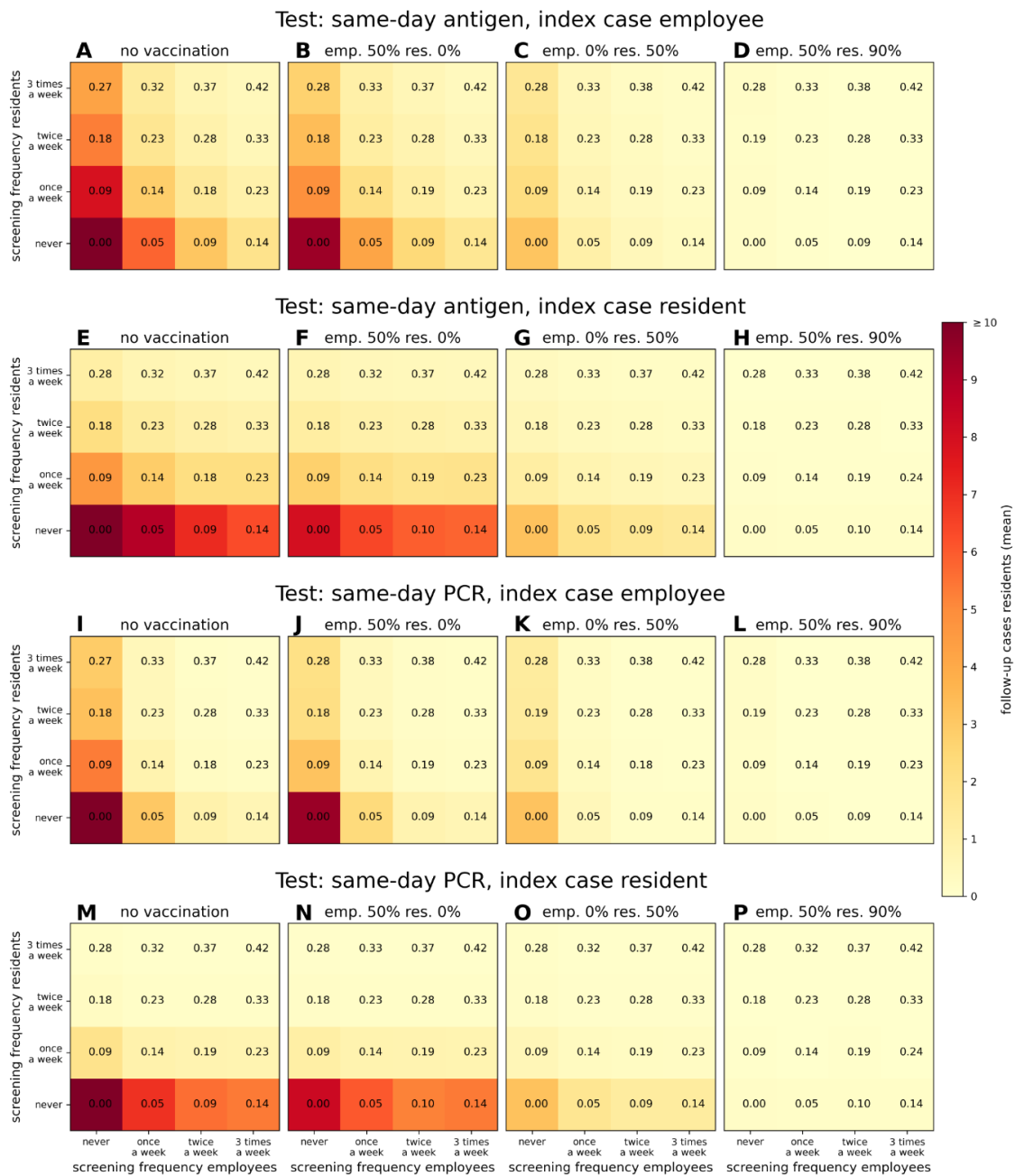

*Figure A12: detailed testing strategy and vaccination scenario results. Preventive screening with antigen tests with same-day turnover panels **(A)** to **(D)** (employee index case) and **(E)** to **(H)** (resident index case). Preventive screening with PCR tests with same-day turnover panels **(I)** to **(L)** (employee index case) and **(M)** to **(P)** (resident index case). Different vaccination scenarios are shown in panels **(A)**, **(E)**, **(I)**, **(M)** (no vaccinations), **(B)**, **(F)**, **(J)**, **(N)** (50% of employees and 0% of residents vaccinated), **(C)**, **(G)**, **(K)**, **(O)** (0% of employees and 50% of residents vaccinated) and **(D)**, **(H)**, **(L)**, **(P)** (50% of employees and 90% of residents vaccinated). Within every panel, color indicates outbreak sizes from small (blue) to large (red). Rows indicate varying employee testing frequency from never to 3 times a week, and columns indicate varying resident testing frequency from never to 3 times a week. Every panel tile is the average outbreak size of 5000 randomly initiated simulation runs.*

# SI note 5: Case studies of outbreaks in Austrian nursing homes

In the following, we describe the detailed chronology of 4 outbreaks in separate housing wards of two Austrian nursing homes[5]. We chose to treat outbreaks in different wards of the same nursing home as distinct outbreaks. We think this choice is warranted since wards are asked to not share staff as precaution to prevent disease spread and residents do not frequently interact with residents of other wards. In rare cases, staff is shared due to logistic issues such as short-term replacements for staff on sick leave and it is likely that in the cases described here, the virus was introduced to another ward of the home this way. Nevertheless, we think that the assumption that contacts between wards are few is justified and introduction of the infection through an employee of another ward can be modelled similarly to an infection by an employee that was infected outside the nursing home.

In addition to the four cases described here, we analyzed four other outbreaks in Austrian nursing homes. In each of those cases, an employee was confirmed to be the index case. By now, some Austrian nursing homes have implemented semi-regular staff screening strategies. Through these screens, 10 previously undetected infections in employees could be uncovered and an outbreak in the respective homes prevented. Given this evidence, we think it is warranted to assume that infections in nursing homes are primarily introduced by employees.

## Case 1

This outbreak occurred in a ward of a nursing home that houses 11 residents. The total number of employees that have regular contact with residents is unknown. Of the 11 residents, 3 and 8 share a table during joint meals, respectively. Residents live in single rooms that share a bathroom with one other resident. The first positive test was recorded on April 3rd 2020. In the ensuing outbreak, 8 out of the 11 residents (73%) and 2 out of an unknown total number of employees were infected. The chronology of the outbreak is described in tab. A4.

| Day since first positive test | Newly positive employees | Newly positive residents |
|---|---|---|
| 0 | 1 | 0 |
| 1 | 0 | 1 |
| 4 | 1 | 3 |
| 6 | 0 | 1 |

---

[5] Tables with the number of infected employees and residents for each of the four outbreaks are also available in the corresponding data repository at https://doi.org/10.17605/OSF.IO/HYD4R.

| | | |
|---|---|---|
| 7 | 0 | 2 |
| 27 | 0 | 1 |
| **sum** | **2** | **8** |

*Table A4: Chronology of an outbreak in a ward of an Austrian nursing home. The index case was most likely introduced by an employee, since the home had implemented a no-visitation policy 21 days prior to the first case. Also three weeks prior to the first positive case, all employees started wearing at least FFP2 masks and gloves when tending to residents. Following the first positive case, all inhabitants were quarantined and confined to their rooms.*

## Case 2

This outbreak occurred in a ward of a nursing home that housed 35 residents at the time and had approximately 18 employees that had regular contact to residents. Tables were shared in groups of 6 (1 table), 5 (2 tables), 4 (1 table), 3 (2 tables) and 2 (2 tables). In addition, there are two single tables and some immobile residents receive their meals in their rooms. Eleven residents live in single-rooms and 24 residents live in double rooms that share a bathroom with another double room. The first positive test was recorded on June 1st 2020. In the ensuing outbreak, 19 out of the 35 residents (54%) and 6 out of 18 employees (33%) were infected. The chronology of the outbreak is described in tab. A5.

| Day since first positive test | Newly positive employees | Newly positive residents |
|---|---|---|
| 0 | 1 | 1 |
| 3 | 0 | 1 |
| 4 | 0 | 9 |
| 5 | 3 | 0 |
| 7 | 0 | 1 |
| 8 | 0 | 1 |
| 9 | 0 | 1 |
| 11 | 1 | 1 |
| 12 | 0 | 1 |
| 14 | 0 | 3 |
| 16 | 1 | 0 |
| **sum** | **6** | **19** |

*Table A5: Chronology of an outbreak in a ward of an Austrian nursing home. The first two positive cases were an employee and a resident, and it is unclear whether an employee or a visitor introduced the virus to the facility. Before the first positive case, employees were wearing surgical face masks. After the first positive case, all employees started wearing at least FFP2 masks and gloves when tending to residents and seating arrangements at joint meals were changed to a maximum of two residents per table.*

## Case 3

This outbreak occurred in a ward of a nursing home with a similar number of residents and staff as well as housing conditions as in the setting described in case 2. Tables were shared in groups of 4 (5 tables), 3 (1 table) and 2 (2 tables). Some immobile residents receive their meals in their rooms. This ward belongs to the same nursing home as the ward described in case 2 and the infection was most likely introduced by employees that had contact with residents in both wards. The first positive test was recorded on June 9$^{th}$ 2020. In the ensuing outbreak, 5 out of the 35 residents (14%) and 1 out of 18 employees (6%) were infected. The chronology of the outbreak is described in tab. A6.

| Day since first positive test | Newly positive employees | Newly positive residents |
|:---:|:---:|:---:|
| 0 | 0 | 1 |
| 1 | 1 | 2 |
| 3 | 0 | 1 |
| 7 | 0 | 1 |
| **sum** | **1** | **5** |

*Table A6: **Chronology of an outbreak in a ward of an Austrian nursing home.** The first positive case was a resident that was most likely infected by an employee from another ward of the same nursing home that experienced an outbreak earlier. Before the first positive case, employees were wearing surgical face masks. After the first positive case, all employees started wearing at least FFP2 masks and gloves when tending to residents and seating arrangements at joint meals were changed to a maximum of two residents per table.*

# Case 4

This outbreak occurred in a ward of a nursing home that housed 34 residents at the time. We do not have information about the number of employees that were in regular contact with the residents and about the seating arrangements at joint meals. All residents live in double-rooms. This ward belongs to the same nursing home as the ward described in case 2 and the infection was most likely introduced by employees that had contact to residents in both wards. The first positive test was recorded on June 10$^{th}$ 2020. In the ensuing outbreak, 6 out of the 34 residents (18%) and 4 employees were infected. The chronology of the outbreak is described in tab. A7.

| Day since first positive test | Positive employees | Positive residents |
|---|---|---|
| 0 | 1 | 0 |
| 2 | 1 | 4 |
| 5 | 0 | 2 |
| 16 | 2 | 0 |
| **sum** | **4** | **6** |

*Table A7: **Chronology of an outbreak in a ward of an Austrian nursing home.** The first positive case was an employee that was most likely infected by an employee from another ward of the same nursing home that experienced an outbreak earlier. Before the first positive case, employees were wearing surgical face masks. After the first positive case, all employees started wearing at least FFP2 masks and gloves when tending to residents.*

# SI note 6: Model results with the strain dominant in Austria in spring 2020.

## No interventions and TTI

In the absence of non-pharmaceutical interventions or containment measures, $R_{eff}$ = 1.60 ± 1.52 if an employee is the index case and $R_{eff}$ = 2.04 ± 1.74 if a resident is the index case. Mean outbreak sizes are 15.9 ± 13.7 if an employee is the index case and 17.8 ± 12.8 if a resident is the index case. In a scenario in which only TTI is implemented, our model yields reproduction numbers of $R_{eff}$ = 1.23 ± 1.30 if an employee is the index case, and $R_{eff}$ = 1.23 ± 1.32 if a resident is the index case and mean outbreak sizes of 5.2 ± 7.1 and 5.7 ± 6.6 for employee and resident index cases, respectively.

## Tables

The tables below are also available in the data repository associated with this work at https://doi.org/10.17605/OSF.IO/HYD4R in expanded form.

| Staff screens per week | Resident screens per week | Outbreak size mean ± std [infected residents] | Outbreak size [10th; 90th] percentile | $R_{eff}$ mean ± std | Test rate mean ± std [tests / day / person] |
|---|---|---|---|---|---|
| Antigen tests with same-day turnover ||||||
| 3 | 1 | 0.4 ± 1.3 | 0.0 [0.0; 1.0] | 0.26 ± 0.59 | 0.23 ± 0.03 |
| 3 | never | 0.7 ± 2.3 | 0.0 [0.0; 1.0] | 0.27 ± 0.61 | 0.14 ± 0.02 |
| 2 | 1 | 0.6 ± 1.7 | 0.0 [0.0; 2.0] | 0.38 ± 0.71 | 0.19 ± 0.03 |
| 2 | never | 1.1 ± 3.0 | 0.0 [0.0; 3.0] | 0.39 ± 0.72 | 0.09 ± 0.02 |
| 1 | 1 | 1.1 ± 2.3 | 0.0 [0.0; 4.0] | 0.62 ± 0.92 | 0.14 ± 0.04 |
| 1 | never | 1.9 ± 4.1 | 0.0 [0.0; 7.0] | 0.62 ± 0.92 | 0.05 ± 0.02 |
| never | 1 | 3.0 ± 4.2 | 1.0 [0.0; 9.0] | 1.27 ± 1.28 | 0.09 ± 0.02 |
| never | never | 5.3 ± 7.1 | 2.0 [0.0; 17.0] | 1.25 ± 1.30 | 0.00 ± 0.01 |
| LAMP tests with same-day turnover ||||||
| 3 | 1 | 0.2 ± 0.8 | 0.0 [0.0; 0.0] | 0.13 ± 0.46 | 0.23 ± 0.03 |
| 3 | never | 0.3 ± 1.5 | 0.0 [0.0; 0.0] | 0.12 ± 0.45 | 0.14 ± 0.02 |
| 2 | 1 | 0.3 ± 1.1 | 0.0 [0.0; 1.0] | 0.22 ± 0.56 | 0.19 ± 0.04 |
| 2 | never | 0.5 ± 2.0 | 0.0 [0.0; 1.0] | 0.22 ± 0.55 | 0.09 ± 0.02 |
| 1 | 1 | 0.7 ± 1.6 | 0.0 [0.0; 2.0] | 0.47 ± 0.80 | 0.14 ± 0.04 |
| 1 | never | 1.3 ± 3.4 | 0.0 [0.0; 5.0] | 0.48 ± 0.83 | 0.05 ± 0.02 |
| never | 1 | 2.5 ± 3.6 | 1.0 [0.0; 7.0] | 1.28 ± 1.32 | 0.09 ± 0.02 |
| never | never | 5.2 ± 7.0 | 1.0 [0.0; 17.0] | 1.26 ± 1.29 | 0.00 ± 0.01 |
| PCR tests with same-day turnover | | | | | |
| 3 | 1 | 0.1 ± 0.6 | 0.0 [0.0; 0.0] | 0.11 ± 0.46 | 0.23 ± 0.04 |
| 3 | never | 0.2 ± 1.4 | 0.0 [0.0; 0.0] | 0.11 ± 0.44 | 0.14 ± 0.02 |
| 2 | 1 | 0.1 ± 0.7 | 0.0 [0.0; 0.0] | 0.14 ± 0.48 | 0.18 ± 0.04 |
| 2 | never | 0.3 ± 1.6 | 0.0 [0.0; 0.0] | 0.15 ± 0.50 | 0.09 ± 0.02 |
| 1 | 1 | 0.4 ± 1.2 | 0.0 [0.0; 2.0] | 0.38 ± 0.75 | 0.14 ± 0.04 |
| 1 | never | 1.0 ± 3.1 | 0.0 [0.0; 3.0] | 0.39 ± 0.77 | 0.05 ± 0.02 |
| never | 1 | 2.1 ± 3.0 | 1.0 [0.0; 6.0] | 1.28 ± 1.32 | 0.09 ± 0.02 |
| never | never | 5.1 ± 7.0 | 1.0 [0.0; 17.0] | 1.24 ± 1.28 | 0.00 ± 0.01 |
| PCR tests with one day turnover | | | | | |
| 3 | 1 | 0.2 ± 0.8 | 0.0 [0.0; 0.0] | 0.12 ± 0.42 | 0.23 ± 0.03 |
| 3 | never | 0.3 ± 1.5 | 0.0 [0.0; 0.0] | 0.11 ± 0.44 | 0.14 ± 0.02 |
| 2 | 1 | 0.3 ± 1.1 | 0.0 [0.0; 1.0] | 0.24 ± 0.57 | 0.19 ± 0.04 |
| 2 | never | 0.6 ± 2.1 | 0.0 [0.0; 1.0] | 0.24 ± 0.58 | 0.09 ± 0.02 |
| 1 | 1 | 0.8 ± 1.8 | 0.0 [0.0; 3.0] | 0.52 ± 0.90 | 0.14 ± 0.04 |
| 1 | never | 1.5 ± 3.6 | 0.0 [0.0; 5.0] | 0.52 ± 0.88 | 0.05 ± 0.02 |
| never | 1 | 2.4 ± 3.5 | 1.0 [0.0; 7.0] | 1.25 ± 1.29 | 0.09 ± 0.02 |
| never | never | 5.2 ± 7.1 | 1.0 [0.0; 17.0] | 1.25 ± 1.30 | 0.00 ± 0.01 |
| PCR tests with two day turnover | | | | | |
| 3 | 1 | 0.4 ± 1.3 | 0.0 [0.0; 1.0] | 0.25 ± 0.58 | 0.23 ± 0.03 |
| 3 | never | 0.6 ± 2.3 | 0.0 [0.0; 1.0] | 0.25 ± 0.59 | 0.14 ± 0.02 |
| 2 | 1 | 0.6 ± 1.6 | 0.0 [0.0; 2.0] | 0.38 ± 0.70 | 0.18 ± 0.04 |
| 2 | never | 1.0 ± 2.9 | 0.0 [0.0; 3.0] | 0.38 ± 0.72 | 0.09 ± 0.02 |
| 1 | 1 | 1.1 ± 2.4 | 0.0 [0.0; 4.0] | 0.62 ± 0.92 | 0.14 ± 0.04 |
| 1 | never | 1.9 ± 4.1 | 0.0 [0.0; 7.0] | 0.64 ± 0.95 | 0.05 ± 0.02 |
| never | 1 | 2.8 ± 4.1 | 1.0 [0.0; 9.0] | 1.25 ± 1.31 | 0.09 ± 0.02 |
| never | never | 5.3 ± 7.1 | 1.0 [0.0; 17.0] | 1.27 ± 1.30 | 0.00 ± 0.01 |

*Table A8: Strain dominant in Austria in spring 2020 with employee index cases. Mean outbreak sizes alongside median, 10th and 90th percentile outbreak ranges, $R_{eff}$ and test rate for scenarios, in which employees undergo preventive testing never, once, two times or three times a week, and residents undergo preventive testing never or once a week. Preventive testing is performed using either antigen tests, RT-LAMP tests or PCR tests, all with same-day results turnover. Values are calculated from simulations with 5000 randomly initialized runs per scenario.*

| Staff screens per week | Resident screens per week | Outbreak size mean ± std [infected residents] | Outbreak size [10th; 90th] percentile | $R_{eff}$ mean ± std | Test rate mean ± std [tests / day / person] |
|---|---|---|---|---|---|
| **Antigen tests with same-day turnover** | | | | | |
| 3 | 1 | 1.9 ± 1.9 | 1.0 [1.0; 4.0] | 0.72 ± 1.06 | 0.24 ± 0.03 |
| 3 | never | 3.6 ± 4.1 | 2.0 [1.0; 10.0] | 1.24 ± 1.35 | 0.14 ± 0.02 |
| 2 | 1 | 2.0 ± 2.0 | 1.0 [1.0; 4.0] | 0.72 ± 1.04 | 0.19 ± 0.03 |
| 2 | never | 3.7 ± 4.2 | 2.0 [1.0; 10.0] | 1.21 ± 1.32 | 0.10 ± 0.01 |
| 1 | 1 | 2.2 ± 2.4 | 1.0 [1.0; 5.0] | 0.76 ± 1.07 | 0.14 ± 0.03 |
| 1 | never | 4.2 ± 4.9 | 2.0 [1.0; 12.0] | 1.23 ± 1.34 | 0.05 ± 0.01 |
| never | 1 | 2.6 ± 3.3 | 1.0 [1.0; 6.0] | 0.71 ± 1.01 | 0.09 ± 0.02 |
| never | never | 5.7 ± 6.8 | 2.0 [1.0; 17.0] | 1.23 ± 1.33 | 0.00 ± 0.01 |
| **LAMP tests with same-day turnover** | | | | | |
| 3 | 1 | 1.6 ± 1.4 | 1.0 [1.0; 3.0] | 0.57 ± 0.94 | 0.24 ± 0.03 |
| 3 | never | 3.4 ± 3.8 | 2.0 [1.0; 9.0] | 1.24 ± 1.33 | 0.14 ± 0.02 |
| 2 | 1 | 1.6 ± 1.5 | 1.0 [1.0; 3.0] | 0.56 ± 0.95 | 0.19 ± 0.03 |
| 2 | never | 3.6 ± 4.1 | 2.0 [1.0; 10.0] | 1.23 ± 1.32 | 0.10 ± 0.02 |
| 1 | 1 | 1.7 ± 1.6 | 1.0 [1.0; 4.0] | 0.58 ± 0.96 | 0.14 ± 0.04 |
| 1 | never | 4.0 ± 4.7 | 2.0 [1.0; 11.0] | 1.24 ± 1.37 | 0.05 ± 0.01 |
| never | 1 | 2.1 ± 2.6 | 1.0 [1.0; 5.0] | 0.57 ± 0.96 | 0.09 ± 0.03 |
| never | never | 5.8 ± 6.8 | 2.0 [1.0; 18.0] | 1.25 ± 1.36 | 0.00 ± 0.01 |
| **PCR tests with same-day turnover** | | | | | |
| 3 | 1 | 1.4 ± 1.2 | 1.0 [1.0; 3.0] | 0.45 ± 0.87 | 0.24 ± 0.03 |
| 3 | never | 3.3 ± 3.7 | 2.0 [1.0; 8.0] | 1.22 ± 1.33 | 0.14 ± 0.02 |
| 2 | 1 | 1.5 ± 1.2 | 1.0 [1.0; 3.0] | 0.46 ± 0.85 | 0.19 ± 0.03 |
| 2 | never | 3.4 ± 4.0 | 2.0 [1.0; 9.0] | 1.25 ± 1.35 | 0.10 ± 0.01 |
| 1 | 1 | 1.5 ± 1.3 | 1.0 [1.0; 3.0] | 0.48 ± 0.89 | 0.14 ± 0.04 |
| 1 | never | 3.7 ± 4.4 | 2.0 [1.0; 10.0] | 1.24 ± 1.35 | 0.05 ± 0.01 |
| never | 1 | 1.8 ± 1.9 | 1.0 [1.0; 3.0] | 0.45 ± 0.86 | 0.09 ± 0.03 |
| never | never | 5.8 ± 6.8 | 2.0 [1.0; 17.0] | 1.23 ± 1.33 | 0.00 ± 0.01 |
| **PCR tests with one day turnover** | | | | | |
| 3 | 1 | 1.7 ± 1.5 | 1.0 [1.0; 3.0] | 0.61 ± 0.98 | 0.23 ± 0.03 |
| 3 | never | 3.3 ± 3.8 | 2.0 [1.0; 9.0] | 1.25 ± 1.32 | 0.14 ± 0.02 |
| 2 | 1 | 1.6 ± 1.4 | 1.0 [1.0; 3.0] | 0.60 ± 0.99 | 0.19 ± 0.03 |
| 2 | never | 3.5 ± 4.0 | 2.0 [1.0; 9.0] | 1.27 ± 1.35 | 0.10 ± 0.01 |
| 1 | 1 | 1.8 ± 1.7 | 1.0 [1.0; 4.0] | 0.59 ± 0.96 | 0.14 ± 0.04 |
| 1 | never | 3.8 ± 4.5 | 2.0 [1.0; 11.0] | 1.19 ± 1.32 | 0.05 ± 0.01 |
| never | 1 | 2.2 ± 2.7 | 1.0 [1.0; 5.0] | 0.61 ± 0.99 | 0.09 ± 0.03 |
| never | never | 5.9 ± 6.9 | 2.0 [1.0; 18.0] | 1.27 ± 1.39 | 0.00 ± 0.01 |
| **PCR tests with two day turnover** | | | | | |
| 3 | 1 | 2.0 ± 2.0 | 1.0 [1.0; 4.0] | 0.76 ± 1.08 | 0.23 ± 0.03 |
| 3 | never | 3.5 ± 4.1 | 2.0 [1.0; 9.0] | 1.23 ± 1.33 | 0.14 ± 0.02 |
| 2 | 1 | 2.0 ± 1.9 | 1.0 [1.0; 4.0] | 0.74 ± 1.07 | 0.19 ± 0.03 |
| 2 | never | 3.7 ± 4.3 | 2.0 [1.0; 10.0] | 1.25 ± 1.37 | 0.09 ± 0.02 |
| 1 | 1 | 2.2 ± 2.4 | 1.0 [1.0; 5.0] | 0.79 ± 1.10 | 0.14 ± 0.03 |
| 1 | never | 4.2 ± 4.9 | 2.0 [1.0; 12.0] | 1.24 ± 1.35 | 0.05 ± 0.01 |
| never | 1 | 2.7 ± 3.4 | 1.0 [1.0; 7.0] | 0.76 ± 1.09 | 0.09 ± 0.03 |
| never | never | 5.7 ± 6.7 | 2.0 [1.0; 17.0] | 1.23 ± 1.34 | 0.00 ± 0.01 |

*Table A9 strain dominant in Austria in spring 2020 with resident index cases:* Mean outbreak sizes alongside median, 10th and 90th percentile outbreak ranges, $R_{eff}$ and test rate for scenarios, in which employees undergo preventive testing never, once, two times or three times a week, and residents undergo preventive testing never or once a week. Preventive testing is performed using either antigen tests, RT-LAMP tests or PCR tests, all with same-day results turnover. Values are calculated from simulations with 5000 randomly initialized runs per scenario.

# Figures

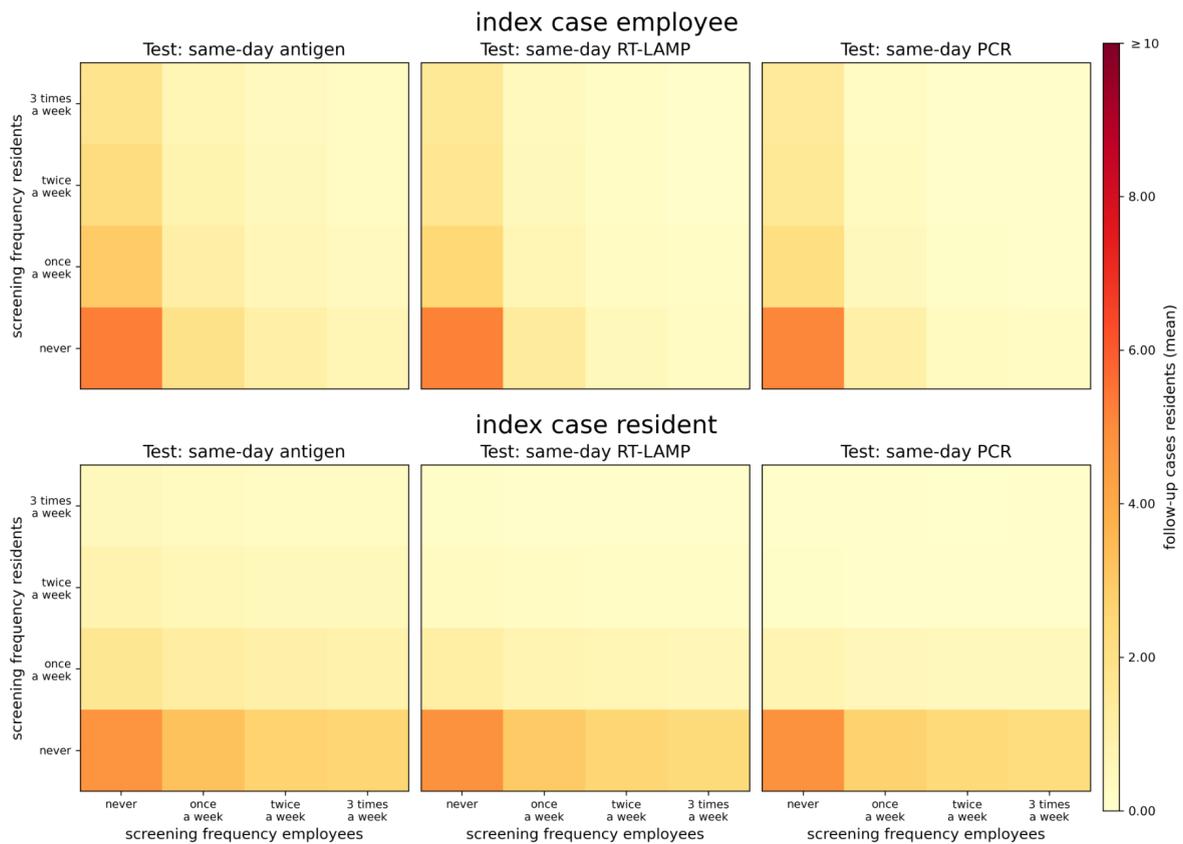

***Figure A13: outbreak sizes for different test technologies and wild-type virus variant.*** *Mean outbreak sizes for a range of testing scenarios in nursing homes, investigating different testing technologies with their characteristic turnover times. Index cases: in the first row, infections are introduced by personnel, in the second row, by residents (typically after seeing visitors). Testing technology: in the first column, antigen tests with same-day turnover are used, in the middle column, RT-LAMP tests with same-day turnover, and in the third column, PCR tests with same-day turnover. Preventive screening frequency: in each heatmap, preventive screening frequency of employees (x-axis) and residents (y-axis) is varied between no screening and one screening every two days. Results represent mean values of 5000 simulation runs per unique configuration.*

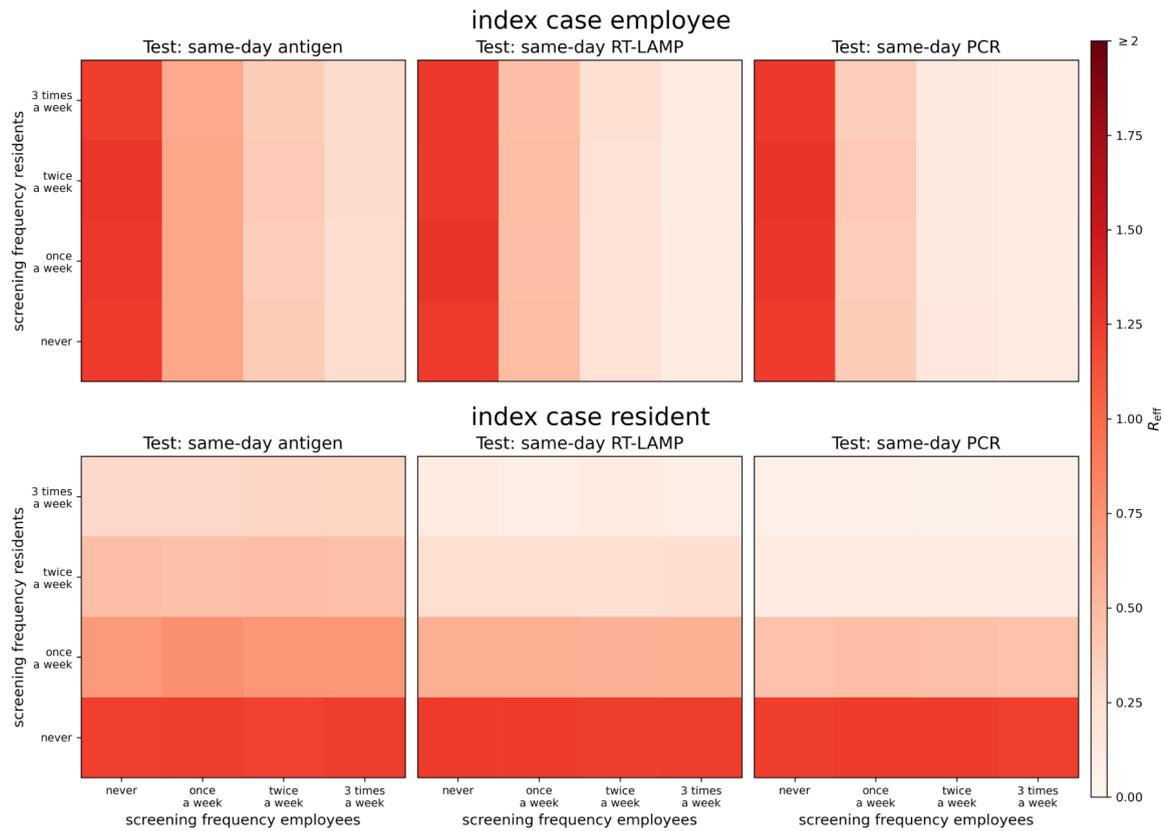

*Figure A14: $R_{eff}$ for different test technologies & strain dominant in Austria in spring 2020. Values of $R_{eff}$ for a range of testing scenarios in nursing homes, investigating different testing technologies with their characteristic turnover times. Index cases: in the first row, infections are introduced by personnel, in the second row, by residents (typically after seeing visitors). Testing technology: in the first column, antigen tests with same-day turnover are used, in the middle column, RT-LAMP test with same-day turnover, and in the third column, PCR tests with same-day turnover. Preventive screening frequency: in each heatmap, preventive screening frequency of employees (x-axis) and residents (y-axis) is varied between no screening and one screening every two days. Results represent mean values of 5000 simulation runs per unique configuration.*

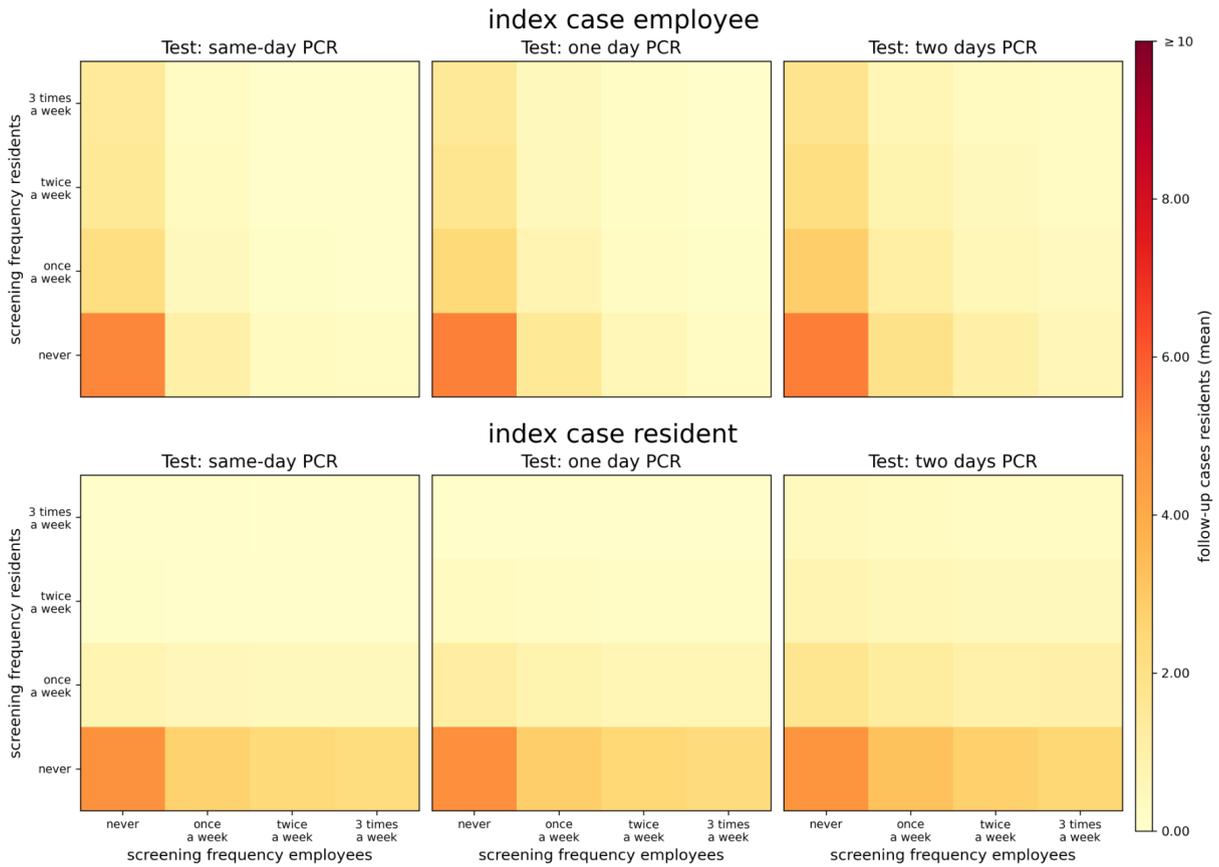

*Figure A15: outbreak sizes for different test turnover times and wild-type virus variant.* Mean outbreak sizes for a range of testing scenarios in nursing homes investigating different test turnover times. Index cases: in the first row, infections are introduced by personnel, in the second row, by residents (resembling visitors). Test turnover time: in all scenarios, PCR tests are used. In the first column, tests have same-day turnover, in the middle column, tests have one-day turnover, and in the third column, tests have two-day turnover. Preventive screening frequency: in each heatmap, preventive screening frequency of employees (x-axis) and residents (y-axis) is varied between no screens and one screen every two days. Results represent mean values of 5000 simulation runs per unique configuration.

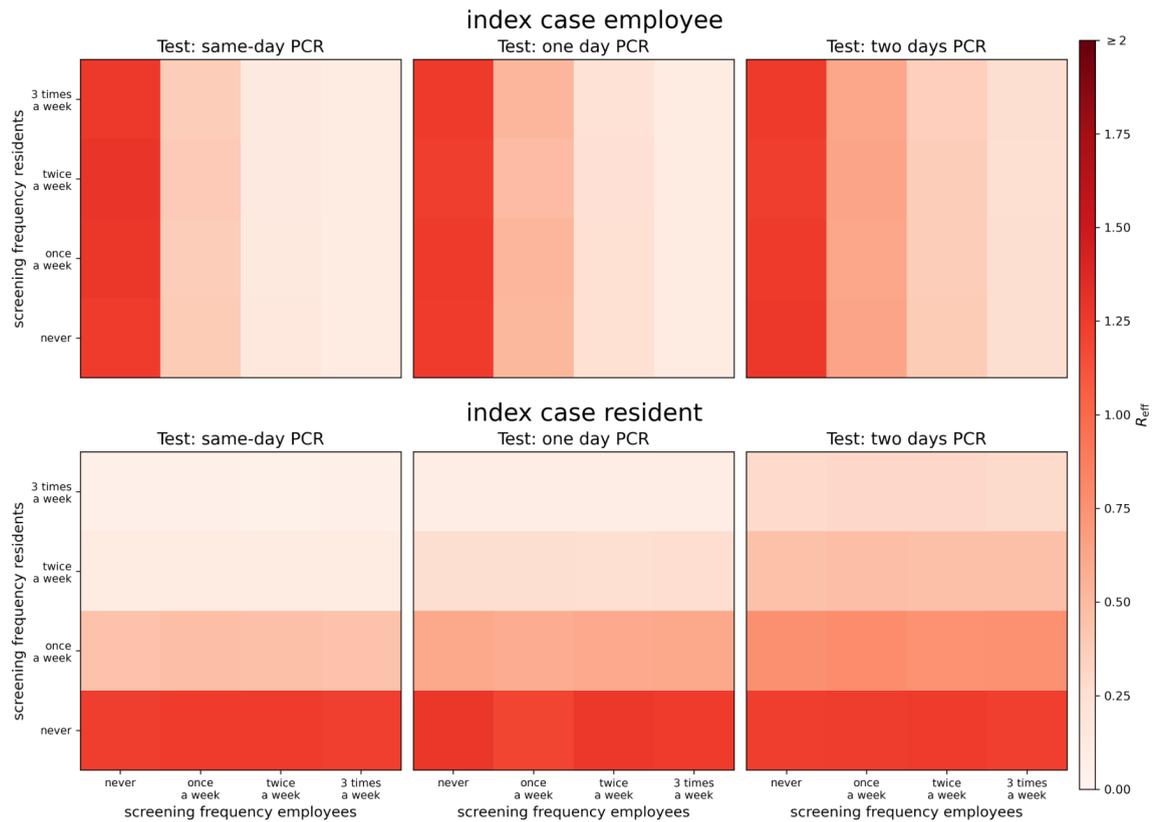

*Figure A16 R<sub>eff</sub> for different test turnover times & wild-type:* Values of $R_{eff}$ for a range of testing scenarios in nursing homes, investigating different test turnover times. Index cases: in the first row, infections are introduced by personnel, in the second row, by residents (resembling visitors). Test turnover time: in all scenarios, PCR tests are used. In the first column, tests have same-day turnover, in the middle column, tests have one-day turnover, and in the third column, tests have two-day turnover. Preventive screening frequency: in each heatmap, preventive screening frequency of employees (x-axis) and residents (y-axis) is varied between no screens and one screen every two days. Results represent mean values of 5000 simulation runs per unique configuration.

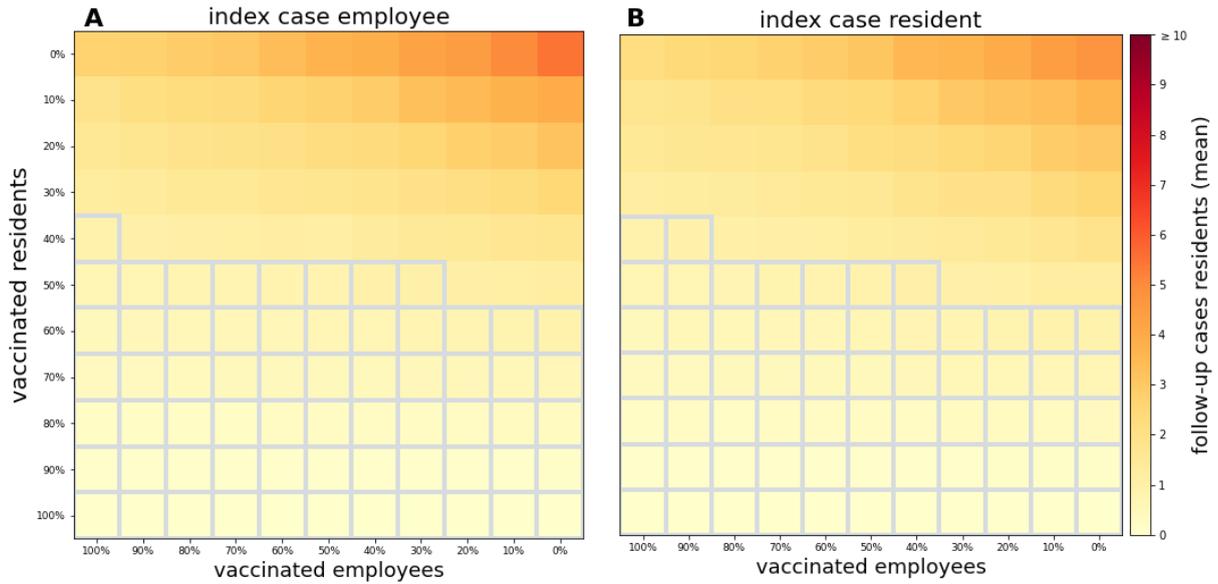

*Figure A17: Outbreak sizes for different ratios of vaccinated employees and residents and wild-type virus variant. Outbreak sizes are indicated from low (blue) to high (red) for (A) employee index cases and (B) resident index cases. Vaccination ratios for which the mean number of resident follow-up cases is < 1 are indicated with grey borders. Outbreak sizes for each combination of (employee, resident) vaccination ratio are averages over 5000 randomly initialised simulation runs each. In addition to vaccinations, the model also implements TTI.*

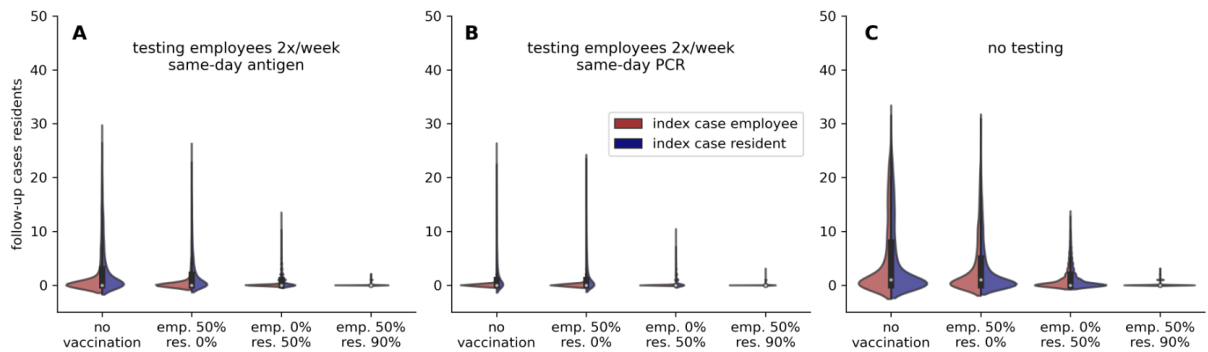

*Figure A18 Strain dominant in Austria in spring 2020: Distributions of the number of infected residents for different vaccination scenarios and testing strategies. The left (red) part of the violins indicates employee index cases, the right (blue) part of the violins indicates resident index cases. (A) the testing strategy consists of TTI and preventive testing of employees two times a week with antigen tests with a same-day turnover. (B) the testing strategy consists of TTI and preventive testing of employees two times a week with PCR tests with same-day turnover. (C) the testing strategy consists of TTI only. For every testing strategy, four vaccination scenarios (no vaccination, 50% of employees, 50% of residents, 50% of employees and 90% of residents) are shown. The plot shows distributions of outbreak sizes from 5000 randomly initialised simulation runs per testing strategy and vaccination scenario.*

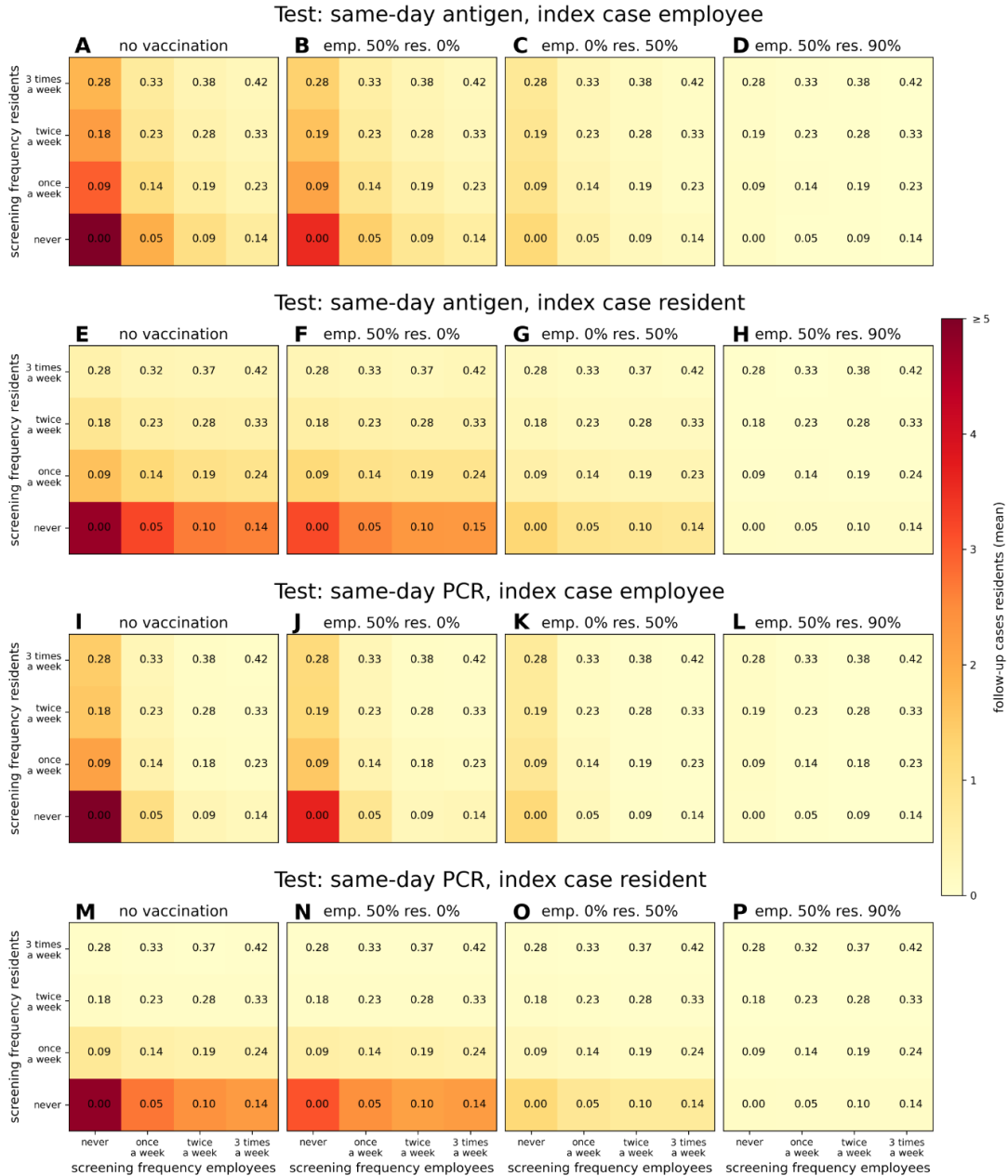

*Figure A19: Detailed testing strategy and vaccination scenario results and wild-type virus variant.* Preventive screening with antigen tests with same-day turnover panels **(A)** to **(D)** (employee index case) and **(E)** to **(H)** (resident index case). Preventive screening with PCR tests with same-day turnover panels **(I)** to **(L)** (employee index case) and **(M)** to **(P)** (resident index case). Different vaccination scenarios are shown in panels **(A)**, **(E)**, **(I)**, **(M)** (no vaccinations), **(B)**, **(F)**, **(J)**, **(N)** (50% of employees and 0% of residents vaccinated), **(C)**, **(G)**, **(K)**, **(O)** (0% of employees and 50% of residents vaccinated) and **(D)**, **(H)**, **(L)**, **(P)** (50% of employees and 90% of residents vaccinated). Within every panel, color indicates outbreak sizes from small (blue) to large (red). Rows indicate varying employee testing frequency from never to 3 times a week, and columns indicate varying resident testing frequency from never to 3 times a week. Every panel tile is the average outbreak size of 5000 randomly initiated simulation runs. Please note the difference in color scale when compared to fig. A12.

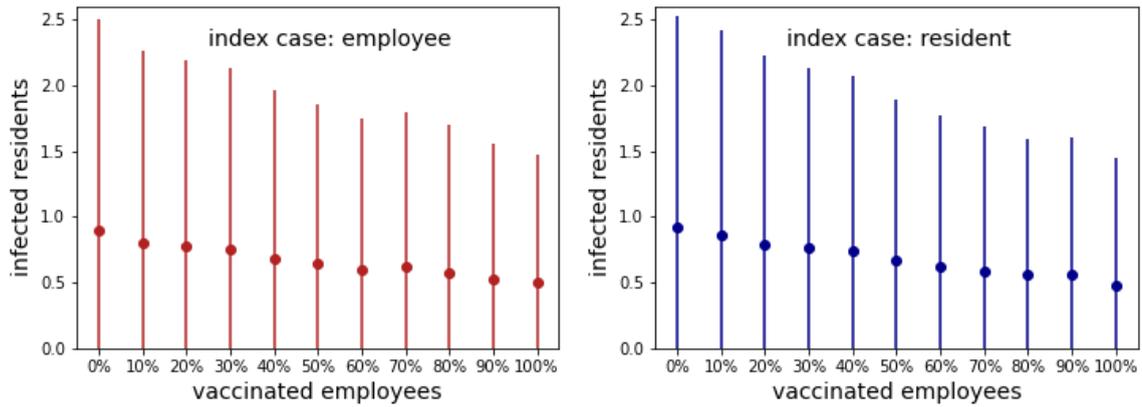

*Figure A20 strain dominant in Austria in spring 2020: impact of increasing employee vaccination rates.* Average number of infected residents at a fixed resident vaccination ratio of 60% (note: in fig. A6, this number is 80%), if employee vaccination ratios are increased from 0% to 100% in 10%-increments. Every data point shows the average outbreak size of 5000 randomly initiated simulation runs and its standard deviation.

# SI note 7: Model results with protective gear for employees and the B.1.1.7 strain

## TTI

In a scenario in which only TTI is implemented, our model yields reproduction numbers of $R_{eff}$ = 0.87 ± 1.03 if an employee is the index case, and $R_{eff}$ = 1.69 ± 1.64 if a resident is the index case and mean outbreak sizes of 5.4 ± 8.4 and 8.7 ± 8.8 for employee and resident index cases, respectively.

## Tables

The tables below are also available in the data repository associated with this work at https://doi.org/10.17605/OSF.IO/HYD4R in extended form.

| Staff screens per week | Resident screens per week | Outbreak size mean ± std [infected residents] | Outbreak size [10th; 90th] percentile | $R_{eff}$ mean ± std | Test rate mean ± std [tests / day / person] |
|---|---|---|---|---|---|
| \multicolumn{6}{c}{Antigen tests with same-day turnover} ||||||
| 3 | 1 | 0.4 ± 1.5 | 0.0 [0.0; 1.0] | 0.18 ± 0.47 | 0.23 ± 0.03 |
| 3 | never | 0.9 ± 3.5 | 0.0 [0.0; 1.0] | 0.18 ± 0.46 | 0.14 ± 0.02 |
| 2 | 1 | 0.6 ± 2.0 | 0.0 [0.0; 1.0] | 0.26 ± 0.57 | 0.19 ± 0.04 |
| 2 | never | 1.3 ± 4.2 | 0.0 [0.0; 3.0] | 0.27 ± 0.57 | 0.09 ± 0.02 |
| 1 | 1 | 1.0 ± 2.6 | 0.0 [0.0; 3.0] | 0.42 ± 0.72 | 0.14 ± 0.04 |
| 1 | never | 2.2 ± 5.5 | 0.0 [0.0; 9.0] | 0.41 ± 0.73 | 0.05 ± 0.02 |
| never | 1 | 2.3 ± 4.2 | 0.0 [0.0; 8.0] | 0.83 ± 1.01 | 0.09 ± 0.03 |
| never | never | 5.1 ± 8.2 | 0.0 [0.0; 20.0] | 0.84 ± 1.02 | 0.00 ± 0.01 |
| \multicolumn{6}{c}{LAMP tests with same-day turnover} ||||||
| 3 | 1 | 0.1 ± 0.9 | 0.0 [0.0; 0.0] | 0.08 ± 0.34 | 0.23 ± 0.04 |
| 3 | never | 0.3 ± 2.0 | 0.0 [0.0; 0.0] | 0.07 ± 0.31 | 0.14 ± 0.02 |
| 2 | 1 | 0.3 ± 1.2 | 0.0 [0.0; 1.0] | 0.16 ± 0.45 | 0.18 ± 0.04 |
| 2 | never | 0.6 ± 3.0 | 0.0 [0.0; 0.0] | 0.14 ± 0.42 | 0.09 ± 0.02 |
| 1 | 1 | 0.6 ± 1.7 | 0.0 [0.0; 2.0] | 0.32 ± 0.65 | 0.14 ± 0.04 |
| 1 | never | 1.6 ± 4.7 | 0.0 [0.0; 4.0] | 0.34 ± 0.66 | 0.05 ± 0.02 |
| never | 1 | 1.9 ± 3.3 | 0.0 [0.0; 6.0] | 0.86 ± 1.05 | 0.09 ± 0.03 |
| never | never | 5.3 ± 8.3 | 0.0 [0.0; 20.0] | 0.84 ± 1.01 | 0.00 ± 0.01 |
| \multicolumn{6}{c}{PCR tests with same-day turnover} ||||||
| 3 | 1 | 0.1 ± 0.6 | 0.0 [0.0; 0.0] | 0.07 ± 0.33 | 0.23 ± 0.03 |
| 3 | never | 0.3 ± 2.0 | 0.0 [0.0; 0.0] | 0.07 ± 0.32 | 0.14 ± 0.02 |
| 2 | 1 | 0.1 ± 0.7 | 0.0 [0.0; 0.0] | 0.09 ± 0.36 | 0.19 ± 0.04 |
| 2 | never | 0.4 ± 2.2 | 0.0 [0.0; 0.0] | 0.09 ± 0.37 | 0.09 ± 0.02 |
| 1 | 1 | 0.4 ± 1.3 | 0.0 [0.0; 1.0] | 0.27 ± 0.60 | 0.14 ± 0.04 |
| 1 | never | 1.2 ± 4.2 | 0.0 [0.0; 2.0] | 0.26 ± 0.60 | 0.05 ± 0.02 |
| never | 1 | 1.5 ± 2.7 | 0.0 [0.0; 5.0] | 0.85 ± 1.04 | 0.09 ± 0.03 |
| never | never | 4.9 ± 8.0 | 0.0 [0.0; 19.0] | 0.84 ± 1.03 | 0.00 ± 0.01 |
| PCR tests with one day turnover | | | | | |
| 3 | 1 | 0.1 ± 0.9 | 0.0 [0.0; 0.0] | 0.08 ± 0.35 | 0.23 ± 0.04 |
| 3 | never | 0.4 ± 2.3 | 0.0 [0.0; 0.0] | 0.08 ± 0.34 | 0.14 ± 0.02 |
| 2 | 1 | 0.3 ± 1.2 | 0.0 [0.0; 1.0] | 0.15 ± 0.44 | 0.18 ± 0.04 |
| 2 | never | 0.7 ± 3.0 | 0.0 [0.0; 1.0] | 0.15 ± 0.44 | 0.09 ± 0.02 |
| 1 | 1 | 0.7 ± 1.9 | 0.0 [0.0; 2.0] | 0.36 ± 0.70 | 0.14 ± 0.04 |
| 1 | never | 1.6 ± 4.6 | 0.0 [0.0; 5.0] | 0.34 ± 0.67 | 0.05 ± 0.02 |
| never | 1 | 1.9 ± 3.3 | 0.0 [0.0; 6.0] | 0.87 ± 1.02 | 0.09 ± 0.02 |
| never | never | 5.0 ± 8.0 | 0.0 [0.0; 20.0] | 0.84 ± 1.03 | 0.00 ± 0.01 |
| PCR tests with two day turnover | | | | | |
| 3 | 1 | 0.3 ± 1.4 | 0.0 [0.0; 1.0] | 0.17 ± 0.45 | 0.23 ± 0.03 |
| 3 | never | 0.9 ± 3.5 | 0.0 [0.0; 1.0] | 0.17 ± 0.48 | 0.14 ± 0.02 |
| 2 | 1 | 0.6 ± 2.0 | 0.0 [0.0; 1.0] | 0.26 ± 0.58 | 0.18 ± 0.04 |
| 2 | never | 1.1 ± 3.9 | 0.0 [0.0; 2.0] | 0.25 ± 0.56 | 0.09 ± 0.02 |
| 1 | 1 | 1.0 ± 2.7 | 0.0 [0.0; 3.0] | 0.42 ± 0.74 | 0.14 ± 0.04 |
| 1 | never | 2.2 ± 5.5 | 0.0 [0.0; 9.0] | 0.44 ± 0.75 | 0.05 ± 0.02 |
| never | 1 | 2.5 ± 4.3 | 0.0 [0.0; 8.0] | 0.87 ± 1.03 | 0.09 ± 0.02 |
| never | never | 5.2 ± 8.2 | 0.0 [0.0; 20.0] | 0.84 ± 1.01 | 0.00 ± 0.01 |

*Table A10: B.1.1.7 and protective gear for employees, with employee index cases.* Mean outbreak sizes alongside median, 10th and 90th percentile outbreak ranges, $R_{eff}$ and test rate for scenarios, in which employees undergo preventive testing never, once, two times or three times a week, and residents undergo preventive testing never or once a week. Preventive testing is performed using either antigen tests, RT-LAMP tests or PCR tests, all with same-day results turnover. Values are calculated from simulations with 5000 randomly initialized runs per scenario.

| Staff screens per week | Resident screens per week | Outbreak size mean ± std [infected residents] | Outbreak size [10th; 90th] percentile | $R_{eff}$ mean ± std | Test rate mean ± std [tests / day / person] |
|---|---|---|---|---|---|
| **Antigen tests with same-day turnover** | | | | | |
| 3 | 1 | 2.8 ± 3.1 | 1.0 [1.0; 7.0] | 0.99 ± 1.28 | 0.23 ± 0.03 |
| 3 | never | 6.4 ± 7.0 | 3.0 [1.0; 18.0] | 1.65 ± 1.65 | 0.14 ± 0.02 |
| 2 | 1 | 2.8 ± 3.2 | 1.0 [1.0; 7.0] | 0.99 ± 1.29 | 0.19 ± 0.03 |
| 2 | never | 6.5 ± 7.2 | 3.0 [1.0; 19.0] | 1.60 ± 1.60 | 0.09 ± 0.01 |
| 1 | 1 | 3.0 ± 3.6 | 1.0 [1.0; 8.0] | 0.98 ± 1.29 | 0.14 ± 0.03 |
| 1 | never | 7.3 ± 7.8 | 3.0 [1.0; 20.0] | 1.66 ± 1.61 | 0.05 ± 0.01 |
| never | 1 | 3.5 ± 4.3 | 1.0 [1.0; 9.0] | 0.98 ± 1.26 | 0.09 ± 0.02 |
| never | never | 8.5 ± 8.8 | 4.0 [1.0; 23.0] | 1.64 ± 1.61 | 0.00 ± 0.01 |
| **LAMP tests with same-day turnover** | | | | | |
| 3 | 1 | 2.1 ± 2.3 | 1.0 [1.0; 5.0] | 0.77 ± 1.15 | 0.23 ± 0.03 |
| 3 | never | 6.2 ± 6.9 | 3.0 [1.0; 18.0] | 1.65 ± 1.63 | 0.14 ± 0.02 |
| 2 | 1 | 2.1 ± 2.3 | 1.0 [1.0; 5.0] | 0.78 ± 1.19 | 0.19 ± 0.03 |
| 2 | never | 6.4 ± 7.1 | 3.0 [1.0; 18.0] | 1.64 ± 1.59 | 0.09 ± 0.01 |
| 1 | 1 | 2.3 ± 2.6 | 1.0 [1.0; 5.0] | 0.74 ± 1.14 | 0.14 ± 0.04 |
| 1 | never | 6.8 ± 7.5 | 3.0 [1.0; 19.0] | 1.66 ± 1.64 | 0.05 ± 0.01 |
| never | 1 | 2.5 ± 2.9 | 1.0 [1.0; 6.0] | 0.79 ± 1.17 | 0.09 ± 0.03 |
| never | never | 8.7 ± 8.9 | 4.0 [1.0; 23.0] | 1.69 ± 1.66 | 0.00 ± 0.01 |
| **PCR tests with same-day turnover** | | | | | |
| 3 | 1 | 1.7 ± 1.7 | 1.0 [1.0; 3.0] | 0.62 ± 1.09 | 0.23 ± 0.03 |
| 3 | never | 6.0 ± 6.8 | 3.0 [1.0; 17.0] | 1.61 ± 1.57 | 0.14 ± 0.02 |
| 2 | 1 | 1.8 ± 1.7 | 1.0 [1.0; 4.0] | 0.60 ± 1.06 | 0.19 ± 0.03 |
| 2 | never | 6.1 ± 6.8 | 3.0 [1.0; 17.0] | 1.68 ± 1.64 | 0.09 ± 0.01 |
| 1 | 1 | 1.8 ± 1.8 | 1.0 [1.0; 4.0] | 0.60 ± 1.05 | 0.14 ± 0.04 |
| 1 | never | 6.6 ± 7.3 | 3.0 [1.0; 19.0] | 1.65 ± 1.60 | 0.05 ± 0.01 |
| never | 1 | 2.0 ± 2.3 | 1.0 [1.0; 4.0] | 0.61 ± 1.06 | 0.09 ± 0.03 |
| never | never | 8.8 ± 9.0 | 4.0 [1.0; 23.0] | 1.68 ± 1.62 | 0.00 ± 0.01 |
| **PCR tests with one day turnover** | | | | | |
| 3 | 1 | 2.2 ± 2.4 | 1.0 [1.0; 5.0] | 0.82 ± 1.21 | 0.23 ± 0.03 |
| 3 | never | 6.2 ± 6.9 | 3.0 [1.0; 17.0] | 1.65 ± 1.65 | 0.14 ± 0.02 |
| 2 | 1 | 2.1 ± 2.4 | 1.0 [1.0; 5.0] | 0.80 ± 1.16 | 0.19 ± 0.03 |
| 2 | never | 6.3 ± 7.0 | 3.0 [1.0; 18.0] | 1.63 ± 1.62 | 0.09 ± 0.02 |
| 1 | 1 | 2.4 ± 2.6 | 1.0 [1.0; 5.0] | 0.83 ± 1.22 | 0.14 ± 0.03 |
| 1 | never | 6.7 ± 7.4 | 3.0 [1.0; 19.0] | 1.64 ± 1.62 | 0.05 ± 0.01 |
| never | 1 | 2.6 ± 3.1 | 1.0 [1.0; 6.0] | 0.81 ± 1.22 | 0.09 ± 0.02 |
| never | never | 8.3 ± 8.7 | 4.0 [1.0; 23.0] | 1.63 ± 1.63 | 0.00 ± 0.01 |
| **PCR tests with two day turnover** | | | | | |
| 3 | 1 | 2.8 ± 3.2 | 1.0 [1.0; 7.0] | 1.03 ± 1.33 | 0.23 ± 0.03 |
| 3 | never | 6.5 ± 7.1 | 3.0 [1.0; 19.0] | 1.66 ± 1.60 | 0.14 ± 0.02 |
| 2 | 1 | 2.9 ± 3.3 | 1.0 [1.0; 7.0] | 1.03 ± 1.34 | 0.18 ± 0.03 |
| 2 | never | 6.6 ± 7.2 | 3.0 [1.0; 19.0] | 1.66 ± 1.63 | 0.09 ± 0.02 |
| 1 | 1 | 3.1 ± 3.6 | 1.0 [1.0; 8.0] | 1.03 ± 1.32 | 0.14 ± 0.03 |
| 1 | never | 7.1 ± 7.6 | 3.0 [1.0; 20.0] | 1.64 ± 1.63 | 0.05 ± 0.01 |
| never | 1 | 3.5 ± 4.2 | 1.0 [1.0; 9.0] | 1.04 ± 1.31 | 0.09 ± 0.02 |
| never | never | 8.7 ± 8.9 | 4.0 [1.0; 23.0] | 1.64 ± 1.59 | 0.00 ± 0.01 |

*Table A11 B.1.1.7 and protective gear for employees, with resident index cases:* Mean outbreak sizes alongside median, 10th and 90th percentile outbreak ranges, $R_{eff}$ and test rate for scenarios, in which employees undergo preventive testing never, once, two times or three times a week, and residents undergo preventive testing never or once a week. Preventive testing is performed using either antigen tests, RT-LAMP tests or PCR tests, all with same-day results turnover. Values are calculated from simulations with 5000 randomly initialized runs per scenario.

# Figures

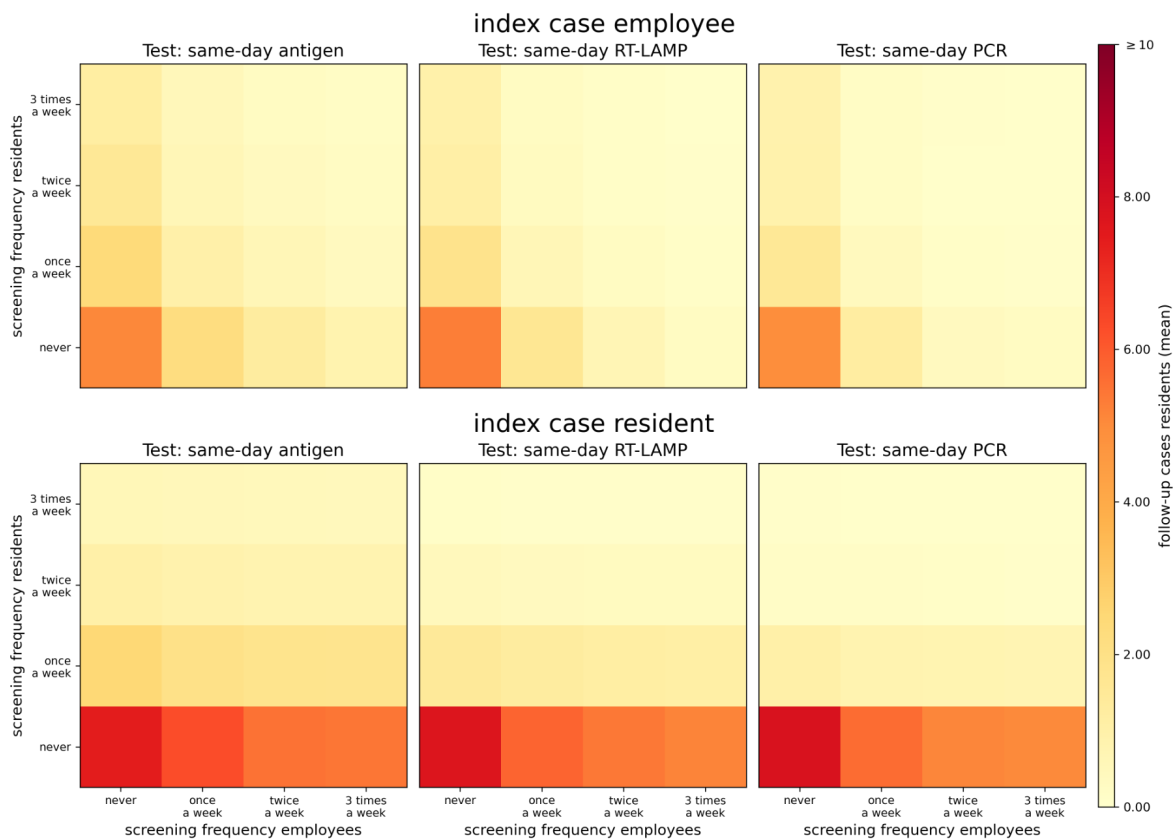

***Figure A21: outbreak sizes for different test technologies, B.1.1.7 and protective gear for employees.*** *Mean outbreak sizes for a range of testing scenarios in nursing homes, investigating different testing technologies with their characteristic turnover times. Index cases: in the first row, infections are introduced by personnel, in the second row, by residents (typically after seeing visitors). Testing technology: in the first column, antigen tests with same-day turnover are used, in the middle column, RT-LAMP tests with same-day turnover, and in the third column, PCR tests with same-day turnover. Preventive screening frequency: in each heatmap, preventive screening frequency of employees (x-axis) and residents (y-axis) is varied between no screening and one screening every two days. Results represent mean values of 5000 simulation runs per unique configuration.*

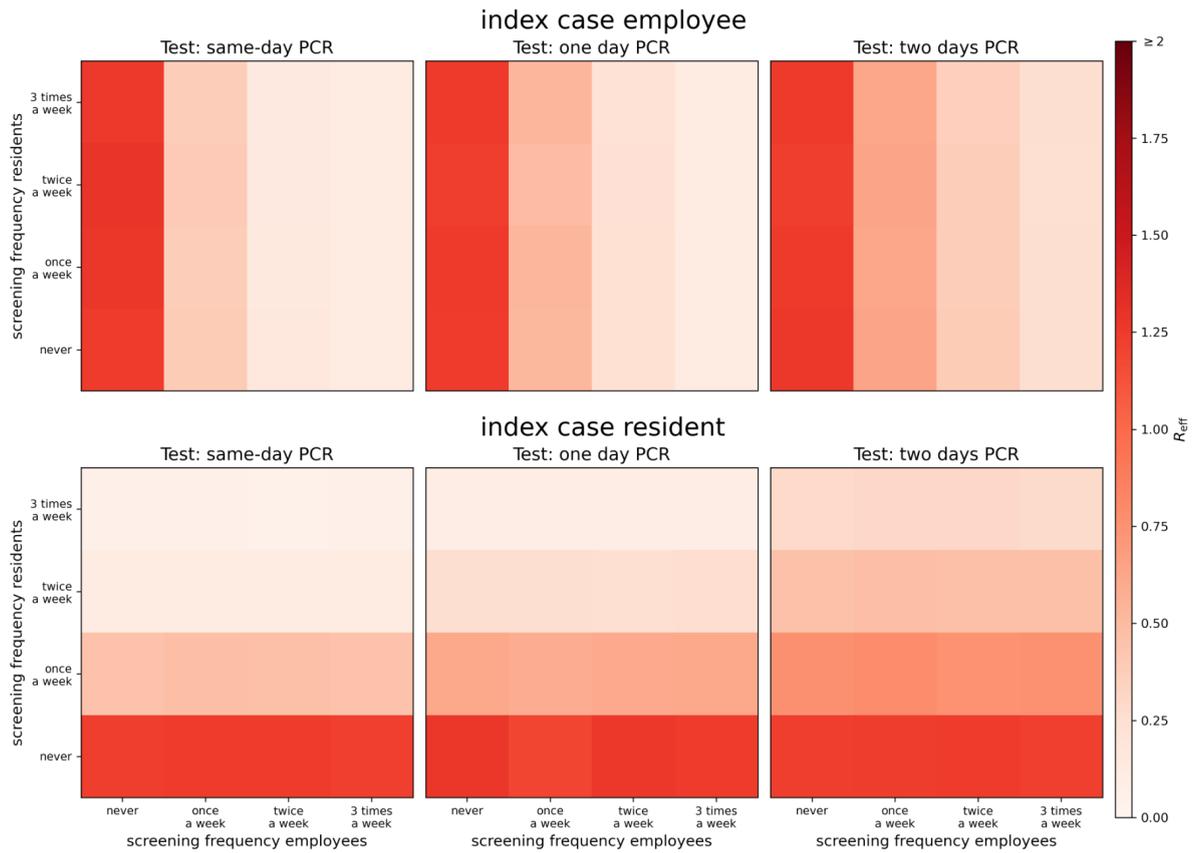

*Figure A22 $R_{eff}$ for different test technologies, B.1.1.7 and protective gear for employees.* Values of $R_{eff}$ for a range of testing scenarios in nursing homes, investigating different testing technologies with their characteristic turnover times. Index cases: in the first row, infections are introduced by personnel, in the second row, by residents (typically after seeing visitors). Testing technology: in the first column, antigen tests with same-day turnover are used, in the middle column, RT-LAMP test with same-day turnover, and in the third column, PCR tests with same-day turnover. Preventive screening frequency: in each heatmap, preventive screening frequency of employees (x-axis) and residents (y-axis) is varied between no screening and one screening every two days. Results represent mean values of 5000 simulation runs per unique configuration.

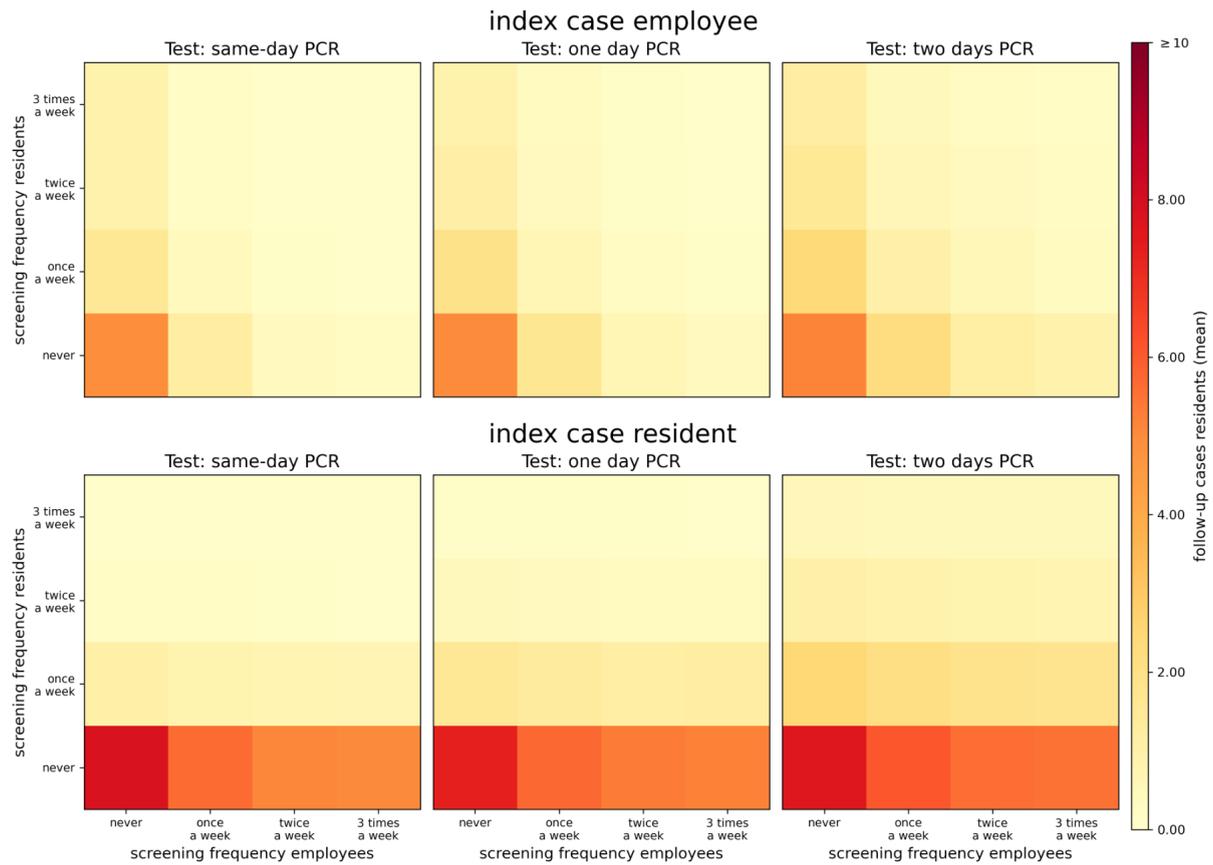

*Figure A23: outbreak sizes for different test turnover times, B.1.1.7 and protective gear for employees.* Mean outbreak sizes for a range of testing scenarios in nursing homes, investigating different testing technologies with their characteristic turnover times. Index cases: in the first row, infections are introduced by personnel, in the second row, by residents (typically after seeing visitors). Testing technology: in the first column, antigen tests with same-day turnover are used, in the middle column, RT-LAMP tests with same-day turnover, and in the third column, PCR tests with same-day turnover. Preventive screening frequency: in each heatmap, preventive screening frequency of employees (x-axis) and residents (y-axis) is varied between no screening and one screening every two days. Results represent mean values of 5000 simulation runs per unique configuration.

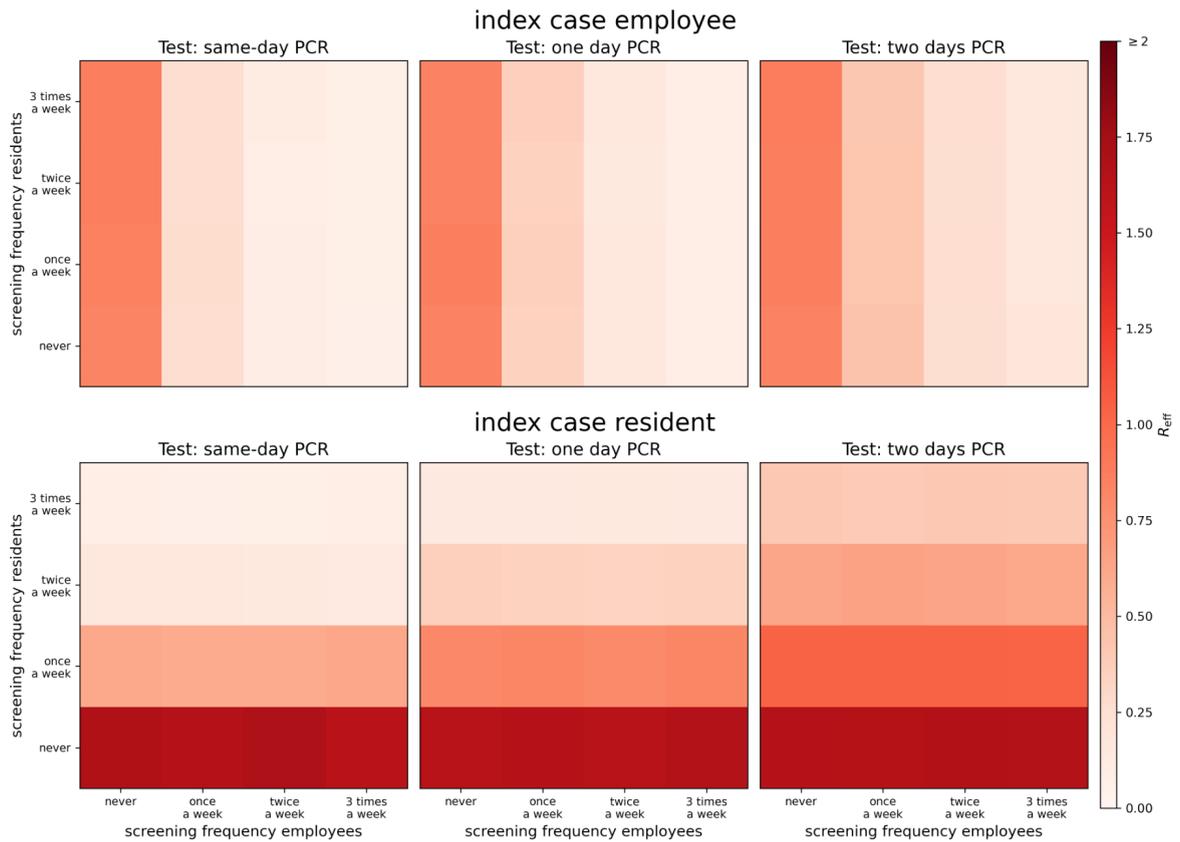

*Figure A24 $R_{eff}$ for different test turnover times, B.1.1.7 and protective gear for employees.* Values of $R_{eff}$ for a range of testing scenarios in nursing homes, investigating different testing technologies with their characteristic turnover times. Index cases: in the first row, infections are introduced by personnel, in the second row, by residents (typically after seeing visitors). Testing technology: in the first column, antigen tests with same-day turnover are used, in the middle column, RT-LAMP test with same-day turnover, and in the third column, PCR tests with same-day turnover. Preventive screening frequency: in each heatmap, preventive screening frequency of employees (x-axis) and residents (y-axis) is varied between no screening and one screening every two days. Results represent mean values of 5000 simulation runs per unique configuration.

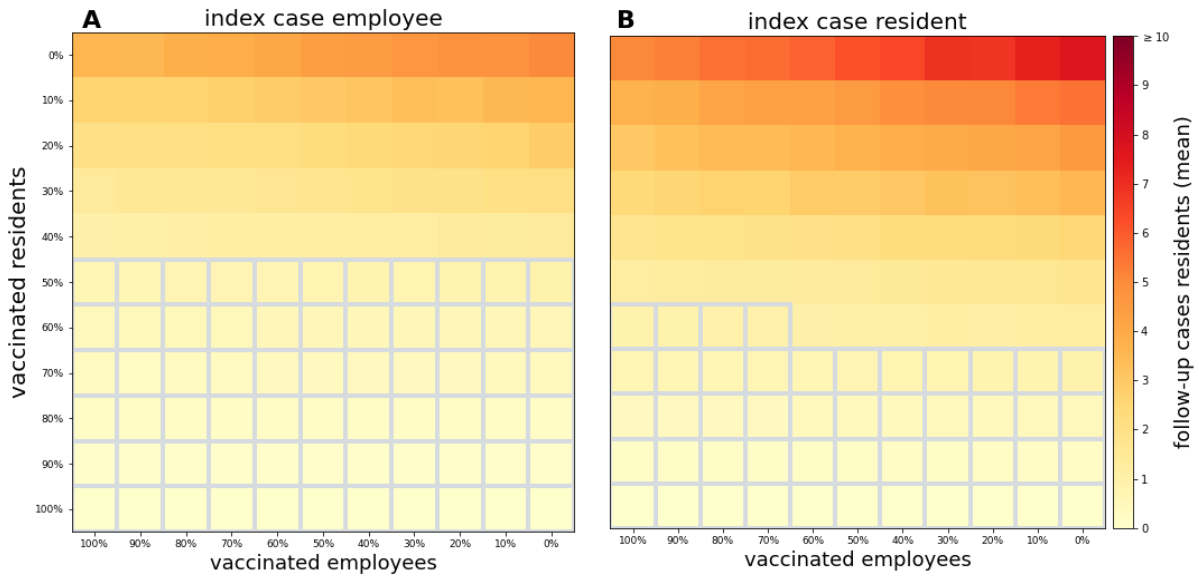

*Figure A25: Outbreak sizes for different ratios of vaccinated employees and residents with the B.1.1.7 variant and protective gear for employees. Outbreak sizes are indicated from low (blue) to high (red) for **(A)** employee index cases and **(B)** resident index cases. Vaccination ratios for which the mean number of resident follow-up cases is < 1 are indicated with grey borders. Outbreak sizes for each combination of (employee, resident) vaccination ratio are averages over 5000 randomly initialised simulation runs each. In addition to vaccinations, the model also implements TTI.*

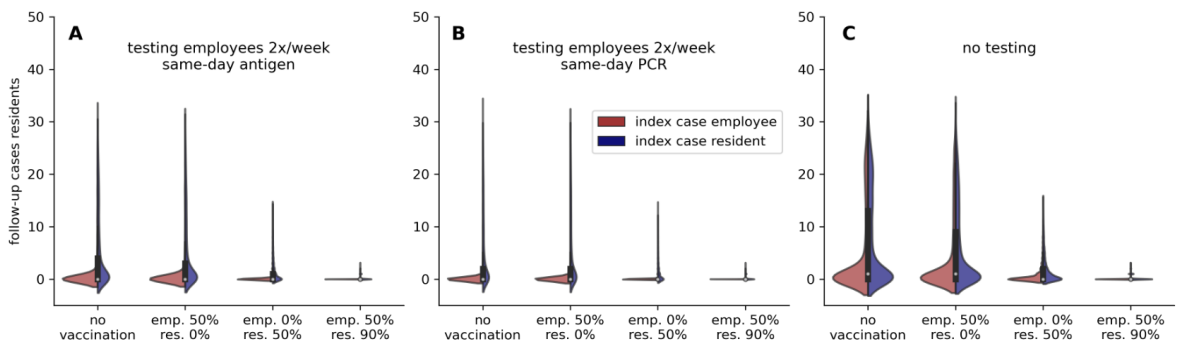

*Figure A26 B.1.1.7 and protective gear for employees: Distributions of the number of infected residents for different vaccination scenarios and testing strategies. The left (red) part of the violins indicates employee index cases, the right (blue) part of the violins indicates resident index cases. **(A)** the testing strategy consists of TTI and preventive testing of employees two times a week with antigen tests with a same-day turnover. **(B)** the testing strategy consists of TTI and preventive testing of employees two times a week with PCR tests with same-day turnover. **(C)** the testing strategy consists of TTI only. For every testing strategy, four vaccination scenarios (no vaccination, 50% of employees, 50% of residents, 50% of employees and 90% of residents) are shown. The plot shows distributions of outbreak sizes from 5000 randomly initialised simulation runs per testing strategy and vaccination scenario.*

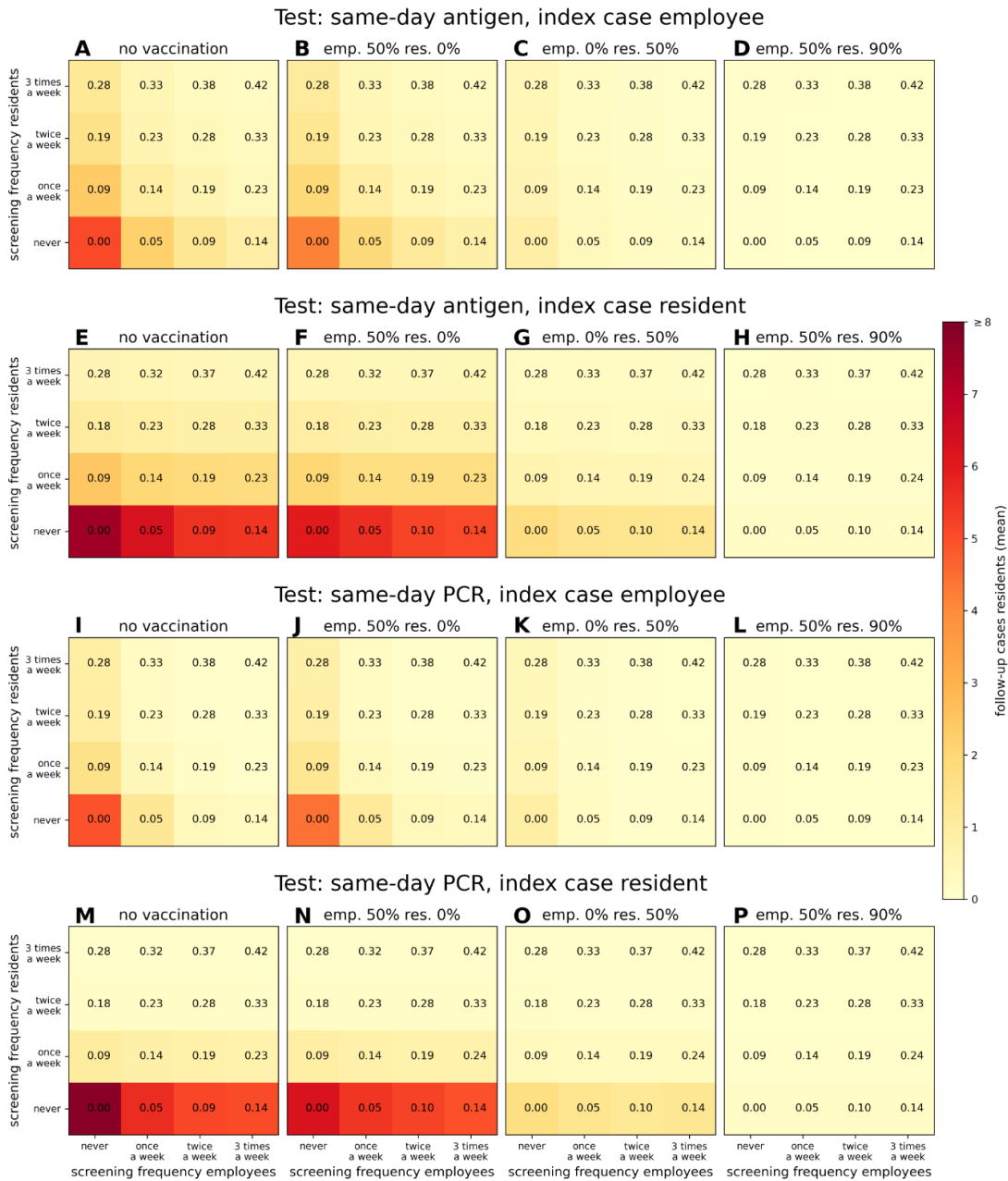

*Figure A27: Detailed testing strategy and vaccination scenario results with the B.1.1.7 variant and protective gear for employees.* Preventive screening with antigen tests with same-day turnover panels *(A)* to *(D)* (employee index case) and *(E)* to *(H)* (resident index case). Preventive screening with PCR tests with same-day turnover panels *(I)* to *(L)* (employee index case) and *(M)* to *(P)* (resident index case). Different vaccination scenarios are shown in panels *(A)*, *(E)*, *(I)*, *(M)* (no vaccinations), *(B)*, *(F)*, *(J)*, *(N)* (50% of employees and 0% of residents vaccinated), *(C)*, *(G)*, *(K)*, *(O)* (0% of employees and 50% of residents vaccinated) and *(D)*, *(H)*, *(L)*, *(P)* (50% of employees and 90% of residents vaccinated). Within every panel, color indicates outbreak sizes from small (blue) to large (red). Rows indicate varying employee testing frequency from never to 3 times a week, and columns indicate varying resident testing frequency from never to 3 times a week. Every panel tile is the average outbreak size of 5000 randomly initiated simulation runs. Please note the difference in color scale when compared to figures A7 and A12.

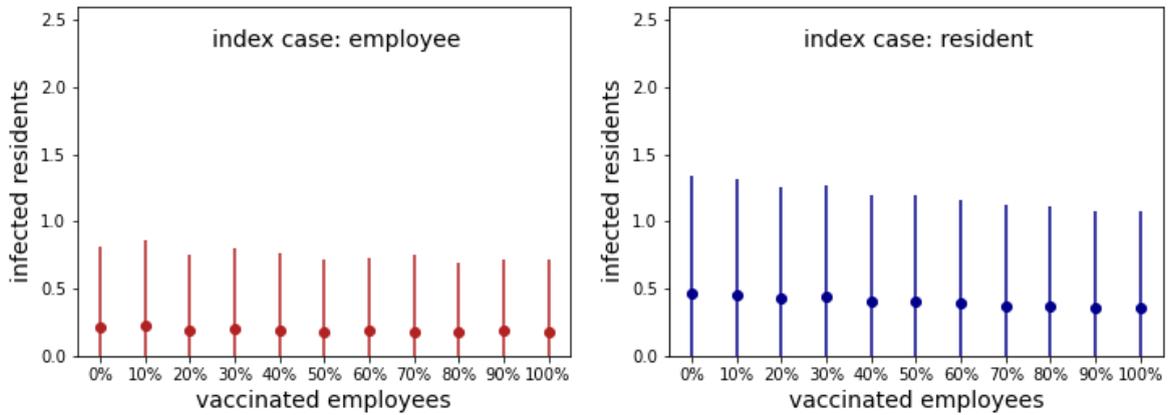

*Figure A28: Impact of increasing employee vaccination rates with the B.1.1.7 variant and protective gear for employees.* *Average number of infected residents at a fixed resident vaccination ratio of 60% (note: in fig A6, this number is 80%), if employee vaccination ratios are increased from 0% to 100% in 10%-increments. Every data point shows the average outbreak size of 5000 randomly initiated simulation runs and its standard deviation.*